\documentclass[a4paper,fleqn,usenatbib]{aa}

\usepackage{txfonts}
\usepackage{pdflscape}

\usepackage{ae,aecompl}

\usepackage{graphicx}	
\usepackage{amsmath}	
\usepackage{amssymb}	
\usepackage{orcidlink}

\newcommand{\nustar}{\textit{NuSTAR}~}
\newcommand{\swift}{\textit{Swift}~}
\newcommand{\eps}{erg s$^{-1}$~}

\newcommand{\pcm}{cm$^{-2}$~}

\newcommand{\kms}{km s$^{-1}$~}
\newcommand{\nh}{$N_{\rm H}$~}

\newcommand{\ed}{$\lambda_{\rm Edd}$~}

\begin{document}


\title{Investigating Changing-Look Active Galactic Nuclei with Long-term Optical and X-Ray Observations}

\author{A. Jana\inst{1, 2}$^{\orcidlink{0000-0001-7500-5752}}$
\thanks{E-mail: arghajit.jana@mail.udp.cl},
C. Ricci\inst{1, 3}$^{\orcidlink{0000-0001-5231-2645}}$,
M. J. Temple\inst{4,1}$^{\orcidlink{0000-0001-8433-550X}}$,
H.-K. Chang\inst{2}$^{\orcidlink{0000-0002-5617-3117}}$,
E. Shablovinskaya\inst{1}$^{\orcidlink{0000-0003-2914-2507}}$, \\
B. Trakhtenbrot\inst{5}$^{\orcidlink{0000-0002-3683-7297}}$,
Y. Diaz\inst{1}$^{\orcidlink{0000-0002-8604-1158}}$, 
D. Ilic\inst{6, 7}$^{\orcidlink{0000-0002-1134-4015}}$,
P. Nandi\inst{8}$^{\orcidlink{0000-0003-3840-0571}}$,
M. Koss\inst{9, 10}$^{\orcidlink{0000-0002-7998-9581}}$
}

\institute{$^{1}$Instituto de Estudios Astrof\'isicos, Facultad de Ingenier\'ia y Ciencias, Universidad Diego Portales, Av. Ej\'ercito Libertador 441, Santiago, Chile \\ 
$^{2}$Institute of Astronomy, National Tsing Hua University, Hsinchu, 300044, Taiwan\\
$^{3}$Kavli Institute for Astronomy and Astrophysics, Peking University, Beijing 100871, China\\
$^{4}$Centre for Extragalactic Astronomy, Department of Physics, Durham University, South Road, Durham DH1 3LE, UK\\
$^{5}$School of Physics and Astronomy, Tel Aviv University, Tel Aviv 69978, Israel \\
$^{6}$Department of Astronomy, University of Belgrade - Faculty of Mathematics, Studentski trg 16, 11000, Belgrade, Serbia\\
$^{7}$Hamburger Sternwarte, Universitat Hamburg, Gojenbergsweg 112, D--21029 Hamburg, Germany \\
$^{8}$Indian Center for Space Physics, 466 Barakhola, Netai Nagar, Kolkata 700099, India \\
$^{9}$Eureka Scientific, 2452 Delmer Street Suite 100, Oakland, CA 94602-3017, USA \\
$^{10}$Space Science Institute, 4750 Walnut Street, Suite 205, Boulder, Colorado 80301, USA
}

\date{Accepted XXX. Received YYY; in original form ZZZ}




\abstract
{Changing-look active galactic nuclei (CLAGNs) show the appearance and disappearance of broad emission lines in their UV/optical spectra on timescales of months to decades.}
{We investigate here how CL transitions depend on several AGN parameters such as accretion rate, obscuration properties and black hole mass.}
{We study a sample of 20 nearby optically-identified CLAGNs from the BAT AGN Spectroscopic Survey (BASS), using quasi-simultaneous optical and X-ray observations taken in the last $\sim 40$ years.}
{We find that for all CLAGNs, the transition is accompanied by a change in Eddington ratio. The CL transitions are not associated with changes in the obscuration properties of the AGN. CLAGNs are found to have a median Eddington ratio lower than the AGNs in the BASS sample in which CL transitions were not detected. The median of the transition Eddington ratio (Eddington ratio at which AGN changes its state) is found to be $\sim 0.01$ for type\,1 $\leftrightarrow$ 1.8/1.9/2 transition, which is consistent with the hard $\leftrightarrow$ soft state transition in black hole X-ray binaries. Most CL events are constrained to occur within 3--4 years, which is considerably shorter than the expected viscous timescale in AGN accretion disk.}
{The transitions of the optical CLAGNs studied here are likely associated to state changes in the accretion flow, possibly driven by disk-instability.}


\titlerunning{CLAGNs}
\authorrunning{Jana et al.}
\maketitle


\section{Introduction}
\label{sec:intro}
Active galactic nuclei (AGNs) are powered by the accretion of matter onto supermassive black holes (SMBHs) located at the center of galaxies \citep[e.g.,][]{Rees1988}. In the optical/UV, AGNs are generally classified as either type\,1 or type\,2. Type\,1 AGNs show both broad emission lines (BELs; full-width half maxima; FWHM $>1000$ \kms) originating in the broad line region (BLR) and narrow emission lines (NELs; FWHM $<1000$ \kms) originating in the narrow line region (NLR). Type\,2 AGNs show only NELs in their UV/optical spectra. Depending on the strength of the BELs, finer classifications (type\, 1.5, 1.8, and 1.9) can be used \citep[e.g.,][]{Osterbrock1981,Winkler1992}. In the X-rays, on the other hand, AGNs are classified based on their obscuration properties, and in particular by their line-of-sight hydrogen column density ($N_{\rm H}$). AGNs can be usually defined to be obscured if $N_{\rm H}>10^{22}$ \pcm, while they are referred as unobscured if $N_{\rm H}<10^{22}$ \pcm. Furthermore, obscured AGNs can be divided into Compton-thick (CT; $N_{\rm H}>10^{24}$ \pcm) and Compton thin ($N_{\rm H}<10^{24}$ \pcm).

Generally, type\,1 AGNs are found to be unobscured, while type\,2 AGNs are obscured \citep[e.g.,][]{Awaki1991,Koss2017,Ricci2017apjs,Oh2022}. 
The FWHM of the emission lines is in good agreement with the X-ray obscuration, with type\,1--1.8 AGNs having $N_{\rm H}<10^{21.9}$ \pcm, and type\,2 AGNs having $N_{\rm H}>10^{21.9}$ \pcm; however, type\,1.9 AGNs show a range of \nh \citep{Koss2017,koss2022b}.
These different classes of AGN can be explained by the simplified AGN unification model (UM), which is based on the orientation with respect to an anisotropic absorber \citep[e.g.,][]{Urry1995,Antonucci1993,Netzer2015,RamosAlmeida17}. According to this scheme, type\,1s are observed face-on, with BLR and NLR visible to the observer, while type\,2s are observed edge-on, with the view to the BLR covered by the obscuring material, which leaves only the NLR directly visible to the observer. The intermediate classes of AGNs (type\,1.5, 1.8, 1.9) are thought to be seen through the edge of the obscuring material, where the gas is not optically thick enough to block the entire BLR \citep[e.g.,][]{Antonucci1993,Goodrich1995,Runco2016}. 
While the UM provides a good first-order explanation to the different AGN populations, over the past few decades it has been shown that several additional parameters, such as the covering factor of obscuring materials, accretion rate, can affect the probability of an AGN to be observed as obscured or unobscured \citep[e.g.,][]{Elitzur2009,Ricci2017,Ricci2023d}. 

Changing-look AGNs (CLAGNs) are the objects that show drastic optical and X-ray spectral variability on timescales that range from hours to years and can be generally divided into two classes \citep[see][for a recent review]{Ricci2023Nat}. In the UV/optical, CLAGNs transition from type\,1 to type\,2, or vice versa, within timescales of months to decades. Most of these objects can be considered as `changing-state' AGNs (CSAGNs). In X-rays, typically, a different kind of CL event is observed. 
In these objects, the \nh show rapid variability on a timescale of hours to years. We refer to these objects as `changing-obscuration' AGNs (COAGNs).

Over the years, many AGNs, such as NGC\,1566 \citep{Oknyansky2019}, NGC\,3516 \citep{Ilic2020}, Mrk\,1018 \citep{Cohen1986}, Mrk\,590 \citep{Shappee2014}, have been found to show CL transitions on a timescale of months to decades. Many of those sources had undergone such transitions more than once. For example, Mrk\,1018 was turned off to enter the type\,1.9 state in 1984 \citep{Cohen1986} and re-brightened again to transition to type\,1 state in 2008 \citep{Shappee2014}. NGC\,1566 showed CL transition several times in the past 60 years, changing its state between type\,1 and 1.8/1.9 \citep{Shobbrook1966,Pastoriza1970,Alloin1986,Baribaud1992,Oknyansky2020}. NGC\,4151 was initially identified as type\,1 AGN in the 1970s, but it moved to type\,1.8/1.9 in the 1980s with the disappearance of the broad lines \citep{Osterbrock1981,Shapovalova2010}. Later, the source changed back to type\,1 state as it re-gained the broad lines. In addition to the local Seyfert galaxies, several higher redshift quasars have been found to show the CL transitions \citep[e.g.,][]{LaMassa2015,Merloni2015,MacLeod2016}. Recently, \citet{Zeltyn2024} identified 116 CLAGNs in the first year data in the Sloan Digital Sky Survey V (SDSS-V) with repeated spectroscopic observations, of which 107 sources are newly identified CLAGNs. That sample of CLAGNs is the largest reported to date.

The origin of the CS and CO events is still unclear, and many models have been proposed to explain these events. Generally, the COAGNs are linked to obscuration associated with the clumpiness of the BLR or the circumnuclear molecular dusty gas and dust \citep[e.g.,][]{Nenkova2008a,Nenkova2008b,Yaqoob2015,Ricci2016,AJ2020,AJ2022a}. The CSAGNs are believed to be caused by the change in the accretion rate, which is attributed to the local disk instabilities \citep{Stern2018,Noda2018}, or major disk perturbation, such as tidal disruption events \citep[TDEs; e.g.,][]{Merloni2015,Ricci2020}.

Some CS events were explained by moving gas clouds and dust which attenuate the BLR emission \citep[e.g.,][]{Goodrich1989,Goodrich1995,Zeltyn2022}.
However, various problems arise when explaining the CS events with obscuration. One needs a large dusty cloud to cover the BLR efficiently, which would take tens of years, assuming reasonable cloud velocity \citep[e.g.,][]{LaMassa2015}. However, many CS transitions are observed in a much shorter timescale of months to years \citep[e.g.,][]{Denney2014,Trakhtenbrot2019,Oknyansky2019}. Additionally, the signature of the obscuration is not observed in the X-ray spectra during/after the transitions \citep[e.g.,][]{Denney2014}. In fact, many CSAGNs show the same level of obscuration before and after the transition. Furthermore, many CSAGNs are found to be unobscured, discarding obscuration as a reason for CS events \citep[e.g.,][]{Lyu2021,AJ2021}. The optical continuum flux also changes with the BEL flux, indicating the accretion flow as a reason for the CS transitions \citep[e.g.,][]{Ricci2023Nat}. Few AGNs, such as NGC\,1365 and NGC\,7582, showed both CS and CO transitions in the past \citep[e.g.,][]{Risaliti2007,Temple2023,Neustadt2023}. However, those transitions are not correlated and are observed in different timescales, indicating independent transitions. Polarimetric studies also suggest that CS transitions are unlikely due to changes in the obscuration properties of the source \citep[e.g.,][]{Marin2017,Hutsemekers2019,Hutsemekers2020}.

The BELs flux responds to changes in the ionizing luminosity, which is evident from reverberation studies \citep[e.g.,][]{Blandford1982,Peterson1993,Runco2016,Fonseca2020,Feng2021,Oknyansky2023}. The appearance and disappearance of the BELs are examples of the extreme variability of the BLR. In the disk-wind model, the BLR could originate from outflows produced by the accretion disk, which directly connects the BELs with the accretion rate \citep[e.g.,][]{Emmering1992,Nicastro2000,Elitzur2009,Temple2023b}. In this framework, the BLR would not be sustained below a certain luminosity, $L_{\rm crit} < 2.3 \times 10^{40}$ $M_8^{2/3}$ \eps \citep[$M_8$ is black hole mass in $10^{8}$ $M_{\odot}$;][]{Elitzur2009}. This model suggests that the AGN would follow the transition sequence as type\,2 $\rightarrow$ 1.8/1.9 $\rightarrow$ 1.2/1.5 $\rightarrow$ 1.0, with increasing accretion rate \citep{Elitzur2012,Elitzur2014}.

Local disk instabilities in the accretion disk could also explain CS events \citep{Noda2018,Sniegowska2020}. The instabilities can be triggered by various mechanisms on different timescales \citep[e.g.,][]{Ricci2023Nat}. In Mrk\,1018, the CS transition is explained with the disk-instability model \citep{Noda2018}. The CS transition is linked with the soft excess, which is believed to ionize the gas clouds in the BLR. In Mrk\,1018, the BEL disappeared when the Eddington ratio (\ed) decreased from $\sim 0.08$ to $\sim 0.006$, with the primary continuum and soft excess flux reduced by a factor of $\sim 60$ and $\sim 7$, respectively. This transition is tied with the soft-to-hard spectral state transition, similar to black hole X-ray binaries (BHXBs), which occurs at \ed $\simeq 0.01-0.02$ \citep[e.g.,][]{Maccarone2003,Done2007,Yang2015}. Similar behavior is also found in other CSAGNs \citep[e.g.,][]{Ai2020,Ruan2019}.

\begin{figure*}
\centering
\includegraphics[width=17cm]{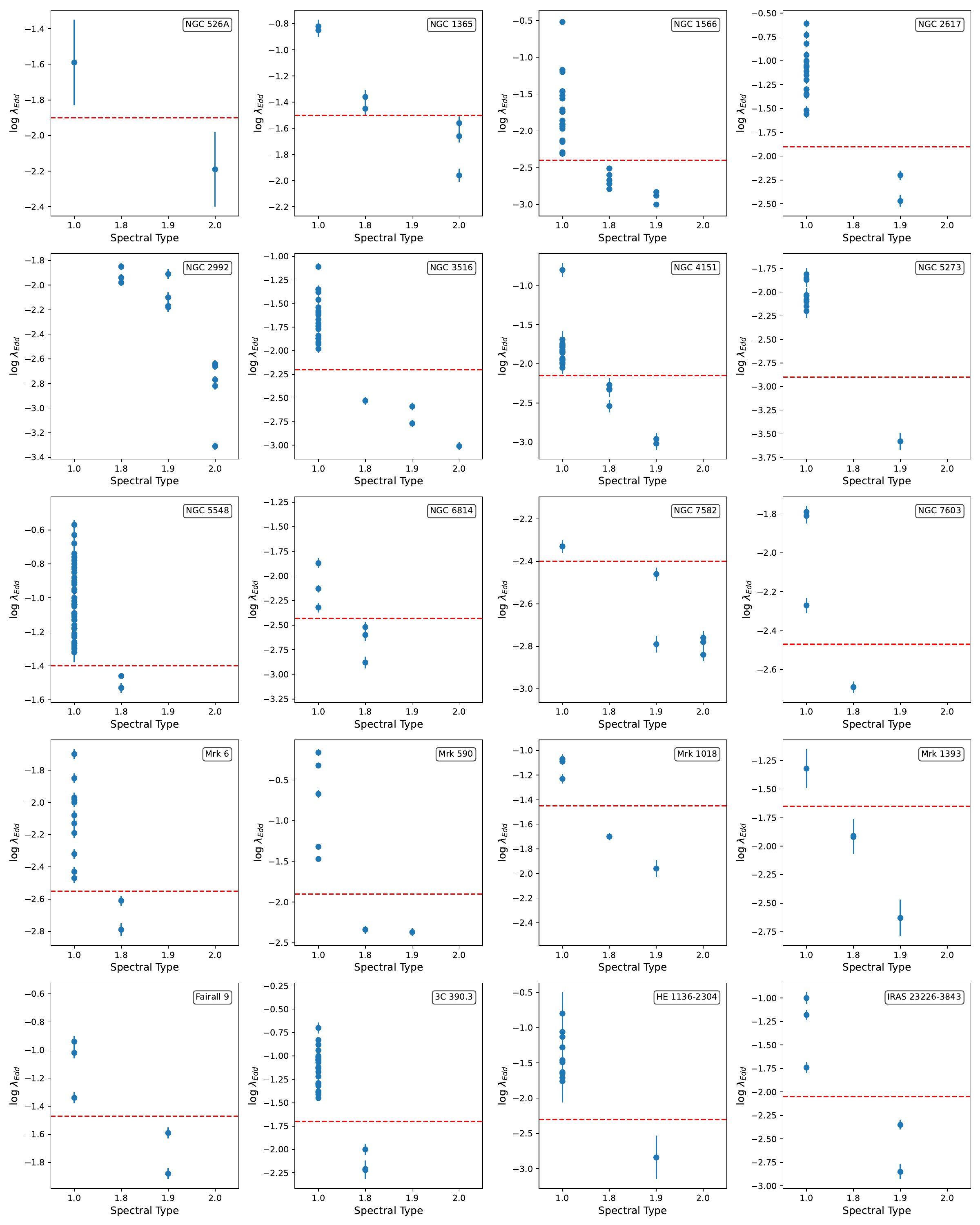}
\caption{Distribution of \ed with the spectral states for each source. The red-dashed horizontal lines in each panel represent the transition Eddington ratio ($\lambda_{\rm Edd}^{\rm tr}$) for each source, except NGC\,2992. NGC\,2992 did not show transition between type\,1 and type\,1.8--2 states.}
\label{fig:ed-state}
\end{figure*}

\clearpage
\subsection*{ }
The timescale of CSAGNs is a concerning factor when comparing it with state transitions in BHXBs. Simple mass-scaling relations indicate the viscous timescale for the AGNs with mass of $\sim 10^{6-8} M_{\odot}$ would be $\sim 10^{4-6}$ years. However, the timescale would reduce as the accretion disk of AGN is radiation pressure-driven, as opposed to the gas pressure-driven accretion disk in BHXBs \citep{Noda2018}. The inclusion of magnetic field would further decrease the timescale \citep{Feng2021mag}. Additionally, various instability mechanisms are suggested to explain the timescale of the CS transitions \citep[e.g.,][]{Sheng2017,Sniegowska2020,Scepi2021}.

Some CSAGNs are associated with external perturbation, such as TDE. In CS quasar SDSS J0159+0033, TDE is believed to cause the CS transition \citep{Merloni2015}. In the local universe, the CS transition in 1ES\, 1927+654 is also found to be caused by TDE \citep{Trakhtenbrot2019,Ricci2020}. The CL event in narrow line Seyfert galaxy SDSS J015804.75--005221.8 can also be associated with the TDE \citep{Petrushevska2023}.

In a sample of CLAGN with \textit{Swift}-BAT  light curves, \citet{Temple2023} found that a majority of CLAGN showed clear changes in their 14-195\,keV X-ray flux at the same time as the change in optical type. This suggests that the majority of CL events in local AGN are not due to changes in obscuration but must instead be driven by changes in the accretion state. However, detailed spectral modeling across the full X-ray energy range is needed to confirm this suggestion, which is one aim of this work.

In this paper, we investigate how the optically identified CL transitions depend on several AGN parameters, such as accretion rate (in terms of \ed), obscuration (in terms of $N_{\rm H}$), and black hole mass. Additionally, we provide constraints on the timescale of the CL transitions with long-term observations. For this purpose, we study a sample of 20 optically-identified CLAGNs using the archival quasi-simultaneous optical and X-ray observations taken in the last $\sim 40$ years. In Section~\ref{sec:analysis}, we present our sample and measurements. In Section~\ref{sec:res}, we present the result of our analysis. In Section~\ref{sec:discus}, we discuss our findings. Finally, in Section~\ref{sec:summary}, we summarize our results. Throughout the paper, we used $\Lambda$CDM cosmology, with the $H_{0}$ = $70$ km s$^{-1}$ Mpc$^{-1}$, $\Omega_{\rm M}$ = $0.3$, $\Omega_{\Lambda}$ = $0.7$.

\begin{figure}
\centering
\includegraphics[width=8.5cm]{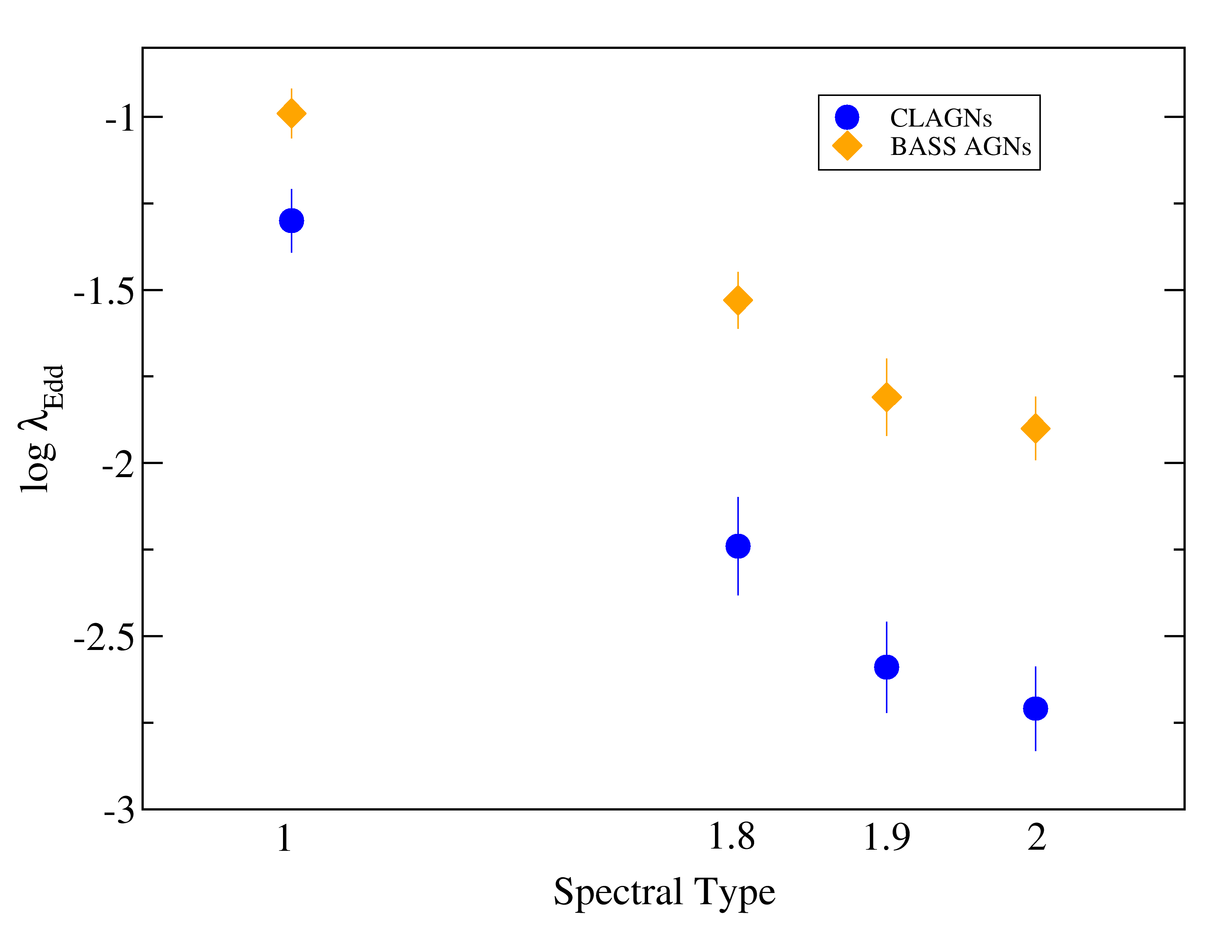}
\caption{Median Eddington ratio in each spectral state. The blue circles represent the median for $\lambda_{\rm Edd}$ for CLAGNs. The orange diamonds represent the median of \ed for the other AGNs from BASS sample for which CL transitions were not detected.} 
\label{fig:ed-med}
\end{figure}

\section{Data and Analysis}
\label{sec:analysis}

\subsection{Sample Selection}
\label{subsec:sample}

We collected our sample of optical CLAGNs from the BAT AGN Spectroscopic Survey (BASS\footnote{\url{https://www.bass-survey.com/}}). \citet{Temple2023} reported 21 CLAGNs by inspecting multi-epoch optical spectra from BASS DR1 \citep{Koss2017} and DR2 \citep{Koss2022}. Of these 21 sources, eight AGNs were reported to be CLAGNs for the first time.
To expand our sample, we conducted an extensive literature search of Swift/BAT AGNs. 
This search revealed an additional 15 CLAGNs that had shown transitions in the last 50 years. Consequently, the total number of optical CLAGNs in our sample increased to 36. We found that all sources have multi-epoch X-ray observations from the HEASARC’s data archive\footnote{\url{https://heasarc.gsfc.nasa.gov/cgi-bin/W3Browse/w3browse.pl}}. 

Next, we checked if those sources have quasi-simultaneous optical observations with the X-ray observations available in different optical states. 
We considered quasi-simultaneous observations if the optical and X-ray observations were taken within one year of each other. In this way, our sample is reduced to 20 sources: 16 sources from \citet{Temple2023} and four CLAGNs from the literature search. The details of the sample are tabulated in Table~\ref{tab:list}.

\subsection{Optical Data \& Classification}
\label{subsec:classification}

We collected the optical data from the literature. The details of the selected optical observation are presented in Appendix~\ref{sec:obs-res-source}. We primarily collected information on the optical spectral state from the literature. The optical classifications were based on the scheme of \citet{Osterbrock1981} and \citet{Winkler1992}, which is based on the variable flux of H$\beta$ BEL. The classifications are based on the ratio of the fluxes of H$\beta$ BEL and [OIII] NEL. i.e. R =f(H$\beta$)/f($OIII$) as follows:
\begin{itemize}
\item Type - 1: $R>2$ 
\item Type - 1.5: $R\sim 0.33-2$ 
\item Type - 1.8: $R<0.33$ 
\item Type - 1.9: No BEL H$\beta$, H$\alpha$ BEL. 
\item Type - 2: No BELs.
\end{itemize}

\noindent When BEL H$\beta$ and NEL [OIII] flux were available, we calculated the ratio to identify the optical state. When it was not available, we used the classification from the literature.

\noindent We note that the optical classifications are not always straightforward, especially when the BELs are weak (type\,1.8/1.9). 
Identifying spectral states in historical data, particularly those classified as type 1.8 or 1.9, can indeed be challenging due to issues
like poor signal-to-noise ratio (SNR) and spectral resolution. In a low flux state (type\,1.8/1.9), the BEL flux are often overestimated 
which led to misclassification of the state \citep[e.g.,][]{Trippe2010}. The type\,1--1.5s states show strong BELs and continuum 
emission and are easily distinguishable from the type\,1.8--2.0 states. In this work, we considered type\,1--1.5 states as type\,1 state.

\begin{table*}
\caption{List of CLAGNs}
\centering
\begin{tabular}{cccccccccccc}
\hline
\hline
No. & Source & BAT & RA & Decl & z & $\log (M_{\rm BH}/M_{\odot})$ & Ref. & CL Transition\\
    &        & ID  &    &       &  &                                &      &   \\
    (1) & (2) & (3) & (4) & (5) & (6) & (7) & (8) & (9) \\    
\hline
\hline
1 & NGC\,526A  & 72 & 20.973 & --35.060 & 0.019  & $8.17\pm0.41$ & 1 & 1.0 (1978) $\rightarrow$ 1.9 (1986--2004) $\rightarrow$  \\
 &           &         &        &        &              &    & &    2.0 (2009) $\rightarrow$ 1.9 (2016--2018) \\
\\
2 & NGC\,1365 & 73 & 53.389 & --36.140 & 0.005 & $6.65\pm 0.09$ & 2 &  2.0 (1978) $\rightarrow$ 1.0 (1979--1993) $\rightarrow$ 1.9 (2009) \\
     &        &   &    &       &  &                              &  &    $ \rightarrow$ 2.0 (2010) $\rightarrow$  1.8 (2012--2013)\\
 &           &         &        &        &      &   &        & $   \rightarrow$ 1.0 (2013--2014) $\rightarrow$ 2.0 (2021)  \\
 & & \\
3 & NGC\,1566  & 216 & 65.002	& --54.938 & 0.0051 & $6.83$ & 1 &  1.0 (1979--1984) $\rightarrow$ 1.8 (1984)   \\ 
 &           &        &  &        &        &              &    & $\rightarrow$ 1.9 (1985) $\rightarrow $1.8 (1985) $ \rightarrow$    \\
  &           &         & &        &        &              &    &  1.0 (1991) $\rightarrow$ 1.8 (1996) $\rightarrow$ 1.0 (2010) \\
  &           &         & &        &        &              &    & $ \rightarrow$ 1.9(2013--2017) $\rightarrow$ 1.0 (2018--2021)   \\

  & \\
4 & NGC\,2617 & 1327  & 128.912 & -4.088 & 0.0142 & $7.32\pm0.08$ & 3 &  1.8 (1994--2003)$\rightarrow$ 1.0 (2013--2022)\\
  &           &         & &        &        &              &    & $  \rightarrow$ 1.9 (2023)   \\
 &           &         &        &        &              &    & $      $ \\
5 & NGC 2992  & 471 & 146.425 & 14.326 & 0.0077 & $7.97$ &  2 &  1.9 (1978--1979) $\rightarrow$ 2.0 (1985--1991)  $\rightarrow$  \\
 &           &      &   &        &        &              &    & 1.9 (1994) $\rightarrow$ 2.0 (1998)$\rightarrow$ 1.9 (1999) \\
  &           &      &   &        &        &              &    &  $\rightarrow$ 2.0 (2006) $\rightarrow$ 1.8 (2014--2021)      \\
 & \\
6 & NGC\,3516 & 530 & 166.698	& 72.569 & 0.0088 & $7.38\pm0.08$ & 2 & 1.0 (1986--2012)$ \rightarrow$ 2.0 (2014--2018)  \\
  &           &      &   &        &        &              &    & $  \rightarrow $ 1.0 (2019--2020)   \\
\\
7 & NGC\,4151 & 595 & 182.636 & 39.405 & 0.0033 & $7.58\pm0.16$ & 2 & 1.0 (1979) $\rightarrow$ 1.8 (1981) $\rightarrow$ 1.9 (1984)  \\
  &           &      &   &        &        &              &    &  $\rightarrow$  1.8 (1985--1987) $ \rightarrow$ 1.0 (1990--1996) \\
 &           &         &        &        &         &     &    & $\rightarrow$1.8 (2000) $\rightarrow$ 1.0 (2001--2021)     \\
 & & \\
8 & NGC\,5273 & 686  & 205.471 & 35.658 & 0.0039 & $6.67\pm0.13$ & 2 &  1.9 (1984--1993) $\rightarrow$ 1.8 (2006) \\
 &           &         &        &        &       &       &    & $   \rightarrow$ 1.0 (2014--2022)  \\
 \\
9& NGC\,5548 & 717  & 214.500 & 25.135 & 0.017  & $7.72 \pm 0.02$& 2 & 1.0 (1978--2001) $\rightarrow$ 1.8 (2005--2007) \\
 &           &         &        &        &        &      &    & $ \rightarrow$ 1.0 (2014--2021) \\
\\
10& NGC\,6814 & 1046 & 295.669 & --10.323& 0.005  & $7.04\pm0.06$& 2 & 1.8 (1975) $\rightarrow$ 1.0 (1979--1984)   \\
  &     &      &         &        &        &              &    & $\rightarrow$ 1.8 (1985) $\rightarrow$ 1.0 (1987) $\rightarrow$ \\ 
  &     &      &         &        &        &              &    & 1.8 (1992) $\rightarrow$ 1.0 (2008--2015) \\
 \\
11& NGC\,7582 & 1188 & 349.598 &--42.370 & 0.0052 & $7.74$  & 2 & 1.8 (1977) $\rightarrow$ 1.0 (1998) $\rightarrow$ 1.9 (1998)\\
 &    &       &         &        &        &         &     &      $ \rightarrow$ 2.0 (2004--2016)  \\
 \\
12& NGC\,7603 & 1189  & 349.738 & --0.244 & 0.029 & $8.59$ &  2 & 1.0 (1974) $\rightarrow$ 2.0 (1975) $\rightarrow $\\
 &      &     &         &        &        &              &    &  1.0 (1976--2009)$\rightarrow$ 1.8 (2012)$\rightarrow $ 1.0 (2019)\\
\\
13& Mrk\,6  & 347   & 103.0501 & 74.427 & 0.018 & $8.1$ & 2 & 1.0 (1976) $\rightarrow$ 1.8 (1977) $\rightarrow$ 1.0 (1981--2008) \\
\\
14& Mrk\,590 & 116  & 33.639 & --0.767 & 0.026 & $7.57$ & 2 & 1.0 (1982--2003) $\rightarrow$ 1.8 (2006) $\rightarrow$  \\
 &           &         &        &        &      &        &    &  1.9 (2013--2014) $\rightarrow$ 1.0 (2017--2018)\\
 &  & \\
15& Mrk\,1018 & 106  & 31.567 & --0.291 & 0.042 & $7.81$ & 2 & 1.9 (1979) $\rightarrow$ 1.0 (1984--2013) $\rightarrow  $  \\
 &           &         &        &        &              &    &   & 1.9 (2015) $\rightarrow$ 1.8 (2019)\\
 \\
16& Mrk\,1393 & 757  & 227.176& --0.183 & 0.054  & $8.61\pm0.30$ & 1 & 1.0 (1984) $\rightarrow$ 1.9 (1993--2001) $\rightarrow  $\\
 &           &         &        &        &              &    &   &  1.8 (2005--2006) $\rightarrow$ 1.0 (2022) \\
\\
17& 3C\,390.3 &994  & 280.553 & 79.774 & 0.056 &   $8.64\pm0.05$ &  1 & 1.0 (1978) $\rightarrow$ 1.8 (1979--1984) \\
  &           &         &        &        &         &     &    & $\rightarrow$ (1985--2014) \\
18& Fairall\,9 & 73 & 20.941 & --58.806 & 0.047 & $8.30\pm0.08$ & 2 & 1.0 (1979--1983) $\rightarrow$ 1.9 (1984)  \\
 &           &         &        &        &              &    &   & $\rightarrow$ 1.0 (1985--2016)\\
 &           &         &        &        &              &    &  \\
\hline
\hline
\end{tabular}
\label{tab:list}
\end{table*}

\begin{table*}
\centering
\begin{tabular}{ccccccccccccc}
\hline
\hline
No. & Source & BAT & RA & Decl & z & $\log (M_{\rm BH}/M_{\odot})$ & Ref. & CL Transition\\
    &        & ID  &    &       &  &                                &      &   \\
(1) & (2) & (3) & (4) & (5) & (6) & (7) & (8) & (9) \\    
\hline
\hline
19& HE\,1136--2304 & 557 & 174.701 & --23.349 & 0.027 & $7.62\pm0.59$ & 1 & 1.9 (1993--2002) $\rightarrow$ 1.0 (2014--2015)\\
\\
20& IRAS\,23226--3843& 1194 & 351.359& --38.471 & 0.035 &  $7.83$ & 2 & 1.9 (1997--2005) $\rightarrow$ 1.0 (2016)  $\rightarrow$ 1.9 \\
 &           &         &        &        &      &        &    & (2017--2019) $\rightarrow$ 1.0 (2019) $\rightarrow$ 1.9 (2020)  \\
\hline
\hline
\end{tabular}
\leftline{Columns: (2) source name, (3) BAT ID of the source, (4) \& (5) source position in J2000 epoch, (6) redshift of the source,}
\leftline{(7) mass of the black hole, (8) references for black hole mass, and (9) information of the optical spectral states.}
\leftline{References for $M_{\rm BH}$: (1) \citet{Koss2017}, (2) \citet{Koss2022}, (3) \citet{Feng2021}.}
\leftline{For the references of CL transition, see Appendix~\ref{sec:obs-res-source}.}
\end{table*}

\subsection{Black hole masses}
\label{subsec:mass}
The mass of supermassive black holes in AGNs has been estimated using BELs and virial prescription.
Often, different methods give a different mass value for a particular AGN. Moreover, the CLAGNs are variable; hence, the question arises if the BLR of these CLAGNs are virialized or if CLAGNs follow the same scaling relation as other AGNs. Hence, it is necessary to use the mass value from the literature carefully. \citet{Jin2022} showed that virial estimation of $M_{\rm BH}$ in the brightest epoch (type-1) is consistent with the $M_{\rm BH}-\sigma_*$ estimation in the faint epochs (type-2) for 26 CLAGNs, suggesting the CLAGNs and AGNs follow the same virial scaling relation. 
\citet{Caglar2023} showed that single epoch virial mass estimation is consistent with the $M_{\rm BH}-\sigma_*$ estimation for type-1 AGNs in BASS, and single epoch measurement are systematically lower by $\simeq 0.12$ dex.

In this work, we use the black hole mass from the BASS DR1 or DR2 catalog for 19 sources in our sample \citep{Koss2017,Koss2022}. The mass estimation is taken from (i) literature measurements with mega-masers, reverberation mapping, or stellar and gas dynamics; (ii) H$\beta$ or H$\alpha$ BELs if $N_{\rm H}<10^{22}$ \pcm \citep[from][]{Mejia-Restrepo2022}; (iii) Stellar velocity dispersion measurements for all Sy1.9 and Sy2 AGNs \citep[from][]{koss2022b}, using $M_{\rm BH}-\sigma_*$ relation \citep[from][]{Kormendy2013}. The mass of NGC 2617 was not available in BASS DR1 or DR2; therefore, we collected the $M_{\rm BH}$ for NGC 2617 from the latest reverberation mapping measurement \citep{Feng2021}.

\subsection{X-Ray Data Analysis}
\label{subsec:x-ray}
In this work, we mainly rely on the X-ray analyses from the literature. However, there are many instances when quasi-simultaneous X-ray observations are available but have yet to be published. We reduced and analyzed those X-ray data, obtained by \swift/XRT, and \emph{XMM-Newton}. In total, we analyzed 99 observations for 17 sources in the current study. The data reduction technique is described in Appendix~\ref{sec:swift}.

The X-ray spectra contain several components: primary continuum, soft excess below 2 keV, and reprocessed emission, which consists of a Fe K$\alpha$ line at $\sim 6.4$~keV and a reflection hump at $\sim 10-40$~keV \citep[e.g.,][]{Ricci2017apjs,Ricci2018}.
For the \emph{Swift}/XRT and \emph{XMM-Newton} spectra in the $0.5-10$~keV energy range, we used an absorbed powerlaw model. We used two absorption components; one is for the Galactic absorption, which is fixed at the Galactic absorption value at the source direction. The Galactic absorption is estimated using \textsc{$N_{\rm H}$} tools of \textsc{ftools} \citep{HI4PI2006}\footnote{\url{https://heasarc.gsfc.nasa.gov/cgi-bin/Tools/w3nh/w3nh.pl}}.
The second component is used for the intrinsic source absorption. We used \textsc{phabs} model for both absorptions components, with \textsc{angr} abundances \citep{Anders1989}, and \textsc{verner} cross-section \citep{Verner1996}. We added a Gaussian line for the Fe K$\alpha$ line at 6.4 keV and a blackbody component for the soft excess if required.

The spectral analysis is carried out in HEASARC's spectral analysis software \textsc{xspec}\footnote{\url{https://heasarc.gsfc.nasa.gov/xanadu/xspec/}}.
We obtained a good fit in all spectra, with $\chi^2/$degrees of freedom $\sim 1$. 
From the X-ray data analysis, we obtained the unabsorbed luminosity ($L_{\rm X}$) and line of sight hydrogen column density ($N_{\rm H}$).
Using the \textsc{clum} task, we calculated unabsorbed luminosity in the $2-10$~keV energy range. The result of the important spectral parameters is tabulated in Appendix~\ref{sec:obs-res-source}.

\subsection{X-ray Luminosity}
\label{subsec:luminosity}

In the present work, we consider absorption-corrected X-ray luminosity in the $2-10$~keV energy range. The $2-10$~keV luminosity ($L_{\rm X}$) is obtained from the literature or the spectral analysis. We only consider the luminosity of the primary continuum emission, which is thought to originate in the X-ray corona \citep[e.g.,][]{T94,CT95,Done2007}.

We consider Hubble constant $H_0$ = $70$ km s$^{-1}$ Mpc$^{-1}$ in the present paper; however, in the literature, various values of $H_0$ are considered. We converted those luminosities to the appropriate luminosity consistent with the cosmological parameters used in the present work. When the unabsorbed X-ray flux ($F_{\rm X}$) was available in the $2-10$ keV energy range, we calculated the luminosity using,
$L_{\rm X}$ = $4\pi d_{\rm L}^2 \frac{F_{\rm X}}{\left(1{+}z\right)^{2-\Gamma}}$. Here, $d_{\rm L}$ and $\Gamma$ are the luminosity distance of the source and photon index of the spectra, respectively. When the $2-10$ keV unabsorbed flux was not available, we estimated the $2-10$ keV unabsorbed flux using \textsc{webpiims}\footnote{\url{https://heasarc.gsfc.nasa.gov/cgi-bin/Tools/w3pimms/w3pimms.pl}} tool, with the corresponding $N_{\rm H}$ and the $\Gamma$, assuming an absorbed powerlaw continuum. When $\Gamma$ was not reported, we assumed $\Gamma=1.8$ for the simulation. In this way, we estimated $2-10$ keV unabsorbed flux for 23 observations for 13 sources when $\Gamma$ was not available.

\subsection{Bolometric Correction and Eddington Ratio}
\label{subsec:edd}

Once we calculated the $L_{\rm X}$, we converted it to the bolometric luminosity ($L_{\rm bol}$) using bolometric correction factors ($k_{\rm bol}$). We used Eddington ratio-dependent bolometric correction from \citet{Gupta2024}. The $2-10$ keV bolometric correction factor is given by,

$\log k_{\rm bol}=C \times (\log \lambda_{\rm Edd})^2 + B\times \log \lambda_{\rm Edd} + A$.

Here, $C=0.054\pm 0.034$, $B=0.309 \pm 0.095$ and $A=1.538\pm 0.063$.
We obtained the Eddington ratios as, $\lambda_{\rm Edd}=k_{\rm bol} \times L_{\rm X}/L_{\rm Edd} = L_{\rm bol}/L_{\rm Edd}$, where, $L_{\rm Edd}=1.3\times 10^{38} (M_{\rm BH}/M_{\odot})$ \eps.

\begin{table*}
\centering
\caption{Mean and median of Eddington ratio (\ed) and Hydrogen column density ($N_{\rm H}$) for CLAGNs and other AGNs from BASS for which CL transitions have not been detected.}
\begin{tabular}{cccccccccc}
\hline
Type &  $\log \lambda_{\rm Edd}$& & $\log (N_{\rm H}/{\rm cm^2})$ & & \\ 
     &  Mean & Median & Mean & Median \\ 
\hline
type\,1  &   $-1.40\pm0.08$ & $-1.30\pm0.09$  & $21.32\pm0.23$ & $21.35\pm0.15$ \\ 
type\,1.8&   $-2.13\pm0.10$ & $-2.24\pm0.14$  & $21.66\pm0.16$ & $21.88\pm0.14$ \\
type\,1.9&   $-2.49\pm0.14$ & $-2.59\pm0.13$  & $21.66\pm0.09$ & $21.61\pm0.08$ \\
type\,2&   $-2.59\pm0.13$ & $-2.71\pm0.12$  & $22.24\pm0.11$ & $22.32\pm0.11$ \\
\hline \hline
\multicolumn{6}{c}{All unbeamed AGNs, from BASS, \citep{Koss2022} } \\ \hline
type\,1    & &  $-0.99\pm0.07$ & & $20.00\pm0.06$  \\
type\,1.8  & &  $-1.53\pm0.08$ & & $21.04\pm0.07$  \\
type\,1.9  & &  $-1.81\pm0.11$ & & $22.28\pm0.13$& \\ 
type\,2  & &  $-1.90\pm0.09$ & & $23.27\pm0.08$& \\ 
\hline
\hline
\end{tabular}
\label{tab:median}
\end{table*}

\section{Results}
\label{sec:res}

\subsection{The relation between Spectral State and Eddington Ratio}
\label{sec:ed-state}

The CLAGNs in our sample have been observed across multiple epochs in both the X-ray and optical wavebands, providing us with information on how these sources evolve over time. However, to ensure consistency in our analysis, we focus only on the epochs where quasi-simultaneous observations ($<1$ years) in both wavebands were obtained. This approach helps minimize the effects of variability, ensuring that the comparison between X-ray and optical properties reflects the true spectral state of the source.

Figure~\ref{fig:ed-state} displays the variation of the Eddington ratio (\ed) with the spectral state for each CLAGN in our sample. 
Our analysis reveals a significant correlation between the optical spectral state and the Eddington ratio for all 20 sources. Specifically, we observe that the CL transitions between type 1 and type 2 states are tightly linked to changes in their accretion rates. 
When the Eddington ratio increases, the AGNs tend to move towards a type\,1 state, characterized by stronger broad emission lines and brighter continuum emission. Conversely, when the Eddington ratio decreases, the sources tend to transition towards a type\,2 state, where broad emission lines are weaker or absent, and the continuum is relatively dimmer. This pattern indicates that the accretion rate plays a crucial role in driving the changing-look phenomenon, with higher accretion rates leading to the type\,1 state and lower accretion rates leading to the the type\,2 state.

To quantify this relationship further, we calculate both the mean and median values of the Eddington ratio (\ed) for each spectral state. This is shown in Figure~\ref{fig:ed-med}, where the blue circles represent the median values of \ed at each state. A clear trend is observed in the figure: both the mean and median Eddington ratios increase as the CLAGNs move towards the type\,1 state and decrease as they move towards the type\,2 state. The mean and median Eddington ratios for each spectral state are provided in Table~\ref{tab:median}.

\begin{figure*}
\centering
\includegraphics[width=17cm]{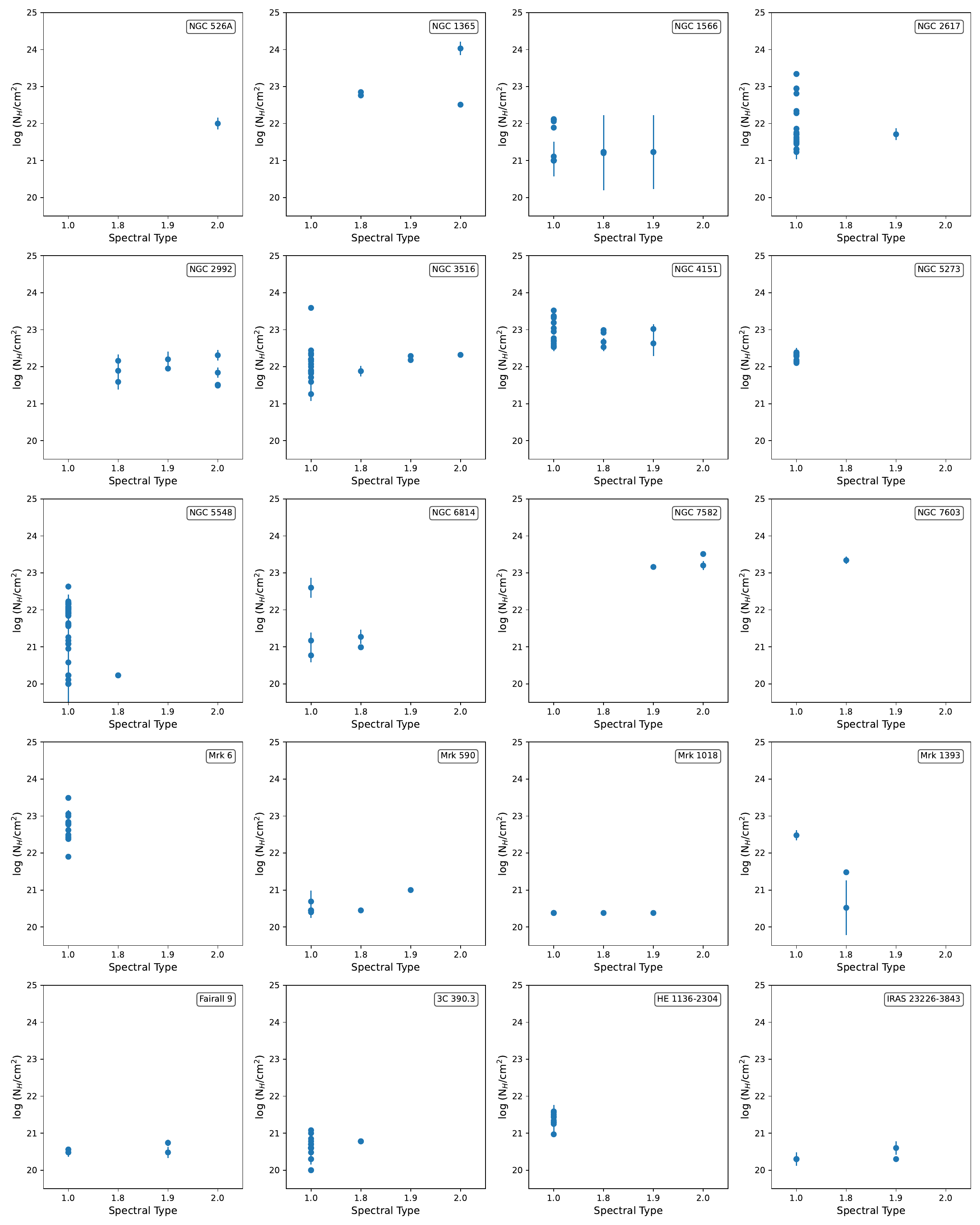}
\caption{Variation of \nh as a function the spectral state for our sample. In NGC 526A, NGC 7603, and HE\,1136--2304, the \nh value is only available for type\, 2, type\,1.8, and type\, 1.0 states, respectively.}
\label{fig:nh-state}
\end{figure*}

\begin{figure}
\centering
\includegraphics[width=8.5cm]{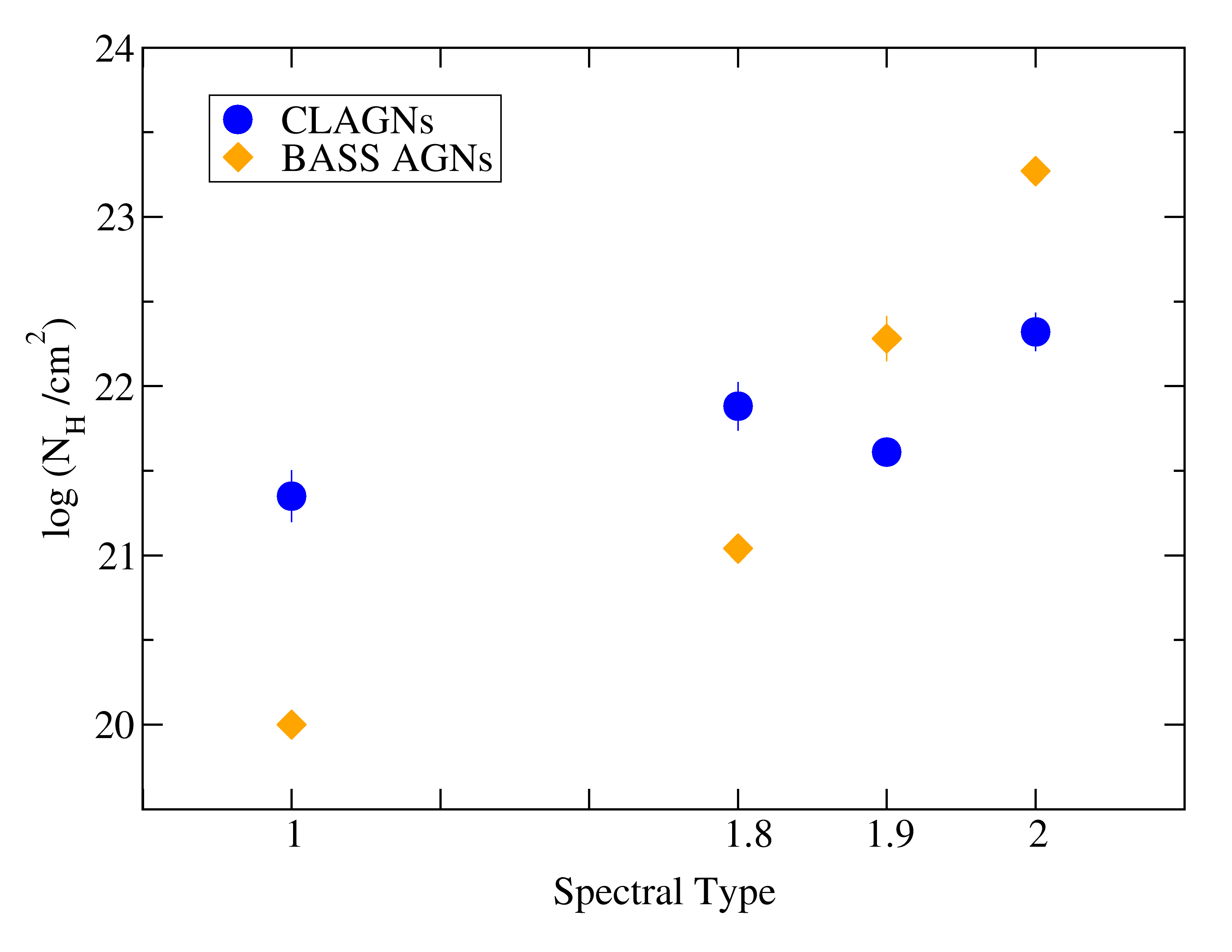}
\caption{Median of \nh in each spectral state. The blue circles and orange diamonds represent the CLAGNs of our sample, and other AGNs from the BASS sample, respectively.}
\label{fig:med-nh}
\end{figure}

\subsection{Relation between Spectral State and X-ray Obscuration}
\label{subsec:nh-state}

The obscuration of AGNs is commonly characterized by the line-of-sight hydrogen column density (\nh) along the line of sight. In Figure~\ref{fig:nh-state}, we examine the variation of \nh as a function of the optical spectral state for each source in our sample.
Interestingly, we did not see any a clear correlation between the spectral state and \nh for most of the sources. Our results suggest that obscuration is not the dominant factor for the majority of the AGNs in our sample.

A notable exception to this trend is NGC\,1365, for which we observe an intriguing behavior. This source appeared to transition into a type\,2 state while simultaneously being in a Compton-thick (CT) X-ray state, characterized by an extremely high column density of obscuring material ($N_{\rm H}>10^{24}$ \pcm). However, NGC\, 1365 shows rapid \nh variability in a timescale of days, which is not related to the optical state transitions. The detailed study found that the CL transition in this source is led by the change of accretion rate, not obscuration (see Section~\ref{subsec:CS-CO} for details).

To further analyze the role of obscuration in AGN transitions, we calculated the median values of \nh for each spectral state. Figure~\ref{fig:med-nh} displays the median \nh for type 1, type 1.8, type 1.9, and type 2.0 states. The blue circles represent the median of \nh for the CLAGNs at each spectral state. The median of \nh is found to be non-variable with respect to the spectral state. The mean and median values of \nh for each spectral state are presented in Table~\ref{tab:median}. The median of \nh was found to be $\log (N_{\rm H}/{\rm cm^2})=21.35\pm0.15$ in type\,1 states. For type\,1.8, type\,1.9, and type\,2 states, the medians are $\log (N_{\rm H}/{\rm cm^2}) = 21.88\pm0.14$, $21.61\pm0.08$, and $22.32\pm0.11$, respectively. 

\begin{figure}
\centering
\includegraphics[width=9cm]{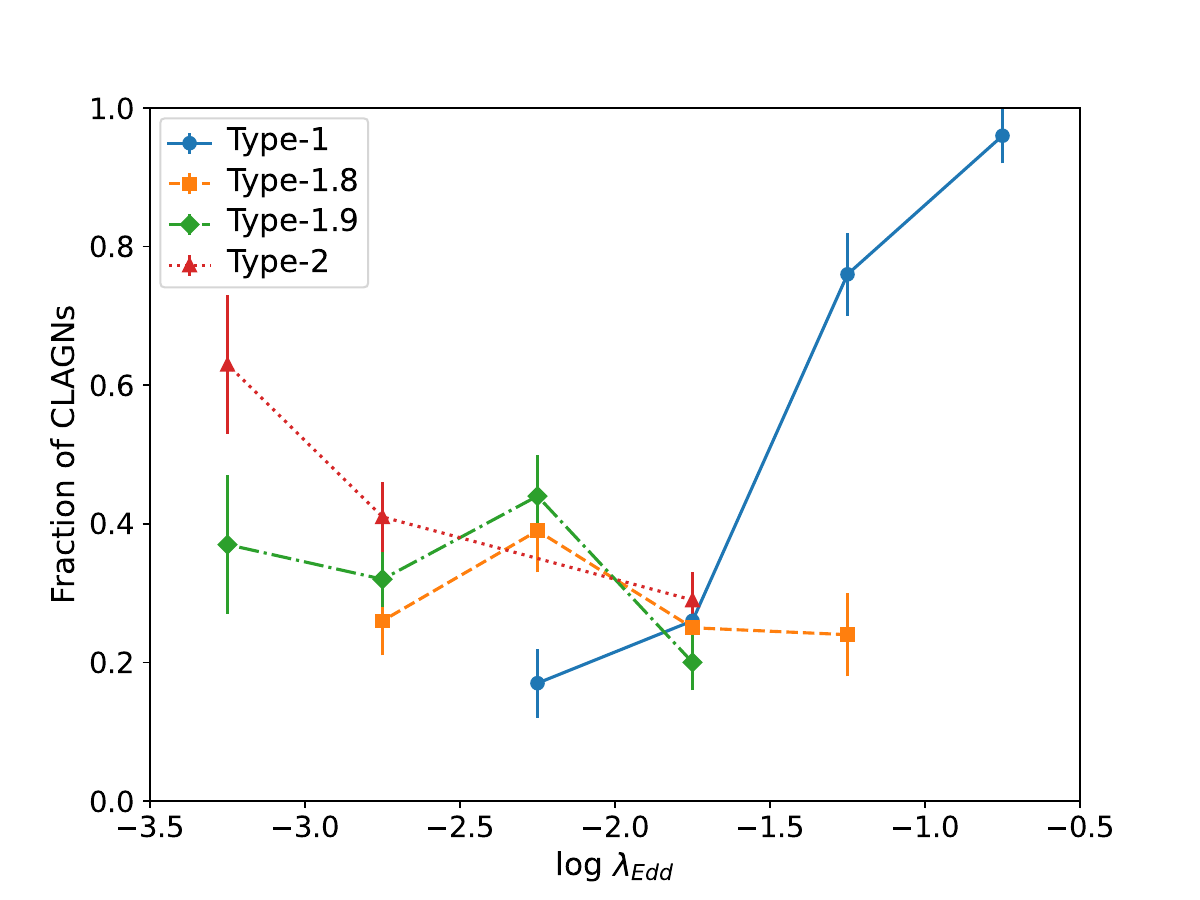}
\caption{Fraction of CLAGNs in different spectral state for different \ed. The blue circles, orange squares, green diamonds, and red triangles represent the fraction of CLAGNs in type\,1, type\,1.8, type\,1.9, and type\,2 states, respectively.}
\label{fig:ed-dist}
\end{figure}

\begin{figure}
\centering
\includegraphics[width=9cm]{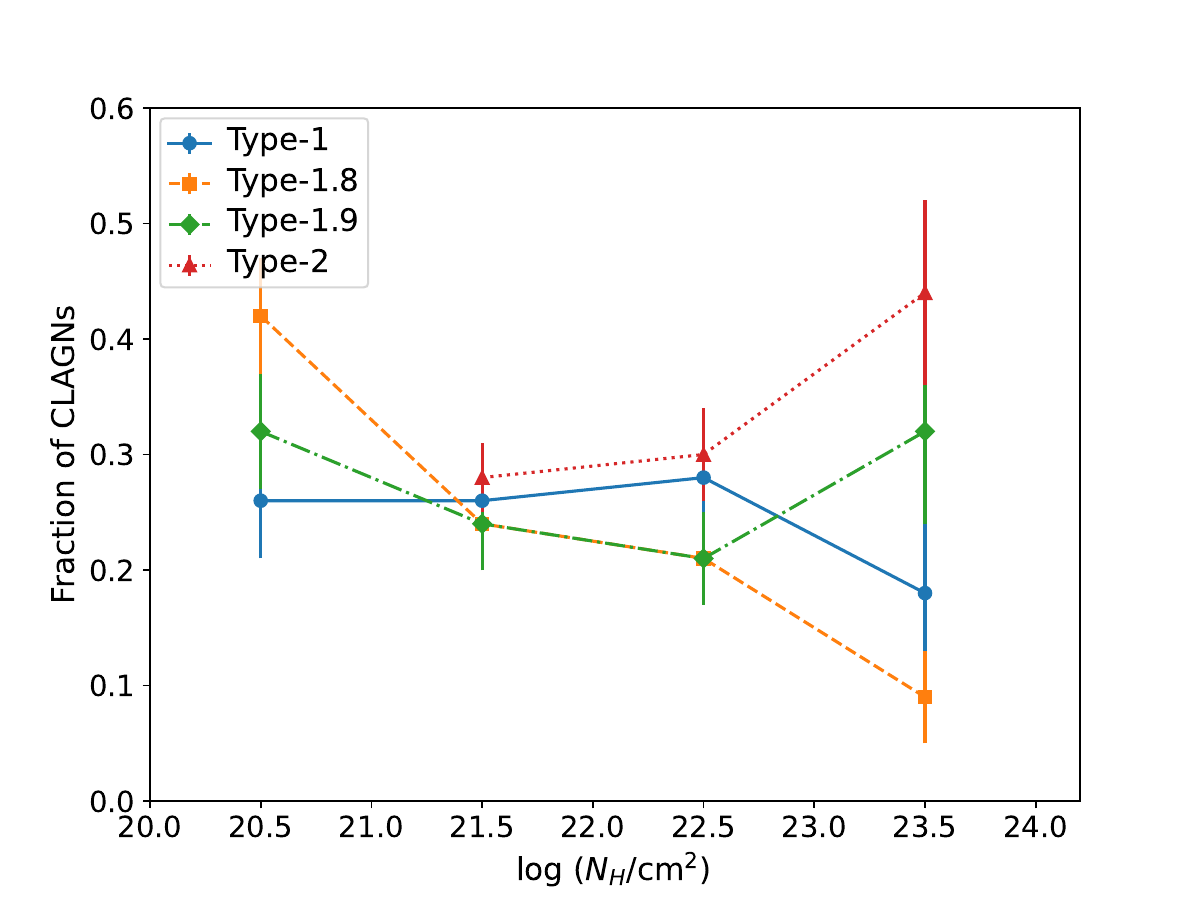}
\caption{Fraction of CLAGNs in different spectral state for different \nh. The blue circles, orange squared, green diamonds and red triangles represent the type\,1, 1.8, 1.9 and 2.0 states, respectively.}
\label{fig:nh-dist}
\end{figure}

\subsection{Distribution of Spectral State}
\label{subsec:distribution}

To further investigate the dependency of spectral state on both the \ed and \nh, we analyzed the distribution function of each spectral state as a function of these parameters. In Figure~\ref{fig:ed-dist}, we present the fraction of CLAGNs in different spectral states as a function of \ed. The blue circles, orange squares, green diamonds, and red triangles represent the distribution function for type\,1 ($f_1$), type\,1.8 ($f_{1.8}$), type\,1.9 ($f_{1.9}$), and type\,2 states ($f_2$), respectively. 
To calculate these distributions, we divided the range of $\log \lambda_{\rm Edd}$ from $-0.5$ to $-3.5$ into bins with a width of $\Delta \log \lambda_{\rm Edd}=0.5$. The fraction of CLAGNs in each spectral state within each bin was then calculated, providing insight into the behavior of AGNs across this range of accretion rates.
The uncertainty in the fraction of CLAGNs for each spectral state was estimated using the 16th and 84th quantiles of a binomial distribution, following the method outlined by \citet{Cameron2011}. 

The $f_1$ displays a clear increase with \ed, while $f_{2}$ shows the opposite behavior. Both $f_{1.9}$ and $f_{\rm 1.8}$ did not show any clear variation. The observed trend of a fraction of CLAGNs in each spectral state also shows that CLAGNs move towards the type\,1 state for increasing \ed.


In Figure~\ref{fig:nh-dist}, we examine the distribution of spectral states as a function of $N_{\rm H}$. Here, the range of $\log N_{\rm H}$ from 20.0 to 24.0 is divided into bins with a width of $\Delta \log (N_{\rm H}/cm^2)=0.5$. Similar to the Eddington ratio distribution, we calculated the fraction of CLAGNs in each spectral state within each bin.

With increasing $\log N_{\rm H}$, we observe a decrease in the fraction of $f_{1}$ and $f_{1.8}$, suggesting that these states are more commonly associated with lower levels of obscuration. Interestingly, the fraction of $f_{1.9}$ does not show any clear variation with \nh.
On the other hand, the fraction $f_2$ increases with increasing \nh, consistent with the idea that type\,2 states are typically more heavily obscured than type\,1 states. However, considering the uncertainties, the $f_1$ and $f_2$ remain constant with \nh.

\begin{figure*}
\centering
\includegraphics[width=12cm]{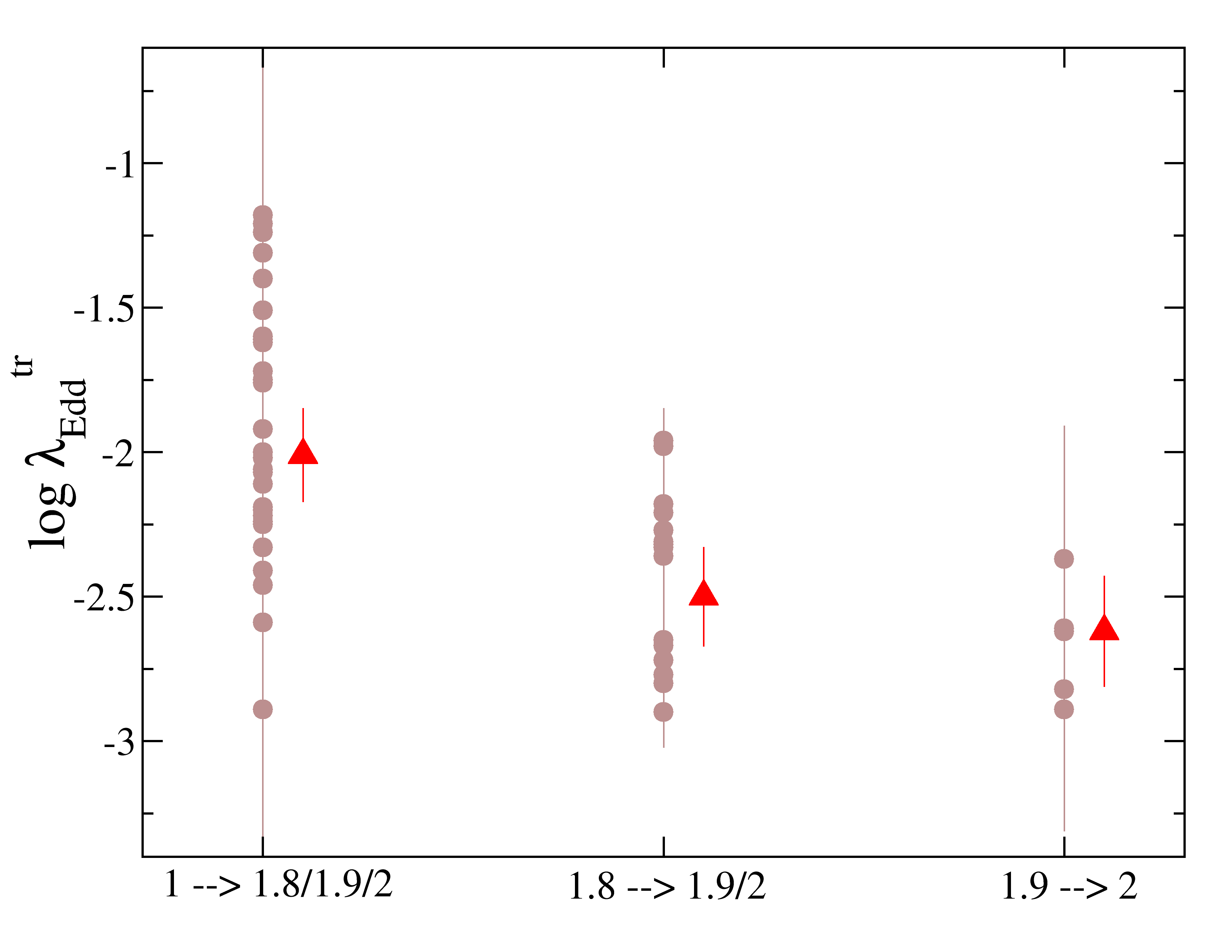}
\caption{Transition Eddington ratio ($\lambda_{\rm Edd}^{\rm tr}$) for different spectral state transitions. The grey circles represent the $\lambda_{\rm Edd}^{\rm tr}$ for individual objects, while the red triangles represent the median value of the $\lambda_{\rm Edd}^{\rm tr}$. }
\label{fig:ed-tr}
\end{figure*}

\begin{table}
\centering
\caption{Median of Transition Eddington ratio ($\lambda_{\rm Edd}^{\rm tr}$)}
\begin{tabular}{cccc}
\hline
\noalign{\smallskip}
& 1 $\rightarrow$ 1.8/1.9/2 & 1.8 $\rightarrow$ 1.9/2 & 1.9 $\rightarrow$ 2.0 \\
\noalign{\smallskip}
\hline
\noalign{\smallskip}
$\log \lambda_{\rm Edd}^{\rm tr}$ & $-2.01\pm 0.23$ & $-2.50\pm0.18$  &  $-2.62\pm0.19$  \\
\noalign{\smallskip}
\hline
\end{tabular}
\label{tab:tr-ed}
\end{table}

\subsection{The transition Eddington Ratio}
\label{subsec:tran-ed}

In our sample of 20 optically-identified CLAGNs, we observe that transitions between spectral states are driven primarily by changes in the accretion rate. To investigate the physical mechanism behind these transitions, we estimated the transition Eddington ratio ($\lambda_{\rm Edd}^{\rm tr}$) for each spectral change in each source. The AGNs undergo transitions between spectral states at $\lambda_{\rm Edd}^{\rm tr}$.

The $\lambda_{\rm Edd}^{\rm tr}$ for each state transition is computed as follows: we first identify the range of Eddington ratios associated with the transition by determining the highest $\lambda_{\rm Edd}$ value of the lower state (type\,2s) and the lowest $\lambda_{\rm Edd}$ value of the higher state (type\,1s). The midpoint of this range is then considered as the transition Eddington ratio, $\lambda_{\rm Edd}^{\rm tr}$, for that specific transition. The uncertainty in $\lambda_{\rm Edd}^{\rm tr}$ is derived by calculating the difference between the highest or lowest values of the range and the mid-point.

To further quantify the transition points between spectral states, we estimate the median of $\lambda_{\rm Edd}^{\rm tr}$, corresponding to different spectral state transitions. We estimated the median by applying bootstrap method. For each bootstrap sample, the median was calculated. This process was repeated 1000 times to generate the distribution of medians. From this distribution, we calculated the median of $\lambda_{\rm Edd}^{\rm tr}$ for each transition.
In Figure~\ref{fig:ed-tr}, we present $\lambda_{\rm Edd}^{\rm tr}$ for the various spectral state transitions observed in our CLAGN sample. The grey circles represent the individual $\lambda_{\rm Edd}^{\rm tr}$ values for each changing-look transition, while the red triangles indicate the median values of $\lambda_{\rm Edd}^{\rm tr}$ for each type of transition. 

For transitions type $1 \leftrightarrow 1.8/1.9/2$, we find that the median value of $\log \lambda_{\rm Edd}^{\rm tr}$ is $-2.01 \pm 0.16$. This suggests that AGNs typically transition from a type 1 state to a type 1.8/1.9/2 state when their accretion rate drops below $\log \lambda_{\rm Edd}^{\rm tr} = -2.01\pm 0.23$.

For the transitions, such as from type $1.8 \leftrightarrow 1.9/2$, and from type $1.9 \leftrightarrow 2.0$, the median values of $\log \lambda_{\rm Edd}^{\rm tr}$ are $-2.50 \pm 0.18$ and $-2.62\pm 0.19$, respectively. Interestingly, the transition Eddington ratios for these states do not differ significantly, and the values are consistent within the uncertainties. 
It has been previously shown that the classifications of type 1.8, type 1.9, and type 2.0 AGNs can be somewhat ambiguous, due to the faintness of the broad emission line (BEL) flux in these states (e.g., \citealp{Trippe2008,Trippe2010}). The spectral lines in these states can be weak and difficult to distinguish, leading to potential misclassifications. As a result, type\,1.8, type\,1.9, and type\,2 states may not always be correctly categorized, which could explain the similar $\lambda_{\rm Edd}^{\rm tr}$ values observed for transitions between these states.
The median values of $\lambda_{\rm Edd}^{\rm tr}$ for each spectral state transition are provided in Table~\ref{tab:tr-ed}.

\begin{figure}
\centering
\includegraphics[width=8.5cm]{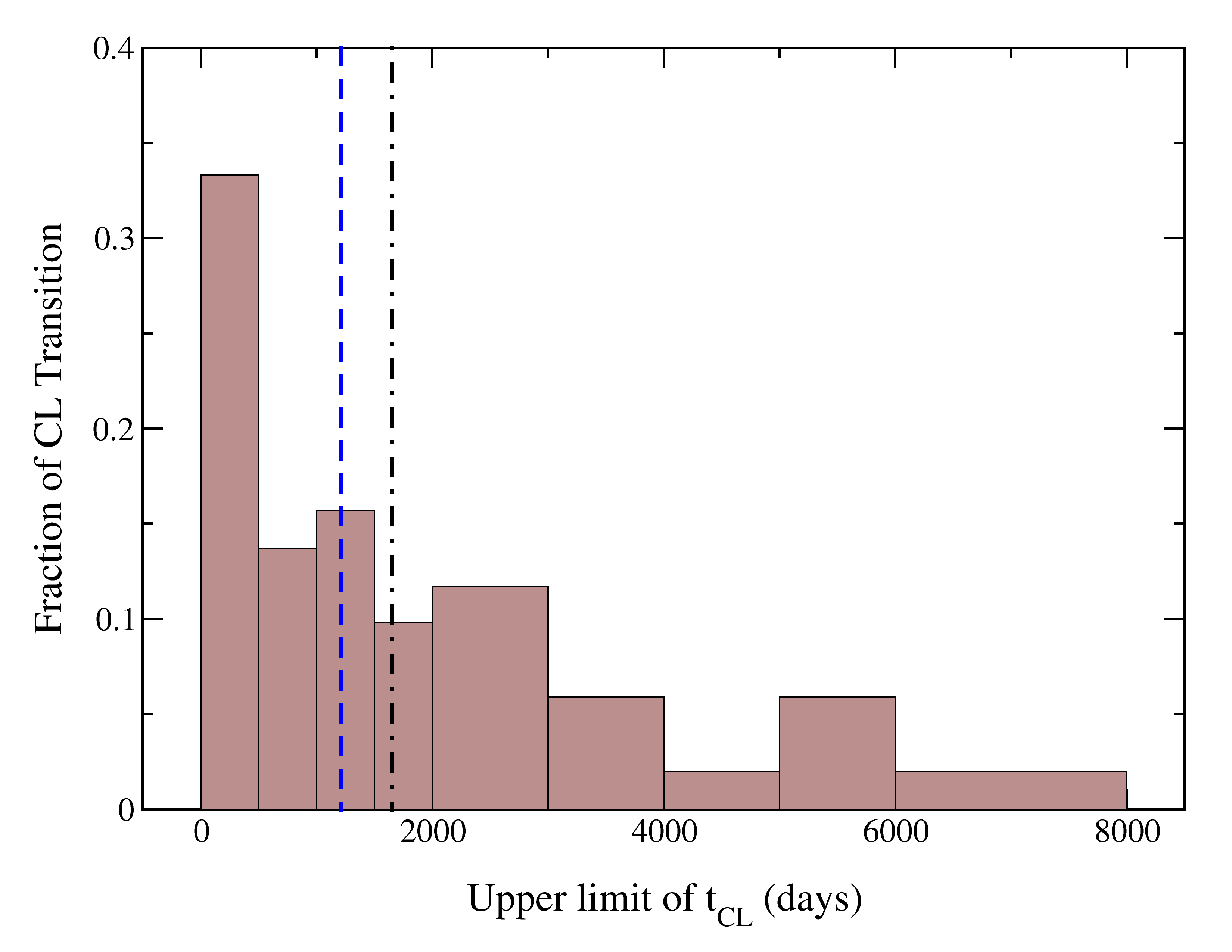}
\caption{Distribution of the upper limit of timescale for the CL events. The vertical blue dashed and black dot-dashed lines represent the median and mean values of the distribution, respectively.}
\label{fig:timescale}
\end{figure}

\begin{figure}
\centering
\includegraphics[width=8.5cm]{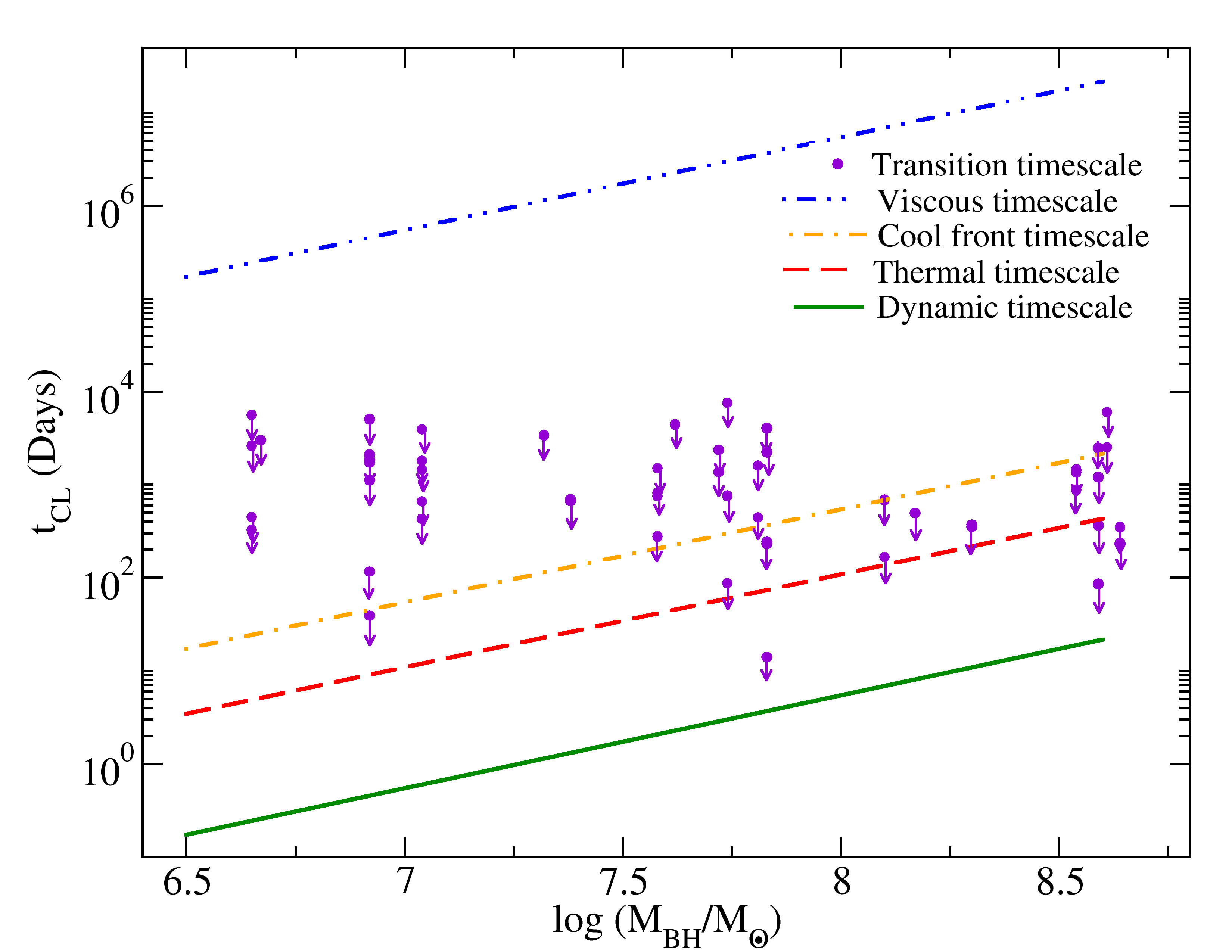}
\caption{Relation of transition timescale ($t_{\rm CL}$) with the black hole mass in logarithm scale ($\log M_{\rm BH}$). The purple down arrow represent the upper limit of the transition time for all CL transitions in our study. The blue dashed-dot-dot-dashed, orange dot-dashed, red dashed, and green solid lines represent the viscous time, cold front propagation time, thermal time, and dynamic time, respectively. The timescales are calculated assuming disk aspect ratio, H/R=0.2, and viscosity parameter, $\alpha=0.1$.}
\label{fig:mbh-clt}
\end{figure}

\begin{figure*}
\centering
\includegraphics[width=15cm]{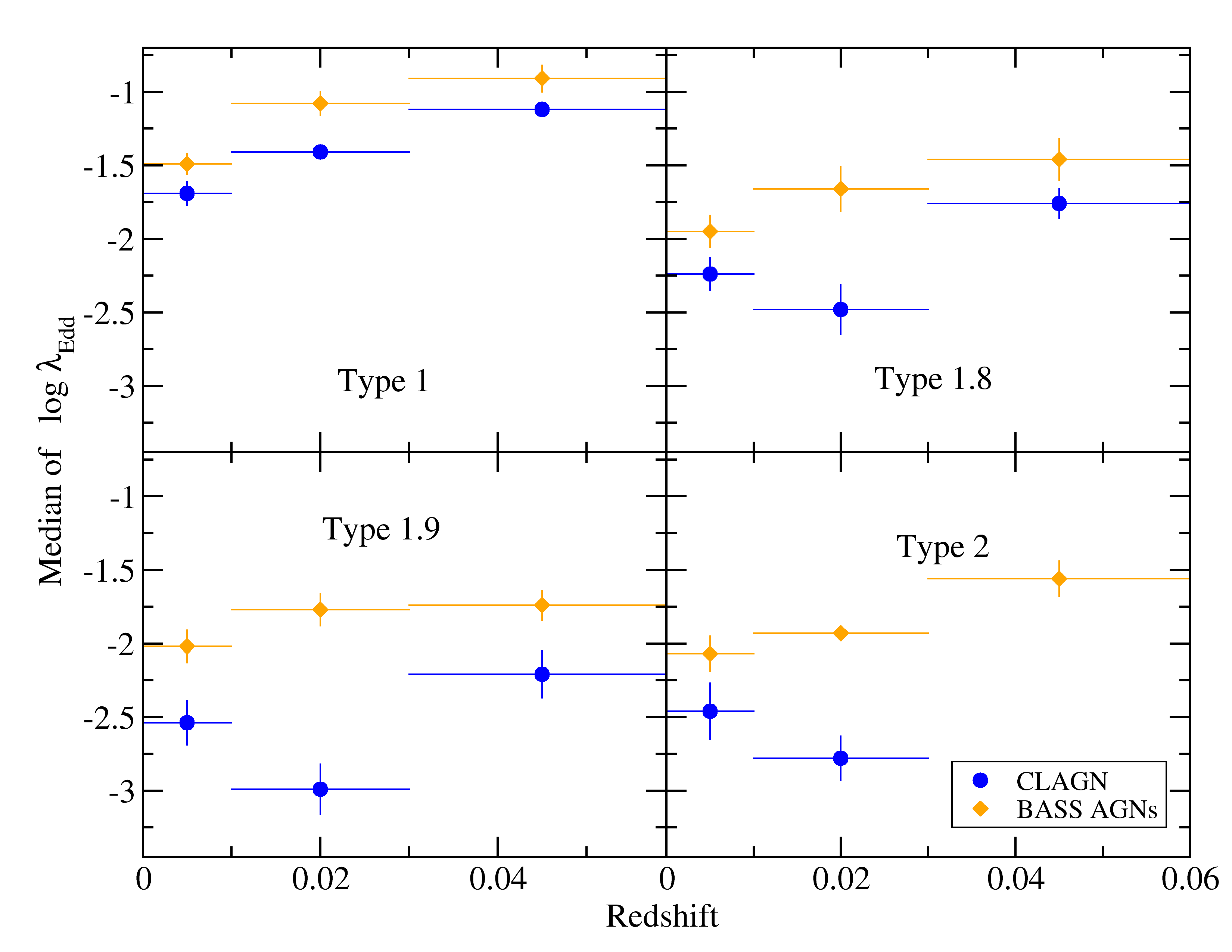}
\caption{Comparison of Median of Eddington ratio ($\lambda_{\rm Edd}$) between CLAGNs and other AGNs in the BASS sample with redshift. Three bins are constructed for redshift, with $\Delta z = 0-0.01$, $0.01-0.03$ and $0.03-0.06$, with the bins contain eight, seven and five CLAGNs, respectively. For the other AGNs from BASS, each bins are constructed by randomly selected the same of number of AGNs (as the number of CLAGNs in the same bin), and bootstrap process with 1000 realizations. The median of $\lambda_{\rm Edd}$ for type\,1, type\,1.8, type\,1.9 and type\,2 with redshift are show in the top left, top right, bottom left and bottom right panels, respectively.}
\label{fig:z-ed}
\end{figure*}

\section{Discussion}
\label{sec:discus}

\subsection{CLAGNs: Accretion vs Obscuration}
\label{subsec:acc-obs}

In our sample of 20 optically-identified CLAGN, we observed a clear correlation between the spectral state of each CLAGN and Eddington ratio (see Figure~\ref{fig:ed-state}). 
We observed that CLAGNs tend to transition to a type 1 spectral state as the Eddington ratio increases, and conversely, they move to a type 2 state as the Eddington ratio decreases. When we calculate the median of \ed at each spectral state, type\,1s states are found to have a higher \ed than type\,2s states (Figure~\ref{fig:ed-dist}).
The distribution function $f_1$ (fraction of CLAGNs in type\,1 state) also shows a clear increase as a function of \ed, indicating that AGNs are more likely to be in a type 1 state when their accretion rate is higher. Conversely, the fraction $f_2$ demonstrates the opposite behavior, with the fraction of type 2 AGNs decreasing as the Eddington ratio increases. This indicates that AGNs are more likely to be in a type 2 state when their accretion rate is lower. This implies that the observed transitions in spectral characteristics are strongly tied to variations in the accretion rate, with AGNs transitioning between spectral states with the change in the accretion rate. 

We also checked the relation between optical spectral state and X-ray obscuration for CLAGNs in our sample. No significant relation was detected between the spectral state and the line of sight column density (see Figure~\ref{fig:nh-state}). While the UM of AGNs posits that type 1 AGNs are typically unobscured (low \nh) and type 2 AGNs are obscured (high \nh), our results show that this relation does not hold for the majority of the CLAGNs in our sample. When we calculate the median of \nh at each spectral state, we did not find a significant variation of \nh with the spectral state (see Figure~\ref{fig:med-nh}).
The distribution of spectral state as a function of \nh (relation between $f_1$, $f_2$ and \nh) might suggest that the optical state could be related to \nh (see Figure~\ref{fig:nh-dist}). This would be consistent with the UM where type\,1s are typically unobscured and type\,2s are typically obscured. However, when we check how the spectral state changes with \nh for individual sources (see Fig~\ref{fig:nh-state}), we clearly see that there is no relation between these two quantities for 19 of the 20 sources of our sample. Moreover, the changes in $f_1$ and $f_2$ with \nh are within uncertainties, indicating that the optical state is not directly tied to the X-ray obscuration. Hence, consistent with the results of \citet{Temple2023}, there is no clear indication of \nh being the driver of the CL transitions.

Instead, the CLAGNs in our sample appear to change their optical and X-ray properties due to intrinsic changes in the accretion flow rather than external factors such as varying obscuration or material along the line of sight. This behavior supports the idea that optically-identified CLAGNs can be classified as changing-state AGNs (CSAGNs), where the optical state transitions are directly linked with the variation of accretion flow around SMBHs.

\subsection{CLAGN with CS and CO Transition}
\label{subsec:CS-CO}

Two CLAGNs in our sample, NGC\,1365 and NGC\,7582, have undergone both changing-state (CS) and changing-obscuration (CO) transitions \citep{Risaliti2005, Piconcelli2007, Bianchi2009, Temple2023}. These two sources provide a unique opportunity to investigate the potential relationship between the CS and CO transitions. In our analysis, we explored whether the CS and CO transitions are connected.

\subsubsection*{NGC\,1365}
NGC\,1365 was observed in the CT state in July 2010, and optical observations carried out in September 2010 revealed that the source was in a type\,2 state, while accreting at $\log \lambda_{\rm Edd}\sim -1.56$ \citep{Brenneman2013}.
In December 2012, the source moved to type\,1.8 state \citep{Lena2016}. The X-ray observation found the source in a Compton thin state at this time, with increasing  $\log \lambda_{\rm Edd} \sim -1.36$ \citep{Walton2013}. From this, it may seem that both \ed and \nh are responsible for the CL transition in NGC\,1365.
However, NGC\,1365 showed rapid absorption variability on a timescale of days \citep{Risaliti2007}. The obscuring clouds are small and found to be located in the BLR, which suggests that the obscuring material cannot block the BLR. \citet{Mondal2022} suggested that the obscuration might be attributed to a failed wind, driven by the variable accretion rate. However, the wind can only contribute to the obscuration of the X-ray source and not affect our view of the BLR.

\subsubsection*{NGC\,7582}
NGC\,7582 showed a variable \nh over the years, with a CT state observed several times \citep[][and references therein]{Lefkir2023}. 
The \nh varied in the $\sim 8-130 \times 10^{22}$ \pcm range over the last $\sim 40$ years, undergoing several CO transitions. 
In 2005, \emph{XMM-Newton} found the source in a CT a state with $N_{\rm H} = (1.3\pm0.1)\times10^{24}$ \pcm \citep{Piconcelli2007}. 
In April 2007, the source was found in Compton thin state with $N_{\rm H}= (3.3\pm0.5)\times 10^{23}$\pcm. Within six months, 
the source transitioned again to a CT state [$N_{\rm H}= (1.2\pm0.2) \times 10^{24}$ \pcm]. In 2012, NGC\,7582 was found in a 
Compton-thin state and moved back to CT state in 2014 \citep{Braito2017}. In 2016, the source was again found in Compton-thin state \citep{Lefkir2023}.

Unfortunately, we do not have simultaneous optical observations during all the CO transitions. However, the \nh variations were observed when the source was in the type\,1.9/2 state, and only \ed was observed to correlate with the optical spectral state in this object, confirming the idea that CO \& CS transitions are independent and that the varying accretion rate drives the optical state transition.

\subsection{Timescale of the CL transitions}
\label{subsec:timescale}

The observed timescales of the CL transitions ($t_{\rm CL}$) challenge our understanding of the accretion properties of SMBHs. Generally, CL transitions are seen on timescales of months to decades (e.g., \citealp{Denney2014,Shapovalova2019,Gezari2017,Trakhtenbrot2019,Ricci2023Nat,Zeltyn2022}). In our current study, we focus on a sample of optically-identified CLAGNs, utilizing optical data accumulated over the past 40 years. By comparing the time intervals between the first and last epochs of observations in different spectral states, we were able to place upper limits on the CL transition timescales for each source. These upper limits provide interesting constraints on the temporal evolution of the accretion processes in AGNs. Figure~\ref{fig:timescale} displays the distribution of our upper limits on the CL transition timescales for our sample. We only considered the timescale for the type\,1 $\leftrightarrow$ 1.8/1.9/2 transition. The timescales cover a range from a few weeks to $\sim 20$ years. We find that the median of the upper limit of the transition timescale is $3-4$ years, which is consistent with previous findings for CLAGNs in BASS \citep{Temple2023}.

The standard thin disk model predicts the radial inflow timescale or viscous timescale to be $t_{\rm vis}\simeq 400~(\frac{H/R}{0.05})^{-2} (\alpha/0.03)^{-1}(R/150~r_{\rm g})^{3/2} M_8$ years \citep{SS73,Noda2018,Stern2018,Ricci2023Nat}. Here, $H$ is the disk height at distance $R$, $\alpha$ is the viscosity parameter, and $M_8$ is the mass of the SMBH in $10^8~M_{\odot}$. For AGNs of mass $\sim 10^{6-8}~M_{\odot}$, $t_{\rm vis}$ would be $\sim 10^{4-6}$ years, i.e. considerably larger than the observed transition time. The dynamical timescale ($t_{\rm dyn}$) is typically shorter than the observed timescale. The dynamic timescale of the gas is related to the orbital motion of the gas around the black hole and is given by $t_{\rm dyn} \simeq 10(R/150~r_{\rm g})^{3/2}~M_8$ days. The thermal timescale ($t_{\rm th}$) or the timescale associated with the heating or cooling of the disk is $t_{\rm th}\simeq t_{\rm dyn}/\alpha \simeq (\alpha/0.03)^{-1}(R/150~r_{\rm g})^{3/2} M_8$ years. Such timescale is generally associated with the stochastic variability of the AGN \citep{kelly2009}. Another relevant timescale is the timescale associated with the radial propagation of the heating and cooling front \citep[$t_{\rm front}$;][]{Osaki1996,Dubus2001}. The cooling front timescale is given by, $t_{\rm front}\simeq 20~ (\frac{H/R}{0.05})^{-1}(\alpha/0.03)^{-1}(R/150~r_{\rm g})^{3/2} M_8$ years \citep{Stern2018}.

In Figure~\ref{fig:mbh-clt}, we show several key theoretical timescales relevant to accretion disk physics, namely, viscous time ($t_{\rm vis}$), thermal time ($t_{\rm th}$), dynamic time ($t_{\rm dyn}$), and heat/cold front timescale ($t_{\rm front}$), along with the upper limit of the observed CL transition time ($t_{\rm CL}$) as a function of $M_{\rm BH}$. For our calculations we considered a disk-aspect ratio $H/R=0.2$, a viscosity parameter $\alpha=0.1$, and $R \sim 150~r_{\rm g}$ (i.e. the typical emission zone for UV-optical continuum \citep[e.g.,][]{Noda2018,Stern2018}. 

Figure~\ref{fig:mbh-clt} clearly shows that all transitions occurred at a shorter timescale than $t_{\rm vis}$. Most transition timescales are consistent with the thermal, cooling front, and dynamic timescales.
These timescales suggest that thermal instabilities or the propagation of heating and cooling fronts in the accretion disk may play a significant role in driving CL transitions. Also, some transition timescales could be consistent with the dynamic time. Here, we also note that one may reduce the $t_{\rm vis}$ if the accretion disk is inflated. This could occur if the total opacity of the disk increases due to heavy elements, which raise both the temperature and scale height \citep{Jiang2016}. Magnetic torques in the inner disk can also heat and expand the disk structure \citep{Agol2000}, while magnetic pressure in the upper layers of the disk contributes to further disk inflation \citep{Dexter2019}. Additionally, magnetically-driven disk winds can remove the angular momentum, further shortening the $t_{\rm vis}$ \citep{Feng2021mag}.

Interestingly, in IRAS\,23226--3843, a transition occurred on a timescale of $\sim 14$ days \citep{Kollatschny2023}, which could be associated with the dynamical time. 
This indicates that, in some cases, CL transitions could be driven by dynamic processes within the inner accretion disk. Such fast transitions are rare but highlight the need to consider multiple physical mechanisms that could influence the CL phenomena. We note that some transitions could be associated with the thermal timescales and others with the dynamic timescales. Establishing precise transition timescales is crucial for identifying the underlying physical processes.

\subsection{Comparing CLAGNs with other AGNs in the BASS}
\label{sec:comparison}

In our sample of CLAGNs, the median value of the $\log$\ed for type\,1, 1.8, 1.9, and 2.0 are $-1.30\pm0.09$, $-2.24\pm0.14$, $-2.59\pm0.13$ and $-2.71\pm0.12$, respectively. For unbeamed (non-blazar) AGNs in the BASS sample in which CL transitions were not detected (hereafter BASS AGNs),
\citet{Koss2022} found the median value of $\log \lambda_{\rm Edd}$ for type\,1, 1.8, 1.9, and 2.0 are $-0.99\pm0.07$, $-1.53\pm0.08$, $-1.81\pm0.11$ and $-1.90\pm0.09$, respectively. In every spectral state, CLAGNs have a lower \ed than AGNs in the BASS (see Figure~\ref{fig:ed-med}).

We also compared the median of \ed for CLAGNs with that of other AGNs from the BASS survey using a redshift-matched sample. 
To do this, we divided the CLAGNs into three redshift bins: $z = 0-0.01, 0.01-0.03$, and $0.03-0.06$, containing eight, seven, and five CLAGNs, respectively. For each bin, we calculated the median \ed of the CLAGNs. Similarly, we constructed three corresponding redshift bins for other BASS AGNs, randomly selecting the same number of AGNs in each bin, i.e., we randomly selected eight, seven, and five AGNs from BASS at redshift bins of $z = 0-0.01, 0.01-0.03$, and $0.03-0.06$, respectively. The median \ed for these AGNs was then estimated using a bootstrapping method with 1000 realizations. This allowed us to determine the median \ed for both the CLAGNs and the other BASS AGNs for each spectral state.
The variation in the median \ed with redshift for both CLAGNs and other AGNs is shown in Figure~\ref{fig:z-ed}. The panels display the median Eddington ratio for type\,1, type\,1.8, type\,1.9, and type\,2 AGNs in the top left, top right, bottom left, and bottom right panels, respectively. In all spectral states and redshift bins, we consistently found that CLAGNs exhibit a lower Eddington ratio compared to other BASS AGNs that did not show CL transitions. We employed the Anderson-Darling (AD) test to compare the distributions of the Eddington ratio for CLAGNs and other BASS AGNs across different spectral states. Our results indicate that the distributions are significantly different in each state, with a p-value of $p_{\rm AD}<0.001$. This finding remained consistent when we repeated the analysis using the redshift-matched distributions for CLAGNs and BASS AGNs.
Our findings that CLAGNs tend to have a lower \ed than other AGNs in the BASS, agree with previous studies \citep{Zeltyn2024,Wang2024}. In the SDSS-V survey, the median of \ed in CLAGNs is found to be $\sim 0.025$, while other AGNs have a median value of \ed $\sim 0.043$ \citep{Zeltyn2024}. The CL quasars are observed to have a lower \ed compared to the general population of the quasar in SDSS \citep{MacLeod2019}. \citet{Temple2023} also found a similar result from the BAT-selected CLAGN sample in the local universe. 

We also obtain a median of \nh for each spectral state for the CLAGNs in our sample. Figure~\ref{fig:med-nh} shows the median of \nh for CLAGNs and other AGNs in each spectral state. For BASS AGNs, the median of \nh increases as the AGNs move towards type\,2 state, which is consistent with the UM of AGNs. Comparing CLAGNs with other AGNs in BASS \citep{Ricci2017apjs}, we find that the median $N_{\rm H}$ for type\,1 and 1.8 for the CLAGNs are higher than other AGNs. For CLAGNs, the median varies in the range of $\log (N_{\rm H}/{\rm cm^2}) = 21.45 - 21.88$. For other AGNs in BASS, the range of the median is $\log (N_{\rm H}/{\rm cm^2}) = 20.00 - 23.27$. The range of \nh suggests that \nh tends to be less variable for CLAGNs than other AGNs in the BASS, indicating that the CL transition does not depend on the obscuration properties. Using the AD test, we found significant differences in the distributions in all the spectral states, with p-values $<0.001$.

\subsection{The physical mechanisms responsible for CL transitions}
\label{subsec:reason}

From our study of 20 optically-identified changing-look AGNs (CLAGNs) using quasi-simultaneous optical and X-ray observations, we find that changes in the accretion flow are the primary driver of CL transitions in all sources, while we did not detect any significant variations in the obscuration properties of these CLAGNs associated with the transitions (see sections ~\ref{sec:ed-state} \& and ~\ref{subsec:nh-state}). This suggests that the optical CLAGNs in our sample can indeed be classified as changing-state AGNs (CSAGNs), where changes in the accretion flow rather than external factors, like obscuration, dictate the transitions between spectral states.

In recent years, several models have been proposed to explain these changing-state (CS) transitions in AGNs \citep{Elitzur2012,Noda2018,Sniegowska2020}. One prominent model is the disk-wind model for the BLR \citep{Nicastro2000,Elitzur2009,Elitzur2014}, which predicts that the BLR should disappear when the AGN luminosity falls below a critical value. This model relies on the idea that radiation pressure driven wind is responsible for the formation of the BLR, and if the bolometric luminosity ($L_{\rm bol}$) falls below a critical value, $L_{\rm crit}=2.3\times 10^{40} M_{8}^{2/3}$ erg/s, the BLR would no longer be sustained. As a result, the BLR vanishes, and the AGN transitions into a type\,2 state. This model provides an effective way to explain why some AGNs undergo transitions from type\,1 to type\,2, linking the appearance of the BLR directly to the strength of the accretion-powered radiation.

However, in our study, we found that this model does not fully explain the observed CL transitions. Specifically, we found that in our sample of 20 optical CLAGNs, all sources have bolometric luminosities well above the critical threshold predicted by the disk-wind model, in their type\,2 states. This suggests that the disk-wind model is not sufficient to explain all CL transitions, particularly those where the AGN retains a high bolometric luminosity. The disappearance of the BLR in these cases likely involves more complex processes tied to the dynamics of the accretion flow or changes in the structure of the central regions of the AGN, rather than simply a drop in luminosity. This challenges the universality of the disk-wind model and points to the need for alternative models that can account for the complex interplay between accretion processes and BLR formation in AGNs undergoing changing-look transitions.

The disk instability model also provides a key framework for understanding the changing-state (CS) transitions in AGNs, linking these transitions to the spectral state changes commonly observed in black hole X-ray binaries (BHXBs; \citealp{Noda2018, Ross2018, Ai2020}). In BHXBs, state transitions are well-studied, and the change in accretion geometry during these transitions leaves a distinct imprint on the correlation between the photon index ($\Gamma$) and the Eddington ratio \citep[\ed, e.g.,][]{Yang2015,Yan2020}. This $\Gamma-\lambda_{\rm Edd}$ correlation acts as a diagnostic of the accretion state, providing valuable insight into the physical mechanisms governing the behavior of both BHXBs and AGNs.

In BHXBs, the $\Gamma-\lambda_{\rm Edd}$ correlation behaves differently depending on whether the source is in a high-soft or low-hard state. During the high-soft state, where the system is dominated by thermal emission from the accretion disk, a positive correlation between $\Gamma$ and \ed is typically observed. In this state, increasing accretion rate leads to efficient cooling in the X-ray corona, which produces softer spectra, leading to an increase in $\Gamma$ \citep{Yang2015,Yan2020,AJ2022b}. 
Conversely, in the low-hard state, a negative correlation between $\Gamma$ and \ed is observed. In this state, the accretion disk recedes, and the inner accretion flow is replaced by a hot radiatively inefficient flow \citep{Zdziarski2014,Yuan2015}. The seed photons are supplied by the synchrotron emission in the hot flow or jets. As the accretion rate decreases, the degree of synchrotron self-absorption decreases, which leads to softer spectra, i.e., increases in $\Gamma$. The critical Eddington ratio at which this correlation flips is found to be around $\lambda_{\rm Edd} \sim 0.01$, marking the transition between the high-soft and low-hard states.

A similar behavior in the $\Gamma-\lambda_{\rm Edd}$ relation is also observed in AGNs, suggesting that the accretion physics in AGNs and BHXBs may be fundamentally similar. In AGNs, this transition in the $\Gamma-\lambda_{\rm Edd}$ correlation typically occurs at $\lambda_{\rm Edd} \approx 0.01-0.02$ \citep{Noda2018,Ruan2019,AJ2023}, which is comparable to the value observed in BHXBs. Studies of changing-look quasars have further supported this connection, showing that quasars evolve from a bright, high-accretion state to a faint, low-accretion state, and vice-versa, with the transition also occurring at $\lambda_{\rm Edd} \approx 0.01$ \citep{Ruan2019,Jin2021}. This suggests that the same underlying physical mechanisms, likely driven by instabilities in the accretion disk, are responsible for the observed state changes.

In the present work, we have found that the median Eddington ratio for CLAGNs during state transitions, $\lambda_{\rm Edd}^{\rm tr} \approx 0.005-0.015$. This value is consistent with the soft-to-hard state transition Eddington ratio seen in BHXBs \citep{Maccarone2003,AJ2022b}, further supporting the hypothesis that disk instabilities are the primary drivers of these transitions in CLAGNs. Specifically, these instabilities likely alter the structure and geometry of the accretion flow, leading to changes in the accretion rate and, consequently, the spectral state of the CLAGNs.

Our sample does not include extreme low accreting (\ed$<10^{-4}$) or high accreting AGNs (super Eddington source, \ed$>1$). For instance, 1ES 1927+654 was found to be accreting at super Eddington rate during the CL transitions \citep{Trakhtenbrot2019,Ricci2020,Ricci2021,Li20221e1927}, with the transition driven by the change in the accretion rate \citep{Li2024b,Li2024a}. On the other hand, several low luminosity changing-look LINERs have been detected \citep{Schimoia2015}. We will study these objects in detail elsewhere.

\section{Summary and conclusions}
\label{sec:summary}

We conducted a comprehensive study of 20 optically identified changing-look active galactic nuclei in the local Universe ($z<0.06$) using quasi-simultaneous optical and X-ray observations from BASS.
This multi-wavelength approach allowed us to explore the connection between changes in the accretion processes and the spectral properties of these AGNs over time. The quasi-simultaneous X-ray and optical data provide crucial insights into the physical mechanisms driving the observed transitions. In this study, we utilized optical and X-ray data from the literature in the last 40\,years. The optical spectral state is classified using optical observations, while the Eddington ratio and line-of-sight hydrogen column density are estimated from X-ray observations. We investigated the dependency of CL transitions on different AGN parameters, such as Eddington ratio, obscuration, and black hole mass. The key findings of our work are summarized below.

\begin{enumerate}
\item The CL transitions are likely caused by changes in the accretion flow. In our sample, all sources show type\,1 $\rightarrow$ 2 transition as \ed decreases,
and vice-versa.
\item The CL transitions are not related to obscuration properties, confirming the idea the CS transitions are solely led by the changes in the accretion flow.
\item CLAGNs are found to have a lower accretion rate than AGNs from the BASS sample for which CL transitions have not been detected. 
\item The median of transition Eddington ratio for type\,1 $\leftrightarrow $ 1.8/1.9/2 is $\log \lambda_{\rm Edd}^{\rm tr} = -2.01\pm0.23$, or $\lambda_{\rm Edd} \approx 0.5-2$\% of Eddington limit. 
The $\lambda_{\rm Edd}^{\rm tr}$ is consistent with the prediction of the disk instability model \citep[e.g.,][]{Noda2018}. The $\lambda_{\rm Edd}^{\rm tr}$ is consistent with the transition Eddington ratio of the soft$\leftrightarrow$hard state transition Eddington ratio in BHXBs.
\item We could only estimate the upper limit of the CL transition times of our sample. We find that the majority of CL transitions in our sample occurred within a timescale of $3-4$ years.
\end{enumerate}

Currently, the main challenge of the study of CLAGNs is low cadence observations, which is not ideal to study the physics underlying the transition mechanism. This will change with the advent of large photometric \citep[LSST;][]{Ivezic2019} and spectroscopic surveys in optical (\citealp[SDSS-V,][]{Kollmeier2017}; \citealp[4MOST,][]{deJong2019}), wide-field surveys in the X-rays (\citealp[with eROSITA,][]{Merloni2020}; and \citealp[Einstein probe,][]{Yuan2015}) and UV wavelength \citep[with ULTRASAT;][]{Shvartzvald2023}. These surveys are expected to identify a large number of new CLAGNs, as well as new transitions of known CLAGNs, which will help to understand the physical mechanism of the spectral transitions in detail. Additionally, in the future, we will also investigate the connection of CLAGNs with state transition in BHXBs using archival multi-wavelength observations.

\section*{Acknowledgements}
We acknowledge the reviewer for their very detailed and helpful comments on the manuscript.
AJ acknowledges support from FONDECYT Postodoctoral fellowship (3230303). 
AJ and HK acknowledge the support of the grant from the National Science and Technology Council of Taiwan with the grand numbers MOST 110-2811-M-007-500, MOST 111-2811-M-007-002, and NSTC 112-2112-M-007-053. 
CR acknowledges support from Fondecyt Regular grant 1230345, ANID BASAL project FB210003 and the China-Chile joint research fund.
MJT acknowledges support from STFC grant ST/X001075/1 and a FONDECYT Postdoctoral fellowship (3220516).
BT is supported by the European Research Council (ERC) under the European Union's Horizon 2020 research and innovation program (grant agreement number 950533) and from the Israel Science Foundation (grant number 1849/19). 
YD is supported by a FONDECYT postdoctoral fellowship (3230310). 
DI acknowledges funding provided by the University of Belgrade - Faculty of Mathematics (the contract \textnumero 451-03-66/2024-03/200104) through the grant of the Ministry of Science, Technological Development and Innovation of the Republic of Serbia.

\section*{DATA AVAILABILITY}
All the data used in the paper are publicly available.

\bibliographystyle{aa}
\bibliography{ref-clagn}


\newpage
\appendix
\newpage

\section{Data Reduction}
In the present study, we reduced the data obtained from the \emph{Swift}/XRT and \emph{XMM-Newton}.

\subsection{Swift/XRT}
\label{sec:swift}
The $0.5-8$ keV \swift/XRT spectra were generated using the standard online tools provided by the UK Swift Science Data Centre \citep{Evans2009}\footnote{\url{https://www.swift.ac.uk/user_objects/}}. We only utilized the data obtained in this work with the `photon counting' mode.

\subsection{XMM-Newton}
\label{sec:xmm}
We used {\it XMM-Newton}/EPIC-PN \citep{Jansen2001} observations in the $0.5-10$ keV energy range in our analysis. The data files were reduced using the Standard Analysis Software (SAS) version 20.0.0. The raw PN event files were processed using \textsc{epchain} task. We checked for particle background flare in the $10-12$ keV energy range. The Good Time Interval file was generated using the \texttt{SAS} task \textsc{tabgtigen}. The source and background spectra were extracted from a circular region of 30$\arcsec$ centered at the position of the optical counterpart and from a circular region of 30$\arcsec$ radius away from the source, respectively. The background region is selected in the same CCD where no other X-ray sources are present. Using \textsc{especget} task, we generated the source and background spectra. We checked for pileup using the \textsc{epatplot} task. We did not find any source that suffered from the pileup.

\section{Observations \& Results}
\label{sec:obs-res-source}

\subsection{NGC\,526A}\label{subsec:ngc526}
Historically, NGC\,526A is known to be a type\,2 AGN. In 1978, the source resembled the spectra of type\,1 state \citep{Griffiths1979}. In 1986 observation, the source lost its broad H$\beta$ line and was classified as type\,1.9 Seyfert \citep{Winkler1992b}. In 2004, the source was still found in type\,1.9 state \citep{Bennert2006}.
However, in July 2009, the broad lines were not observed, making the source again type\,2. In September 2016, a broad H$\alpha$ line reappeared as the spectral state changed to type\,1.9 \citep{Temple2023}. 

NGC\,526A was observed in X-rays over the years. The \nh~ was observed to be in the range of $N_{\rm H}\sim (1-4)\times 10^{22}$ \pcm. We did not have much simultaneous/quasi-simultaneous optical and X-ray data for NCG 526A. We observed $\lambda_{\rm Edd} \sim 0.03$ and $\sim 0.006$ for type\,1 and type\,2 states, respectively. The CL transition in NGC\,526A is likely to be caused by the change in accretion rate.

\begin{table*}
\caption{NGC\,526A}
\centering
\begin{tabular}{cccccccccccc}
\hline
Dates & H$\alpha$ & H$\beta$ & Optical & Optical & $\log L_{\rm X}$ &$\log \lambda_{\rm Edd}$ & $\log N_{\rm H}$ &X--ray & Ref. & $\Delta T$\\
UT & BEL & BEL & type & Inst. & & & $\log({\rm cm}^{-2})$ & Inst.& & (Days) \\
\hline
\hline
1977/12/24& --& --& -- & -- & $43.53\pm0.12$ & $-1.58\pm0.13$ & -- & H-1 & 1 & -- \\
\hline
1978 & Y &Y& 1.0 & AAT & $43.52\pm0.12$ & $-1.59\pm0.17$ & -- & H-1 & 1, 1 & $<365$\\
\hline
1983/09/06& --& --& -- & -- & $43.29\pm0.05$ & $-1.98\pm0.23$ & $22.60\pm0.15$ & EXO & 2 & --\\
\hline
1986/11/19& Y& N& 1.9 & SAAO  & -- & -- & -- & -- & 3 & --\\
1987/07/30& Y& N& 1.9 & SAAO  & -- & -- & -- & -- & 3 & 504\\
1988/12/15& --& --& -- & --   &  $42.98\pm0.04$ & $-2.19\pm0.21$ & $22.00\pm 0.16$ & G & 4 & -- \\
\hline
2003/06/22& --& --& -- & -- &  $43.07\pm0.01$ & $-2.09\pm0.11$ & $22.60\pm0.17$ & RXTE & 5 & 452 \\
2004/09/16& Y& N& 1.9 & NTT & -- & -- & -- & -- & 6 & --\\
\hline
2009/07/20& N& N& 2.0 & CTIO & -- & -- & -- & -- & 7 & -- \\
2011/01/17& --& --& -- & -- & $43.60\pm0.04$ & $-1.49\pm0.11$ & $22.09\pm0.01$ & Suz & 8 & --\\
2013/12/22& --& --& -- & -- & $43.44\pm0.01$ &$-1.68\pm0.11$ & $22.02\pm0.01$& XMM & 9 & --\\
2016/09/12& Y& N& 1.9 & duPont & -- & -- & -- & -- & 7 & --\\
2018/08/04& Y& N& 1.9 & VLT & -- & -- & -- & -- & 7 & --\\
\hline
\end{tabular}
\leftline{References : (1) \citet{Griffiths1979}, (2) \citet{Turner1989}, (3) \citet{Winkler1992b},}
\leftline{(4) \citet{Nandra1994}, (5) \citet{Rivers2013}, (6) \citet{Bennert2006}, (7) \citet{Temple2023},}
\leftline{(8) \citet{Kawamuro2016}, (9) \citet{Laha2020}.}
\leftline{Optical instruments: AAT: AAT/IPCS; LCO: LCO 0.6m/EMI; CTIO: CTIO 1.5m/RC; NTT: ESO-NTT/EMMI;}
\leftline{duPont: duPont/BC; VLT: VLT/Xshooter.}
\leftline{X-ray instruments: H-1: \emph{HEAO--1}; EXO: \emph{EXOSAT}/LE+ME; G: \emph{Ginga}/LAC; XMM: \emph{XMM-Newton}/EPIC-PN; }
\leftline{RXTE: \emph{RXTE}/PCA; Suz: \emph{Suzaku}/XIS.}
\label{tab:ngc526a}
\end{table*}

\subsection{NGC\,1365}\label{subsec:ngc1365}
NGC\,1365 is one of two AGNs that showed both CS and CO events in the past. NGC\,1365 is known to show rapid absorption variability. NGC\,1365 was initially classified as type\,2 with no broad lines in the optical spectra observed in 1978 \citep{Phillips1980}. \citet{Edmunds1982} re-classified the source as type\,1 based on the observations made in November 1979. The source remained as type\,1 AGN as observed in August 1993.

In 1995, the source was found in CT state \citep{Iyomoto1997}, while three years later, the source moved to Compton thin state \citep{Risaliti2000}. In 2002-03, the source showed CO transition within six weeks \citep{Risaliti2005}. In 2007, 1365 showed CO transition in a timescale of days \citep{Risaliti2007}. However, no optical observations were available at that time.

NGC\,1365 was found in type\,1.9 state in January 2009 with a weak BEL H$\alpha$, and NEL H$\beta$ \citep{Trippe2010}. In June 2010, {\it Suzaku} observation of the source found $N_{\rm H}\sim 6\times10^{23}$ \pcm. The source moved to CT state in August 2010, with $N_{\rm H} \sim 1.1 \times 10^{24}$ \pcm, and $\lambda \sim 0.03$ \citep{Brenneman2013}. The \ed was about $\sim 30\%$ lower than the June observation. In September 2010, no broad lines were observed in the spectrum, as the source was in the type\,2 state. This could be related to the CT state observed in August 2010.

NGC\,1365 was found in the CT state with {\it Chandra} observation in April 2012 \citep{Nardini2015}. In July 2012, the source again moved to Compton thin state with $N_{\rm H}\sim 2\times 10^{23}$ \pcm. The source was observed to be in type\,1.8 states in November 2012, with broad lines \citep{Lena2016}. The Eddington ratio also increased to $\lambda_{\rm Edd}\sim 0.04$ in December 2012, as observed by {\it XMM-Newton} and {\it NuSTAR} \citep{Walton2013}. The source remained in the type\,1.8 state in January 2013 \citep{Lena2016}, and the Eddington ratio increased slightly. During this time, the X-ray observations revealed a drop in the \nh, with $\sim 10^{22}$ \pcm \citep{Liu2021}. In this period, the appearance of broad lines could be related to both an increase in the \ed, and a decrease in the \nh. 

In December 2013, the broad lines strengthened as the source moved to type\,1 state. The October 2014 observation also found the source in the same state. The source returned to type\,2 in December 2021 \citep{Temple2023}. Subsequently, the Eddington ratio decreases to $\lambda_{\rm Edd} \sim 0.01$, as found from the Swift/XRT observation in December 2021. The source was found in Compton-thin state at this time with $N_{\rm H}\sim 3\times10^{22}$ \pcm.

We found that NGC\,1365 was in the type\,1 for $\lambda_{\rm Edd}>0.1$. The source was in type\,2 when $\lambda_{\rm Edd}<0.01$. During type\,1.8 state, we found \ed in the range of $\sim 0.03-0.05$. The source is also found in the CT state during the type\,2 state, while it moved to the Compton thin state during the type\,1-1.8 state. It seems that both \nh and \ed drive the CS transition in NGC\,1365. However, the source exhibits rapid variability, which suggests that the location of the obscuring materials is BLR \citep{Walton2013}. 
It is also suggested that the obscuration could be attributed to the failed-wind, driven by accretion rate. Hence, the variable accretion rate is most likely cause the CL transition in the source. Nonetheless, one needs to study the source in more detail to understand the CL transition in this source.

\begin{table*}
\caption{NGC\,1365}
\centering
\begin{tabular}{cccccccccccc}
\hline
Dates & H$\alpha$ & H$\beta$ & Optical & Optical & $\log L_{\rm X}$ &$\log \lambda_{\rm Edd}$ & $\log N_{\rm H}$ &X--ray & Ref. & $\Delta T$\\
UT & BEL & BEL & type & Inst. & & & $\log({\rm cm}^{-2})$ & Inst.& & (Days) \\
\hline
\hline
1978/08/27 & N & N & 2.0 & CTIO1    &  -- & -- & -- & -- & 1 & --\\
1979/08/17 & -- & -- & -- & -- & $42.63\pm0.03$ & $-0.82\pm0.09$ &-- & Ein. & 2 & 91 \\
1979/11/16 & Y & Y & 1.0 & AAT & -- & -- & -- & -- & 3 & --\\
\hline
1980/08/15 & -- &  -- & 1.0 & -- & $42.61\pm0.03$ &  $-0.85\pm0.09$ & -- & Ein  & 2 & --  \\
\hline
1993/08/28 & Y & Y & 1.0 & ESO & -- & -- & -- & -- & 4 & -- \\
1994/08/12 & -- & --  & --      & --  & $42.03\pm0.15$ & $-1.55\pm0.15$ & $20.18^*$& ASCA & 5 &--\\
\hline
2009/01/08 & Y & N & 1.9 & CTIO1 & -- & -- & -- &--& 6 & --\\
2010/07/15 & -- & -- & -- & -- & $42.02\pm0.03$ & $-1.56\pm0.09$ & $24.03\pm0.01$ & Suz & 7 & 64\\
2010/09/17 & N & N & 2.0 & CTIO2 & -- & -- & -- &--& 8 & --\\
\hline
2012/04/09 & -- & -- & -- & -- & $41.99\pm0.03$ & $-1.56\pm0.09$ & $24.30$ & Ch & 9 & --\\
2012/07/25 & -- & -- & -- & -- & $41.98\pm0.03$ & $-1.56\pm0.09$ & $23.39\pm0.05$ & XMM, Nu & 10 & --\\
\hline
2012/11/03 & Y & Y & 1.8 & Gemini & -- & -- & -- & -- & 11 & 51\\
2012/12/24 & -- & -- & -- & -- & $42.19\pm0.02$ & $-1.36\pm0.09$ &$22.85\pm0.04$ & XMM, Nu & 12 & --\\
\hline
2013/01/18 & Y & Y & 1.8 & Gemini & -- & -- & -- & -- & 11 & 5\\
2013/01/23 & -- & -- & -- & -- & $42.12\pm0.01$ & $-1.45\pm0.09$ &$22.76\pm0.08$ & XMM, Nu & 12 & --\\
\hline
2013/12/10 & Y & Y & 1.0 & VLT-X & -- & -- & --& -- & 8 & --\\
2014/10/12 & Y & Y & 1.0 & VLT-M & -- & -- & -- & -- & 13 & --\\
\hline
2021/12/12 & N & N & 2.0 & Mag & -- & -- & -- & -- & 8 & 13\\
2021/12/25 & -- & -- & -- & -- & $41.67\pm0.03$ & $-1.96\pm0.09$ & $22.51\pm0.03$ & Swift & 14 & --\\
\hline
\end{tabular}
\leftline{References: (1) \citet{Phillips1980}, (2) \citet{Maccacaro1982}, (3) \citet{Edmunds1982}, (4) \citet{Schulz1999}.}
\leftline{(5) \citet{Iyomoto1997}, (6) \citet{Trippe2010}, (7) \citet{Brenneman2013}, (8) \citet{Temple2023},}
\leftline{(9) \citep{Nardini2015}, (10) \citep{Rivers2015}, (11) \citet{Lena2016}, (12) \citet{Walton2013},}
\leftline{(13) \citet{Venturi2018}, (14) This work.}
\leftline{Optical instruments: CTIO1: CTIO 1.5m/RCS; AAT: AAT/RGOS; CTIO2: CTIO 2.1m/RCS; ESO: ESO 3.6m/CAS;}
\leftline{Gemini: GMOS 8.1m/IFU; VLT-X: VLT/Xshooter, VLT-M: VLT-MUSE; Mag: Magellan/MagE.}
\leftline{X-ray instruments: Ein.: \emph{Einstein}/MPC; ASCA: \emph{ASCA}/GIS+SIS XMM: \emph{XMM-Newton}/EPIC-PN; Suz: \emph{Suzaku}/XIS;}
\leftline{Nu: \emph{NuSTAR}/FMPA+FMPB; Swift: \emph{Swift}/XRT.}
\label{tab:ngc1365}
\end{table*}

\subsection{NGC\,1566}\label{subsec:ngc1566}

NGC\,1566 is one of the first AGNs that showed optical variability \citep{Pastoriza1970,devaucouleurs1973,Osmer1974}.
In 1979, NGC\,1566 was in a type\,1 state with \ed $\simeq 0.007$. The source remained in type\,1 state until April 1984 \citep{Alloin1986}.
In August 1984, NGC\,1566 was observed to be in type\,1.8 states with $\lambda_{\rm Edd} \sim 0.002$. The source moved to type\,1.9 state in October 1985 with the disappearance of the broad H$\beta$ line. During this observation, The Eddington ratio decreased to $\log \lambda_{\rm Edd} \sim 0.001$.

In November 1985, the broad H$\beta$ line recovered as the source moved to type\,1.8, with the Eddington ratio increased to $\lambda_{\rm Edd} \sim 0.002$. In 1991, NGC\,1566 was observed to be in type\,1 state \citep{Kriss1991}. In the next two decades, NGC\,1566 was found to move between type\,1--1.9 states few times \citep{Aguero2004,Koss2017,Ochmann2024}. During these observations, no simultaneous X-ray observations were available.
NGC\,1566 showed an outburst in June 2018, with the flux in all wavebands increased \citep{Oknyansky2019,Oknyansky2020}. During this time, the source moved to type\,1 state with the $\lambda_{\rm Edd} \sim 0.01-0.3$ \citep{Oknyansky2019,AJ2021,Ochmann2024}. 

NGC\,1566 showed several CL events over the years. We found that the increasing \ed moved the source towards the type\,1 state. The source was generally found in the type\,1 state for $\lambda_{\rm Edd} > 0.005$. NGC\,1566 was observed to be unobscured over the years, with $N_{\rm H}\sim 10^{21}$ \pcm. We did not observe a change in \nh, even though optical classification changed in this source. In the 2018 outburst of the source, an ionized absorber was seen with $N_{\rm H}\sim 10^{22}$ \pcm \citep{AJ2021} when the source was in the high state. In NGC\,1566, the variable \ed is responsible for the optical state change.

\begin{table*}
\caption{NGC\,1566}
\centering
\begin{tabular}{ccccccccccc}
\hline
Dates & H$\alpha$ & H$\beta$ & Optical & Optical &   $\log L_{\rm X}$  &$\log \lambda_{\rm Edd}$ & $\log N_{\rm H}$ &X--ray & Ref. & $\Delta T$\\
UT    & BEL       & BEL      & type    & Inst. & &   &     $\log({\rm cm}^{-2})$ & Inst.&  &  (Days)   \\
\hline
\hline
\hline
1979/08/09 & -- & -- & --   &  -- & 41.77 &  $-2.15\pm0.07$ & -- &Ein. &  1 & 7 \\
1979/08/16 & Y & Y & 1.0 & AAT2 & --     & -- & -- & -- & 2 & --\\
\hline
1984/04/07 & Y & Y & 1.0 & CTIO & --     & -- & -- & -- & 2 & --\\
\hline
1984/08/23 & --   & --  & -- & -- & $41.27\pm0.03$   & $-2.67\pm0.06$  & $21.23\pm0.05$   & EXO & 3 & 2\\
1984/08/25 & Y & Y & 1.8 & ESO2 & -- & --  & -- & -- & 2 & --\\
\hline
1984/10/03 & Y & Y &  1.8 & ESO2  & $41.43\pm0.02$ &  $-2.51\pm0.05$ & $21.20\pm1.00$  &  EXO & 2, 3 & 27 \\
1984/10/30 & Y & Y &  1.8 & ESO2  & $41.34\pm0.03$  & $-2.60\pm0.06$ & $21.23\pm1.00$  & EXO  & 2, 3 &--\\
\hline
1985/02/23 & Y & Y     & 1.8 & ESO2 & -- & -- & -- & -- &2 & --\\
1985/10/11 & Y & N     & 1.9 & ESO2 & -- & -- & -- & -- &3 & 9\\
1985/10/20 & -- & --    & -- & -- & $41.07\pm0.03$  & $-2.88\pm0.06$  &$21.23\pm1.00$  & EXO &  3& --\\
\hline
1985/11/15 & Y & Y & 1.8 & ESO2 & -- & -- & -- & -- & 2, 3 & 1\\
1985/11/16 & -- & -- & -- & -- & $41.22\pm0.03$  & $-2.72\pm0.05$ &$21.23\pm1.00$ & EXO &  3& --\\
1986/01/12 & -- & -- & -- & -- & $40.96\pm0.03$  & $-2.98\pm0.05$ &$21.23\pm1.00$ & EXO &  3& --\\
\hline
1991/02/08 & Y & Y     & 1.0 & HST & -- & -- & -- & -- & 4 & --\\
1992/09    & --  & --  & -- & -- &  $41.29\pm0.06$  & $-2.65\pm0.06$  & $21.23\pm1.00$  & ROSAT &  5 & --\\
1996/11/09 & Y & Y &   1.8 & RCT & -- & -- & -- & -- & 6 & -- \\ 
\hline
2010/09/17 & Y & Y   & 1.0 & CTIO & -- & -- & -- & -- & 7 & -- \\
2012/05/19 & --  & --  & -- & -- &  $40.94\pm0.02$  & $-3.00\pm0.06$  & $22.48\pm0.12$  & Suz &  8& --\\
2013/10/10 & Y & Y & 1.9 & Gemini & -- & --  & -- & -- & 9 & --\\
\hline
2015/09/24 & Y & Y & 1.9 & VLT & -- & --  & -- & -- & 10 & 41 \\
2015/11/05  & --  & --   & -- &  -- & $41.53\pm0.01$  & $-2.40\pm0.05$  &$21.54\pm0.07$  & XMM  & 11 & --\\
\hline
2017/10/23 & Y & Y & 1.9 & VLT & -- & --  & -- & -- & 10 & --\\
\hline
2018/06/26 & -- & -- & -- & SAOO & $43.13\pm0.02$ &$-0.52\pm0.05$ & $21.89\pm0.05$ & XMM, Nu &  11 &  -- \\
\hline
2018/07/17  & --  & --   & -- &  -- & $42.98\pm0.01$  & $-0.72\pm0.03$  &$21.09\pm0.05$  & Swift  & 12 & 3 \\
2018/07/20 & Y & Y & 1.0 & SALT & -- & --  & -- & -- & 10 & --\\
\hline
2018/07/30 & Y & Y & 1.0 & SALT & -- & --  & -- & -- & 10 & --\\
2018/07/31  & --  & --   & -- &  -- & $42.76\pm0.02$  & $-1.00\pm0.03$  &$21.02\pm0.04$  & Swift  & 12 & 3 \\
\hline
2018/08/02 & Y & Y & 1.0 & SAOO & $42.62\pm0.02$  & $-1.17\pm0.05$ & $21.03\pm0.02$& XMM, Nu & 13, 11 & 0\\
2018/10/04 & Y & Y & 1.0 & SALT & $41.96\pm0.01$ & $-1.94\pm0.05$ & $21.02\pm0.43$ & XMM, Nu & 10, 12 & 0\\
2018/11/30 & Y & Y & 1.0 & SAOO & $41.99\pm0.02$ &$-1.91\pm0.05$ & $21.01\pm0.10$ & Swift & 14, 12& 0\\
2018/12/01 & Y & Y & 1.0 & SAOO & $42.14\pm0.02$ & $-1.74\pm0.05$ & $21.01\pm0.07$ &Swift & 14, 12& 0\\
2019/01/09 & Y & Y & 1.0 & SAOO & $42.03\pm0.02$ & $-1.86\pm0.05$ & $21.02\pm0.05$ & Swift & 14, 12& 0\\
2019/03/27 & Y & Y & 1.0 & SAOO & $41.62\pm0.01$ & $-2.31\pm0.05$ & $21.00\pm0.09$ & Swift & 14, 12 & 0\\
2019/08/28 & Y & Y & 1.0 & SAOO & $41.99\pm0.02$ & $-1.91\pm0.05$ & $21.11\pm0.40$&Swift & 14, 12& 0\\
\hline
2019/09/04 & Y & Y & 1.0 & SAOO & -- & -- &  -- & -- & 14 & 1\\
2019/09/05 & --  & --  & -- &  -- & $42.14\pm0.02$  & $-1.74\pm0.03$  &$21.05\pm0.04$  & Swift  & 12 & -- \\
2019/09/23 & Y & Y & 1.0 & SAOO & -- & -- & -- & -- &  14 & 3 \\
2019/09/26 & --  & --  & -- &  -- & $41.93\pm0.02$  & $-1.97\pm0.05$  &$21.01\pm0.05$  & Swift  & 12 & -- \\
\hline
\end{tabular}
\leftline{References: (1) \citet{Maccacaro1987}, (2) \citet{Alloin1986}, (3) \citet{Baribaud1992}, (4) \citet{Kriss1991},}
\leftline{(5) \citet{Ehle1996}, (6) \citet{Aguero2004}, (7) \citet{Koss2017}, (8) \citet{Kawamuro2013}, (9) \citet{dasilva2017},}
\leftline{(10) \citet{Ochmann2024}, (11) \citet{AJ2021}, (12) This Work, (13) \citet{Oknyansky2019}, (14) \citet{Oknyansky2020}.}
\leftline{Optical instruments: MSO: MSO/IT; ESO1: ESO/EC; AAT1: AAT/IDS; AAT2: AAT/IPCS; ESO2: ESO/IDS;}
\leftline{HST: HST/FOS; RCT: RCT/REOSC; CTIO: CTIO 1.5m/RCS; Gemini: GMOS 8.1m/IFU; VLT: VLT/MUSE;}
\leftline{SALT: SALT/RSS; SAOO: SAOO/MASTER-IIs.}
\leftline{X-ray instruments: Ein: \emph{Einstein}/IPC; EXO: \emph{EXOSAT}/ME; ROSAT: ROSAT/PSPC; XMM: \emph{XMM-Newton}/EPIC-PN;}
\leftline{Suz: \emph{Suzaku}/XIS; Nu: \emph{NuSTAR}/FPMA+FPMB; Swift: \emph{Swift}/XRT.}
\label{tab:ngc1566}
\end{table*}

\subsection{NGC\,2617}\label{subsec:ngc2617}
NGC\,2617 was found in type\,1.8 state in 1994--2003 \citep{Moran1996,Jones2004,Kollatschny2008}. The source moved to type\,1 state between 2003 and 2013 \citep{Shappee2014,Oknyasky2017}. The $\lambda_{\rm Edd}$ increased to $\sim 0.1$ in 2013, from $\lambda_{\rm Edd}\sim 0.01$ in 1992. The source remained in type\,1 state until 2023, with $\lambda_{\rm Edd}\sim 0.05-0.2$. In October 2023, NGC\,2617 lost its broad H$\beta$ line, and source moved to type\,1.9 state with $\lambda_{\rm Edd}\sim 0.003$ \citet{Oknyansky2023ATel}.

In our study, we found a clear relation between spectral state and \ed. The source was found in type\,1.8--1.9 state when $\lambda_{\rm Edd}<0.01$. 
For $\lambda_{\rm Edd}>0.03$, the source was observed in type\,1 state. NGC\,2617 was found to be an unobscured state ($N_{\rm H}\sim 10^{21}$ \pcm) 
in all spectral states. This suggests that the change in the accretion rate drives the optical state change in NGC\,2617.

\begin{table*}
\caption{NGC\,2617}
\centering
\begin{tabular}{ccccccccccc}
\hline
Dates & H$\alpha$ & H$\beta$ & Optical & Optical &   $\log L_{\rm X}$  &$\log \lambda_{\rm Edd}$  & $\log N_{\rm H}$ &X--ray & Ref. & $\Delta T$\\
UT    & BEL       & BEL      & type    & Inst.  &   &       &    $\log({\rm cm}^{-2})$ & Inst.&  &  (Days)   \\
\hline
\hline
1992       & -- & -- & -- &   -- & $41.74 $  & $-2.60\pm0.09$   & --  & ROSAT  & 1 & --\\
1994       & Y & Y & 1.8 & KPNO &  -- & --  & --  & -- & 2 & -- \\ 
1998/03/01 & Y & Y & 1.8 & CA &  -- & --  & --  & -- & 1 & -- \\
2003/12/30 & Y & Y & 1.8 & UKS  & --     & --    & -- & -- & 3 & --\\
\hline
2013/04/25 & Y & Y & 1.0 &  APO     & --    & -- & -- & -- & 4 & 2 \\
2013/04/27 &-- & -- & -- &   --           & $43.15\pm0.01$  & $-1.01\pm0.08$   & $21.14\pm0.07$  & XMM  & 5 & --\\
2013/04/29 & Y & Y & 1.0 & APO & --& -- & --     & -- & 4  & 3\\
\hline
2013/05/12 & Y & Y & 1.0 & MDM2 &  $43.11\pm0.02$ & $-1.06\pm0.08$   & $21.12\pm0.10$ & Swift & 4, 6 & 0\\
2013/05/19 & Y & Y & 1.0 & MDM2 &  $43.07\pm0.02$  &  $-1.11\pm0.08$  & $21.23\pm0.20$  &Swift & 4, 6 & 0\\
2013/05/20 & Y & Y & 1.0 & MDM2 &  $43.12\pm0.02$ & $-1.05\pm0.08$ & $21.28\pm0.01$ &Swift & 4, 6 & 0\\
2013/05/21 & Y & Y & 1.0 & MDM2 &  $43.16\pm0.01$ & $-1.00\pm0.08$  & $21.24\pm0.01$ &  Swift & 4, 6 & 0 \\
2013/05/24 & --  & --   & -- & -- & $43.46\pm0.01$ &$-0.61\pm0.06$ & $21.04\pm0.07$& XMM & 5 & 3\\
2013/05/27 & Y & Y & 1.0 & APO &  $43.37\pm0.02$ & $-0.73\pm0.05$  & $21.15\pm0.02$   & Swift  & 4, 6 & 0\\
\hline
2013/06/01 & Y & Y & 1.0 & MDM1 &  $42.86\pm0.01$   &  $-1.36\pm0.05$  & $21.07\pm0.03$   & Swift & 4, 6 & 0\\
2013/06/02 & Y & Y & 1.0 & MDM1 &  $42.87\pm0.02$ & $-1.35\pm0.07$ & $21.16\pm0.05$& Swift & 4, 6 & 0 \\
2013/06/04 & Y & Y & 1.0 & MDM1 &  $42.88\pm0.02$  & $-1.35\pm0.07$  & $21.12\pm0.02$ & Swift & 4, 6 & 0 \\
\hline
2014/01/04 & Y & Y & 1.0 & MDM1 &  $43.30\pm0.01$  & $-0.82\pm0.08$ & $21.24\pm0.24$ & Swift   & 7, 6 & 0\\
2014/01/05 & Y & Y & 1.0 & MDM1 &  $43.21\pm0.01$  & $-0.94\pm0.08$ & $21.13\pm0.26$& Swift   & 7, 6 & 0\\
2014/01/07 & Y & Y & 1.0 & MDM1 &  $43.10\pm0.01$   & $-1.07\pm0.08$ & $21.18\pm0.20$ &Swift  & 7, 6 & 0\\
2014/01/09 & Y & Y & 1.0 & MDM1 &  $43.00\pm0.02$  & $-1.20\pm0.08$ & $21.13\pm0.22$ & Swift & 7, 6 & 0\\
2014/01/10 & Y & Y & 1.0 & MDM1 &  $43.04\pm0.02$  & $-1.15\pm0.07$   & $21.11\pm0.16$  & Swift & 7, 6 & 0 \\
2014/01/11 & Y & Y & 1.0 & MDM1 &  $43.00\pm0.02$   & $-1.20\pm0.08$   & $21.23\pm0.26$ &Swift & 7, 6 & 0\\
2014/03/02 & Y & Y & 1.0 & MDM1 &  $42.91\pm0.02$  & $-1.30\pm0.08$   & $21.35\pm0.15$ & Swift  & 7, 6 & 0\\
\hline
2019/11/13 & Y & Y & 1.0 & Lijang & $42.69\pm0.02$ & $-1.56\pm0.08$ & $21.31\pm0.16$    & Swift   & 8, 6 & 0 \\
\hline
2021/10/30 & Y & Y & 1.0 & CMO & $42.54\pm0.02$ & $-1.74\pm0.08$ & $21.31\pm0.16$    & Swift   & 9, 6 & 0 \\
2022/01/30 & Y & Y & 1.0 & WIRO & $42.91\pm0.02$ & $-1.30\pm0.08$ & $21.31\pm0.16$    & Swift   & 9, 6 & 0 \\
\hline
2021/02/19 & Y & Y & 1.0 & CMO & -- & -- & --    & -- & 8 & 11 \\
2022/03/02 & -- & -- & -- & -- & $42.73\pm0.02$ & $-1.52\pm0.07$ & $21.31\pm0.16$    & Swift   &  6 & -- \\
\hline
2023/10/08* & Y & N & 1.9 & WIRO & $41.87\pm0.04$ & $-2.47\pm0.07$ & $21.31\pm0.16$    & Swift   & 10, 6 & $<30$ \\
\hline
\hline
\end{tabular}
\leftline{* Exact date of optical observation is not known.}
\leftline{Reference: (1) \citet{Kollatschny2008}, (2) \citet{Moran1996}, (3) \citet{Jones2004}, (4) \citet{Shappee2014}, }
\leftline{(5) \citet{Giustini2017}, (6) This Work, (7) \citet{Fausnaugh2017}, (8) \citet{Feng2021}, (9) \citet{Oknyansky2023}, }
\leftline{(10) \citet{Oknyansky2023ATel}.}
\leftline{Optical instruments: CA: CAO 2.2m/CAFOS; KPNO: KPNO 2m/GCS; UKS: UK Schmidt 1.2m (6dFS); }
\leftline{APO: APO 3.5m/DIS; MDM2: MDM 2.4m/OSMOS; MDM1: MDM 1.3m/CCDS; Lijang: Lijang 2.4m/YFOSC;}
\leftline{CMO: CMO 2.5m; WIRO: WIRO 2.3m. }
\leftline{X-ray instruments: ROSAT: \emph{ROSAT}/PSPC; XMM: \emph{XMM-Newton}/EPIC-PN; Swift: \emph{Swift}/XRT.}
\label{tab:ngc2617}
\end{table*}

\subsection{NGC 2992}\label{subsec:ngc2992}
NGC 2992 changed its spectral type several times in the last $\sim 40$ years \citep{Guolo2021}. Here, we briefly outlined the CL events of this source. In 1978 and 1979, the optical spectra showed that the source was in a type\,1.9 state, with a weak broad H$\alpha$ line in the spectra. It lost its broad H$\alpha$ line in 1985, as it was in the type\,2 state. The X-ray luminosity decreased as the $\lambda_{\rm Edd}$ changed from $\sim 0.01$ to $\sim 0.002$ in 1985. In 1994, the broad H$\alpha$ line appeared as it transitioned to type\,1.9 state \citep{Allen1999}. NGC\,2992 moved to type\,2 state in April 1998, and moved back to type\,1.9 state in November 1998.

In 2006, NGC 2992 lost its broad H$\alpha$ line again and moved to type\,2 state \citep{Trippe2008}. The X-ray flux was low with the $\lambda_{\rm Edd} \sim 0.002$ at that time. In 2014, both H$\alpha$ and H$\beta$ lines showed broad components, as the source moved to type\,1.8 \citep{Guolo2021}. The 2021 observations revealed that the source remained in type\,1.8 state \citep{Guolo2021}. The $\lambda_{\rm Edd}$ was observed to $\sim 0.01$ in 2021. \citet{Guolo2021} studied the source using long-term data, and they found that the spectral state is correlated with the X-ray flux. They showed that broad H$\alpha$ line appeared if $L_{\rm 2-10~keV} > 2.6\times10^{42}$ \eps, i.e., $\lambda_{\rm Edd}>0.01$. Our study also found the same result, i.e., the source moved to type\,2, if $\lambda_{\rm Edd}<0.01$. However, we could not distinguish the type\,1.8 and type\,1.9 states from the $\lambda_{\rm Edd}$ variation. It is possible that the old data did not detect the H$\beta$ BEL due to low SNR.

NGC 2992 showed a variable absorption over the years, with \nh varied in the range of $N_{\rm H} \sim (0.3-2) \times 10^{22}$ \pcm. We did not find any relation of \nh with the spectral state change. The change in the accretion rate is likely to be the reason for the CL transition in NGC 2992.

\begin{table*}
\caption{NGC 2992}
\centering
\begin{tabular}{ccccccccccc}
\hline
Dates & H$\alpha$ & H$\beta$ & Optical & Optical &   $\log L_{\rm X}$  &$\log \lambda_{\rm Edd}$  & $\log N_{\rm H}$ &X--ray & Ref. & $\Delta T$\\
UT    & BEL       & BEL      & type    & Inst.  &  $\log({\rm erg~s^{-1}})$ &       &    $\log({\rm cm}^{-2})$ & Inst.&  &  (Days)   \\
\hline
1978/11/22& --& -- & --  & --       & $42.86\pm0.06$ &  $-2.10\pm0.08$ & --  & Ein. & 1 & 32\\
1978/12/24& Y &  N & 1.9 & Lick & -- & -- & -- & -- & 2 & --\\
\hline
1979/05/24& Y &  N & 1.9 & ESO1   & -- & --  & -- & -- & 3 & 9\\
1979/06/02& --& -- & -- & --  & $43.09\pm0.04$ &   $-1.85\pm0.06$ & $22.16\pm0.17$ & Ein.  & 4 & --\\
\hline
1985/01/06& --& -- & --  & --      & $42.32\pm0.06$     & $-2.67\pm0.08$ & $22.39\pm0.15$ & EXO  & 5 & 66\\
1985/03/13& N & N  & 2.0 & CNPq  & --  & -- & -- & --& 6  & --\\
\hline
1988/03/12& N & N  & 2.0 & ESO2  & --  & -- & -- & --& 7  & --\\
1991/03/17& N & N  & 2.0 & ESO3  & --  & -- & -- & --& 7  & --\\
\hline
1994/03/05& Y & N  & 1.9 & ANU1 & -- & --   & -- &--& 8 & 92\\
1994/06/05& --& -- & --  & -- & $41.68\pm0.03$ & $-3.31\pm0.06$ &$21.84\pm0.14$  & ASCA & 9 & --\\
\hline
1997/12/11& --& -- & --       & -- & $41.99\pm0.10$   &  $-3.00\pm0.11$    & $22.15\pm0.12$& BS & 10 & 114\\
1998/04/04& N & N  & 2.0 & ESO3 & -- & -- & -- &-- & 10 & --\\
\hline
1998/11/25& --& -- & --  & -- & $43.04\pm0.05$ &  $-1.91\pm0.07$ & $21.95\pm0.01$   & BS & 10 & 48\\
1999/01/12& Y & N  & 1.9 & ESO4 & -- & -- &--&-- &  10 & --\\
\hline
2005/12/28& --& -- & --   & -- & $42.18\pm0.02$ &   $-2.82\pm0.05$ & $21.49\pm0.10$  &  RXTE  & 11 & 12\\
2006/01/09& N & N  & 2.0 & CTIO & -- & --  & -- & -- & 12 & --\\
\hline
2006/01/11& N & N  & 2.0 & CTIO & -- & --  & --&-- & 12 & 1\\
2006/01/12& -- & --  & --      & -- & $42.33\pm0.02$&  $-2.66\pm0.06$ & $21.51\pm0.10$ & RXTE & 11 & --\\
\hline
2006/01/28& --& -- & --        & -- & $42.34\pm0.02$ &   $-2.65\pm0.05$ & $21.51\pm0.10$ &RXTE  & 11 & 5 \\
2006/02/02& N & N  & 2.0 & CTIO& -- & -- & -- & -- &12 & --\\
\hline
2014/04/05& Y & Y & 1.8 & ANU2 & -- & -- & -- & -- &13 & --\\
\hline
2021/03/05& --& --& --  & --  & $42.97\pm0.03$ &   $-1.91\pm0.05$    & $21.59\pm0.21$   & Swift & 14 & 3 \\
2021/03/08& Y & Y & 1.8 & SOAR & -- & -- & -- & -- &15 & --\\
\hline
2021/04/14& Y & Y & 1.8 & SOAR&  $43.01\pm0.03$  & $-1.98\pm0.06$ & $21.89\pm0.16$ &  Swift &  15, 14 & 3 \\
2021/04/17& Y & Y & 1.8 & Gemini  & -- & -- & -- &-- & 15 & --\\
\hline
\end{tabular}
\leftline{Reference: (1) \citet{Elvis1990}, (2) \citet{Shuder1980}, (3) \citet{Durret1988}, (4) \citet{Maccacaro1982},}
\leftline{(5) \citet{Turner1989}, (6) \citet{Busko1990}, (7) \citet{Marquez1998}, (8) \citet{Allen1999},} 
\leftline{(9) \citet{Weaver1996}, (10) \citet{Gilli2000}, (11) \citet{Murphy2007}, (12) \citet{Trippe2008},}
\leftline{(13) \citet{Dopita2015}, (14) This Work, (15) \citet{Guolo2021}. }
\leftline{Optical instruments: Lick: Lick 3m/Shane; ESO1: ESO 3.6m/BCS; CNPq: CNPq 1.6m; ESO2: ESO 2.2m/BCS;}
\leftline{ESO3: ESO 1.52m/BCS; ANU: ANU 2.3m/DBS; ESO4: ESO/NTT; CTIO: CTIO 1.5m/RCS; ANU: ANU 2.3m/WiFeS;}
\leftline{SOAR: SOAR/SIFS ; Gemini: Gemini 8.1m/GMOS-IFU.}
\leftline{X-ray instruments: Ein: \emph{Einstein}/MPC; EXO: \emph{EXOSAT}/ME; ASCA: \emph{ASCA}/SIS+GIS;  BS: \emph{BeppoSAX}/PDS; }
\leftline{RXTE: \emph{RXTE/}PCA; Swift: \emph{Swift}/XRT.}
\label{tab:ngc2992}
\end{table*}

\subsection{NGC\,3516}\label{subsec:ngc3516}

NGC\,3516 was classified as a type\,1 AGN based on the optical observation in 1986 \citep{Pogge1989}. 
The source remained in type\,1 state until 2012 \citet{Popovic2002,Shapovalova2019}. During this period, the Eddington ratio was observed to be $\lambda_{\rm Edd}\sim 0.01-0.05$. Later, the \ed decreased to $\sim 0.002$ in November 2013. The broad lines vanished in 2014 when the optical flux reached its minimum, with the source moved to type\,2 state \citep{Shapovalova2019}. The Eddington ratio was observed to be $\lambda_{\rm Edd}\sim 0.001$ at that time. 

In 2014--2018, the source generally remained in a low state, showing occasional flares \citep{Ilic2020}. 
The Eddington ratio was found to be \ed $\sim 0.001-0.008$. 
The optical observations taken in 2019-2020 showed the source was in type\,1 state \citep{Ilic2020}.
During this time, the X-ray observations showed \ed in the range of $\sim 0.01-0.02$.

We found a correlation between spectral state and \ed in NGC\,3516. Generally, the source was in the type\,1 state for $\lambda_{\rm Edd}>0.01$. 
The source moved to type\,2 state for $\lambda_{\rm Edd}< 0.008$. 
NGC\,3516 was observed to have variable absorption properties with a warm absorber was observed. The \nh of the ionized absorber varied in the range of $10^{21-23}$ \pcm in the type\,1 state. The high \nh ($\sim 10^{23}$ \pcm) is explained with the ionized outflow from the warm absorber \citep{Mehdipour2022}.
In type\,1.8, 1.9, or type\,2 states, the associate \nh was $\sim 10^{22}$ \pcm. Overall, the CL transition in NGC\,3516 is attributed to the change in the accretion rate, not the obscuration properties.

\begin{table*}
\caption{NGC\,3516}
\centering
\begin{tabular}{ccccccccccc}
\hline
Dates & H$\alpha$ & H$\beta$ & Optical & Optical &   $\log L_{\rm X}$  &$\log \lambda_{\rm Edd}$  & $\log N_{\rm H}$ &X--ray & Ref. & $\Delta T$\\
UT    & BEL       & BEL      & type    & Inst.  &  $\log({\rm erg~s^{-1}})$ &       &    $\log({\rm cm}^{-2})$ & Inst.&  &  (Days)   \\
\hline
1985/12/06& --& --& --       & --  & $42.42\pm0.01$ &$-1.94\pm0.04$ & $22.41\pm0.23$ &   EXO & 1 & 61 \\
1986/02/05& Y & Y & 1.0      & Lick &--             & --        & -- & --   & 2 & --\\
\hline
1998/04/13& Y & Y & 1.0      &  HST     & $42.91\pm0.01$ &$-1.38\pm0.05$  &-- & ASCA  & 3, 4 & 0\\
\hline
2001/04/10& --& --& --       & --  & $42.70\pm0.02$&   $-1.62\pm0.05$   & $22.20\pm0.21$ & XMM & 5 & 4 \\
2001/04/14&  Y&  Y& 1.0     & SAO    & -- & --  & --  & --  & 6 & --\\
\hline
2001/11/11& --& --& --       & --  & $42.72\pm0.03$ &  $-1.60\pm 0.06$   & $22.19\pm0.21$ &  Ch & 5 & 11\\
2001/11/22&  Y&  Y& 1.0    & GHO    & --  & -- & --   & -- & 6 & --\\
\hline
2005/10/12& --& --  & --  & --  & $42.77\pm0.01$ &   $-1.54\pm0.04$  & $22.47\pm0.16$   & Suz  & 7 & 47\\
2005/11/28&  Y&  Y& 1.0   & GHO   & --   & --  & --  & -- & 6 & --\\
\hline
2012/04/08&  Y&  Y& 1.0      & MDM  & $42.48\pm0.02$  & $-1.87\pm0.05$ & $21.88\pm0.11$ & Swift  & 8, 9 & 0\\
2012/04/16&  Y&  Y& 1.0      & MDM & $42.38\pm0.02$  & $-1.98\pm0.05$  & $22.00\pm0.11$ &Swift & 8, 9 & 0\\
2012/04/28&  Y&  Y& 1.0      & MDM & $42.51\pm0.02$  & $-1.84\pm0.05$  & $21.88\pm0.12$ & Swift & 8, 9 & 0\\
\hline
2013/11/04& --& --& --    & -- & $41.64\pm0.02$   &  $-2.77\pm0.05$ & $22.18\pm0.05$   & Suz & 10 & 0\\
2014/03/23&  N&  N& 2.0        & SAO     & --  & --  & -- & --  & 6 & 15\\
2014/04/07& --& --& --  & --  & $41.39\pm0.02$   & $-3.01\pm0.05$  & $22.32\pm0.05$ & Suz & 10 & --\\
\hline
2017/04/02&  N&  N& 2.0        & SAO    & --  & -- & --  & -- & 6& 249\\
2017/12/07& --& --  & --&--   & $42.11\pm0.02$ &  $-2.28\pm0.08$ & $22.00\pm0.05$   & Swift, Nu  & 9 & --\\
2017/12/26& --& -- & -- & --   & $42.27\pm0.02$  &   $-2.10\pm0.07$  & $22.02\pm0.07$   & Swift, Nu  & 9 & 45 \\
2018/02/09&  N&  N& 2.0      & SAO       & -- & -- & --   & --  & 6& --\\
\hline
2019/12/07&Y & Y & 1.0       & CMO        & --      & -- & -- & --    & 11& --\\
\hline
2020/02/10&Y & Y & 1.0        & CMO      & --  & -- & -- & --   & 12& 9\\
2020/02/19& --& --& --       & --     & $42.48\pm0.02$   &   $-1.82\pm0.05$ & $21.71\pm0.18$  & Swift& 9 & --\\
\hline
2020/03/25&Y & Y & 1.0      &  CMO         & --  & -- & --  & --   & 12 & 4\\
2020/03/29& --& --& --       & --   & $42.66\pm0.02$ &  $-1.87\pm0.05$& $21.84\pm0.10$   &  Swift & 9 & --\\
\hline
2020/04/14& Y & Y& 1.0       & WIRO           &   --  & --  & --    & --  &  12 & 2\\
2020/04/16& --& --& --       & --   & $42.57\pm0.01$  & $-1.77\pm0.04$  & $21.90\pm0.10$   & Swift & 9 & --\\
\hline
2020/04/25&Y & Y & 1.0   & CMO    & $43.53\pm0.08$   & $-1.82\pm0.08$ & $21.79\pm0.08$  & Swift  &12, 9 & 0\\
\hline
2020/05/21& Y & Y& 1.0      & WIRO    &  --  & --  & --  & --  &  12 & 1\\
2020/05/22& --& --&  --  &--    & $42.49\pm0.01$   & $-1.86\pm0.05$ & $21.61\pm0.16$ & Swift & 9 & --\\
\hline
2020/06/03&Y & Y & 1.0       & CMO        & $42.45\pm0.05$ & $-1.91\pm0.07$ & $21.82\pm0.12$ & Swift & 12, 9 & 18\\
2020/06/21& --& --& --       & --  & $42.43\pm0.02$ & $-1.93\pm0.06$ & $21.59\pm0.17$& Swift& 9 & --\\
2020/06/22&Y & Y & 1.0   & CMO        & -- & -- & --   & --  & 12 & 1\\
\hline
\end{tabular}
\leftline{References: (1) \citet{Ghosh1991}, (2) \citet{Pogge1989}, (3) \citet{Popovic2002}, (4) \citet{Nandra1999}, }
\leftline{(5) \citet{Turner2005}, (6) \citet{Shapovalova2019}, (7) \citet{Noda2013}, (8) \citet{DeRosa2018}, (9) This Work, }
\leftline{(10) \citet{Noda2016}, (11) \citet{Ilic2020}, (12) \citet{Oknyansky2021}. }
\leftline{Optical instruments: Lick: Lick 3m/Shane; HST: HST/STIS; SAO: SAO 1m/UAGS; GHO: GHO 2.1m/BCS;}
\leftline{MDM: MDM 1.3m/B\&C; CMO: CMO/SAI 2.5m; WIRO: WIRO 2.3m.}
\leftline{X-ray instruments: EXO: \emph{EXOSAT}/ME; XMM: \emph{XMM-Newton}/EPIC-PN; Ch: \emph{Chandra}/ACIS; Suz: \emph{Suzaku}/XIS;}
\leftline{Swift: \emph{Swift}/XRT, Nu: \emph{NuSTAR}/FPMA+FPMB.}
\label{tab:ngc3516}
\end{table*}

\subsection{NGC\,4151}\label{subsec:ngc4151}

NGC\,4151 is a well-known Seyfert galaxy that shows high variability. Over the years, it has gone through several CL events. 
Here, we discuss the brief history of the source in the last 40 years. In the late 1970s, the source was classified as the type\,1 
AGN \citep{Antonucci1983}. From the mid-1980s, the broad lines weakened as the source transitioned to type\,1.8 state \citep{Antonucci1983}. 
In April 1984, the source lost its broad H$\beta$ line as it moved to type\,1.9 state \citep{Kielkopf1985}. Simultaneous observation 
by {\it EXOSAT} revealed a very low X-ray flux with $\lambda_{\rm Edd}\sim 0.001$ \citep{Pounds1986}. 

The January 1985 observation showed that the source recovered its broad H$\beta$ lines and moved to type\,1.8 state \citep{Peterson1988}. 
The \ed also increased and was observed to be in the range of $\lambda_{\rm Edd} \sim 0.003-0.005$ between 1985-1987 \citep{Yaqoob1989}. 
During this time, the source was found to be in type\,1.8 state \citep{Peterson1988}. 

From 1990, the broad lines strengthened as it moved to type\,1 state and remained there until January 1999 \citep{Sergeev2001,Shapovalova2008}. 
During this period, the $\lambda_{\rm Edd}$ was observed to be in the range of $\sim 0.01-0.16$. In December 2000, the broad lines weakened,
as it entered in type\,1.8 state \citep{Shapovalova2008}. The X-ray flux was declined this time, with $\lambda_{\rm Edd}\sim 0.003$. 
It recovered again in November 2001, transitioning to a type\,1 state. For the next two decades, the source was found to be type\,1 state \citep{Shapovalova2008,Chen2023,Li2022}. The \ed was found to be in the range of $\lambda_{\rm Edd} \sim 0.01-0.02$ during this period.

NGC\,4151 showed high absorption variability over the years, with the \nh changed more than an order in a timescale of days \citep{Puccetti2007}. 
However, NGC\,4151 was not observed in a CT state. We did not observe any relation of the \nh with the optical spectral state in the source. 
We observed a clear relation of the spectral state and \ed in NGC\,4151. Our study found that it transitioned several times between type\,1, 1.8, 
and 1.9 states. The source was in type\,1.9 state when $\lambda_{\rm Edd}\sim 0.001$. The type\,1.8 state was found for $\lambda_{\rm Edd}\sim 0.002-0.005$. 
The type\,1 state was found when $\lambda_{\rm Edd}>0.009$. This implied that the variable accretion rate in the source causes the CL transition.

\begin{table*}
\caption{NGC\,4151}
\centering
\begin{tabular}{ccccccccccc}
\hline
Dates & H$\alpha$ & H$\beta$ & Optical & Optical &   $\log L_{\rm X}$  &$\log \lambda_{\rm Edd}$  & $\log N_{\rm H}$ &X--ray & Ref. & $\Delta T$\\
UT    & BEL       & BEL      & type    & Inst.  &  $\log({\rm erg~s^{-1}})$ &       &    $\log({\rm cm}^{-2})$ & Inst.&  &  (Days)   \\
\hline
1979/05/31& --& --& --       & -- & $42.52\pm0.02$ &   $-2.05\pm0.08$ & -- &   Ein.  & 1 & 1\\
1979/06/01& Y& Y& 1.0 & Lick & -- & -- & --& --   &  2 & --\\
\hline
1981/07/06& Y& Y& 1.8 & Lick &  -- &  --& -- & --  &  2& --\\
\hline
1984/04/18&  Y& N & 1.9      & KPNO   & $41.58\pm0.04$ &  $-3.02\pm0.08$ & $22.63\pm0.34$ &  EXO & 3, 4 & 0\\
\hline
1985/01/20& Y& Y& 1.8 & Lowell & --& -- & -- & --   & 5 & 7 \\
1985/01/27& --& --& --       & --  & $42.27\pm0.05$   &  $-2.32\pm0.07$ & $22.67\pm0.10$  & EXO  & 6 & --\\
\hline
1985/05/15& Y& Y& 1.8 & Lowell  & $42.22\pm0.04$& $-2.37\pm0.08$ & $22.78\pm0.14$& EXO & 5, 6  & 0\\
1986/03/01& Y& Y& 1.8 & Lowell  & $42.17\pm0.03$& $-2.43\pm0.08$ & $22.91\pm0.14$ &EXO & 5, 6 & 0\\
1986/03/03& Y& Y& 1.8 & Lowell  & $42.16\pm0.05$& $-2.44\pm0.09$ & $22.69\pm0.30$& EXO & 5, 6 & 0\\
\hline
1987/12/17& --& --& --       & --  & $42.32\pm0.06$  & $-2.27\pm0.09$ & $22.92\pm0.05$    & Ginga & 7 & 3\\
1987/12/20&Y& Y& 1.8 & WO & --  & -- & --   & -- & 8 & --\\
\hline
1990/01/06& --& --& --   & --  &  $42.72\pm0.04$  & $-1.83\pm0.07$ & $22.52\pm0.11$ & Ginga & 9 & 67\\
1990/03/14&Y& Y& 1.0 & CrAO & --  & -- & --  & --       & 10 & --\\
\hline
1990/05/15& --& --& --       & -- & $42.91\pm0.06$  & $-1.61\pm0.08$ & $22.08\pm0.30$ &  Ginga & 9 & 72\\
1990/07/27&Y& Y& 1.0         & Oky &  --   & -- & --   & --  & 11 & --\\
1990/11/26& --& --& --       & --  &$42.96\pm0.05$  &  $-1.55\pm0.10$ & $22.86\pm0.09$   & Ginga & 9 & --\\
\hline
1991/01/15& --& --& --       & --   &$43.00\pm0.03$  &  $-1.55\pm 0.07$ & $22.74\pm0.12$  &  Ginga  & 9 & 3\\
1991/01/18&Y& Y& 1.0 & CrAO  & -- & --   &--   & --   & 10 & --\\
\hline
1993/05/24& --& --& -- & -- & $42.69\pm0.11$  & $-1.86\pm0.08$ & $22.95\pm0.03$  & ASCA  & 12 & 53\\
1993/07/16&Y& Y& 1.0 & CrAO & --  & -- & --   &  -- & 10 & --\\
\hline
1993/11/05& --& --& --  & --      & $42.74\pm0.09$ & $-1.81\pm0.10$ & $22.34\pm0.03$  & ASCA &  12 & 9\\
1993/11/14&Y& Y& 1.0  & CrAO & --  & -- &  --   & -- &  10 & -- \\
\hline
1993/12/04& Y& Y& 1.0 & CrAO  & $42.84\pm0.14$  & $-1.69\pm0.11$ &  $22.70\pm0.08$ &ASCA  & 10, 13 & 0\\
\hline
1995/05/10& --& --& --       & --   & $43.58\pm0.09$  & $-0.80\pm 0.09$ & $22.57\pm0.06$  & ASCA  & 14 & 63\\
1995/07/12&Y& Y& 1.0 & CrAO & --  & -- & --     & --  & 10 & --\\
\hline
1996/07/06& --& --& --       & --  &$42.61\pm0.10$   & $-1.95\pm0.09$ & $23.34\pm0.12$  & BS & 15 & 5\\
1996/07/11&Y& Y& 1.0 & SAO & --  & -- & --   & -- &  16 & --\\
\hline
1996/11/15& Y& Y& 1.0 & SAO  & --  & --     & --&--& 16 & 19\\
1996/12/04& --& --& --       & --  & $42.78\pm0.06$ & $-1.76\pm 0.09$ & $23.04\pm0.18$  &  BS  & 15 & --\\
\hline
1999/01/04& --& --& --       & --  & $42.61\pm0.07$  & $-1.95\pm0.08$ & $22.53\pm0.11$ &BS  & 17 & 8\\
1999/01/12& Y& Y& 1.0 & SAO  & --  & -- & --    & -- & 16 & --\\
\hline
2000/12/20& Y& Y& 1.8 & SAO  & --  & --   & --& -- & 16 & 3\\
2000/12/23& --& --& --       & -- & $42.06\pm0.04$ & $-2.54\pm0.08$ & $22.53\pm0.11$ & XMM & 18 & --\\
\hline
2001/11/25& Y& Y& 1.0 & SAO  & --  & -- &--&--& 16 & 23 \\
2001/12/18& --& --& --       & --  & $42.71\pm0.08$ & $-1.84\pm 0.09$ & $22.54\pm0.11$ &BS  & 15 & --\\
\hline
2005/12/26& --& --& --       & --    & $42.70\pm0.06$  & $-1.85\pm 0.09$& $22.77\pm0.03$    &Swift & 19 & 5\\
2005/12/31& Y& Y& 1.0 & FAST & -- & -- & -- & --  & 20 & --\\
\hline
2019/07/20&Y& Y& 1.0 & Hale & -- & --& --& --  & 21 & 4\\
2019/07/24&--& -- & --  & -- & $42.63\pm0.02$  & $-1.93\pm 0.08$ & $23.32\pm0.06$ &Nu  & 19 & --\\
\hline
2020/01/20& Y& Y& 1.0 & FAST & --  & --& -- &-- & 20 & 8\\
2020/01/28& --& --& --       & --   & $42.58\pm0.02$  & $-1.98\pm0.08$& $23.52\pm0.03$   & Swift  & 19 & -- \\
\hline
2020/11/06&--& --& --       & --  & $42.76\pm0.08$ &  $-1.78\pm0.09 $& $23.19\pm0.13$  &Swift & 19 & 3\\
2020/11/09&Y& Y& 1.0 & Lijang & -- & -- & --& --  &  22 & --\\
\hline
2020/12/07& Y& Y& 1.0 & FAST & $42.53\pm0.10$ & $-2.00\pm 0.09$ & $23.26\pm0.11$   & Swift  & 20, 19 & 0 \\
\hline
2021/05/14&--& --& --       & --  & $42.78\pm0.06$  & $-1.76\pm 0.08 $ & $23.37\pm0.12$ &Swift  & 19 & 2 \\
2021/05/16&Y& Y& 1.0 & Lijang & --& --  & --& --  & 22 & -- \\
\hline
2021/11/12&--& --& --       & --   & $42.80\pm0.05$&  $-1.74\pm0.08$& $22.68\pm0.04$ &Swift  & 19 & 5\\
2021/11/17& Y& Y& 1.0 & FAST & --  & -- & --   & -- & 20 & --\\
\hline
\hline
\end{tabular}
\leftline{References : (1) \citet{Perola1982}, (2) \citet{Antonucci1983}, (3) \citet{Kielkopf1985}, (4) \citet{Pounds1986}, }
\leftline{(5) \citet{Peterson1988}, (6) \citet{Yaqoob1989}, (7) \citet{Yaqoob1991}, (8) \citet{Maoz1991}, }
\leftline{(9) \citet{Yaqoob1993}, (10) \citet{Sergeev2001}, (11) \citet{Ayani1991}, (12) \citet{Weaver1994}, }
\leftline{(13) \citet{Yaqoob1995}, (14) \citet{Zdziarski2002}, (15) \citet{deRosa2007}, (16) \citet{Shapovalova2008}, }
\leftline{(17) \citet{Schurch2002}, (18) \citet{Schurch2003}, (19) This Work, (20) \citet{Chen2023}, }
\leftline{(21) \citet{Oh2022}, (22) \citet{Li2022}. }
\leftline{Optical instruments: Lick: Lick-Shane 3m; KPNO: KPNO 2.1m/Goldcam; Lowell: Lowell/IDS; WO: Wise 1m; }
\leftline{CrAO: CrAO 2.6m; Oky: Okayama 188cm; SAO: SAO 1m; FAST: FLWO 1.5m/FAST; Hale: Hale-200in/DBSP;}
\leftline{Lijang: Lijang 2.4m/YFOSC; }
\leftline{X-ray instruments: Ein.: \emph{Einstein}/MPC; EXO: \emph{EXOSAT}/ME; Ginga: \emph{Ginga}/LAC; ASCA: \emph{ASCA}/SIS+GIS;}
\leftline{ BS: \emph{BeppoSAX}/PDS; XMM: \emph{XMM-Newton}/EPIC-PN; Swift: \emph{Swift}/XRT; Nu:\emph{NuSTAR}/FPMA+FPMB.}
\label{tab:ngc4151}
\end{table*}

\subsection{NGC\,5273}\label{subsec:ngc5273}
NGC\,5273 was originally classified as a type\,1.9 galaxy in 1984 \citep{Ho1995}. The optical spectrum revealed that the source was still in type\,1.9 state in June 1993 \citep{Koratkar1995}. The June 1992 observation showed that it was a low-luminosity AGN with $\lambda_{\rm Edd}\sim 0.0003$. The source was found to move to type\,1.8 state in 2006 \citep{Koss2017}. Eventually, the X-ray flux also increased over the years \citep{Neustadt2023}. NGC\,5273 was found to move to type\,1 states in May 2014 \citep{Neustadt2023}. We observed the $\lambda_{\rm Edd} \sim 0.006$ from the quasi-simultaneous X-ray observations with \nustar \citep{Pahari2017}. 

We found a correlation of \ed and spectral states in NGC\,5273. The source was found in the type\,1 state when $\lambda_{\rm Edd}>0.006$. The type\,1.9 state was observed at low \ed, at $\sim 0.0003$. NGC\,5273 is an obscured AGN with the \nh varied in the range of $\sim 1-4 \times 10^{22}$ \pcm \citep{Neustadt2023}, with correlation with the CL transition. Hence, the change in the \ed is the most probable reason for the CL event in NGC\,5273.

\begin{table*}
\caption{NGC\,5273}
\centering
\begin{tabular}{ccccccccccc}
\hline
Dates & H$\alpha$ & H$\beta$ & Optical & Optical &   $\log L_{\rm X}$  &$\log \lambda_{\rm Edd}$  & $\log N_{\rm H}$ &X--ray & Ref. & $\Delta T$\\
UT    & BEL       & BEL      & type    & Inst.  &  $\log({\rm erg~s^{-1}})$ &       &    $\log({\rm cm}^{-2})$ & Inst.&  &  (Days)   \\
\hline
1984/02/13& Y& N& 1.9 & Palomer & -- & -- & --  & --  & 1  & -- \\
\hline
1992/06/26& --& --& --       & --  & $40.09\pm0.12$ &$-3.58\pm 0.09$ & -- & ROSAT  & 2 & 367 \\
1993/06/28& Y& N& 1.9 & Palomer & --  & -- & -- & -- & 2 & --\\
\hline
2002/06/14& --& --& --       & --   & $41.33\pm0.01$  & $-2.35 \pm 0.07$ & $22.30\pm0.01$  & xmm & 3 & --\\
2006/03/05& Y& N& 1.8 & SDSS & --  & -- & --      & -- & 4  & -- \\
2013/07/16& --& --& --       & --   & $41.04\pm0.01$  & $-2.65 \pm 0.07$ & $22.30\pm0.01$  & Suz & 3 & --\\
\hline
2014/06/30& Y& Y& 1.0 & APO & --  & --& --   & -- & 5 & 14\\
2014/07/14& --& --& --       & --  &$41.47\pm0.01$ & $-2.20\pm 0.07$  & $22.39\pm0.03$&   Nu & 6 & --\\
\hline
2022/03/23& Y& Y& 1.0 & LBT & --  & --    & --   & --  & 3 & 2\\
2022/03/25& --& --& --       & --   & $41.63\pm0.04$  & $-2.03\pm 0.07$  & $22.10\pm0.08$  &Swift & 3 & --\\
\hline
2022/03/30& Y& Y& 1.0 & Keck & --  & -- & -- & -- & 3 & 1\\
2022/03/31& --& --& --       & --   & $41.62\pm0.05$  & $-2.39\pm 0.07$ & $22.25\pm0.07$ &Swift & 3 & --\\
\hline
2022/04/25& --& --& --       & --  & $41.79\pm0.03$   & $-1.85\pm 0.07$ & $22.34\pm0.13$&Swift  & 3 & 5 \\
2022/04/30& Y& Y& 1.0 & Keck & --  & --   & -- & --  & 3 & --\\
\hline
2022/05/25& --& --& --       & --  &$41.52\pm0.03$ & $-2.15 \pm 0.07$ & $22.37\pm0.14$  &Swift  & 3 & 2\\
2022/05/27& Y& Y& 1.0 & UH88 & --  & -- & -- & --  & 3 & --\\
\hline
2022/07/09& Y& Y& 1.0 & UH88 & $41.83\pm0.04$  & $-1.81 \pm 0.07$ & $22.14\pm0.10$  &Swift &  3 & 0 \\
\hline
2022/08/14& --& --& --       & --  & $41.56\pm0.05$  & $-2.10\pm 0.07$ & $22.28\pm0.14$ & Swift &  3 & 5\\
2022/08/19& Y& Y& 1.0 & UH88 & --  & -- & -- & --  & 3 & -- \\
\hline
2022/08/22& Y& Y& 1.0 & Keck & --  & -- &-- & --  & 3 & 2\\
2022/08/24& --& --& --       & -- & $41.58\pm0.04$  & $-2.08\pm 0.07$ & $22.30\pm0.16$  & Swift &  3 & --\\
\hline
\end{tabular}
\leftline{References : (1) \citet{Ho1995}, (2) \citet{Koratkar1995},  (3) \citet{Neustadt2023}, (4) \citet{Koss2017},}
\leftline{(5) \citet{Bentz2014}, (6) \citet{Pahari2017}.}
\leftline{Optical instruments: Palomer: Palomer; APO: APO 3.5m/DIS; LBT: LBT 8.4m/MODS; Keck: Keck 10m/LRIS;}
\leftline{UH88: UH 88inch/SNIFS.}
\leftline{X-ray instruments: ROSAT: \emph{ROSAT}/HRI; Suz: \emph{Suzaku}/XIS, Nu: \emph{NuSTAR}/FPMA+FPMB; Swift: \emph{Swift}/XRT.}
\label{tab:ngc5273}
\end{table*}

\subsection{NGC\,5548}\label{subsec:ngc5548}

NGC\,5548 is a well-studied Seyfert galaxy. The source is generally found in the type\,1 state; however, it has transited to type\,1.8 several times. 
From the late-1970s, the source was observed to be in type\,1 state with \ed in the range of $\lambda_{\rm Edd} \sim 0.04-0.3$ \citep{Shapovalova2004,Sergeev2007}. 

NGC\,5548 moved to type\,1.8 state in April 2005 with weakening broad lines. The X-ray flux also decreased, with $\lambda_{\rm Edd}\sim 0.03$. 
The optical spectra taken in June and July 2007 showed that the source remained in the type\,1.8 state \citep{Koss2017}. The \ed was found to 
be $\lambda_{\rm Edd}\sim 0.03$ \citep{Liu2010}. Subsequently, the source was observed to recover its broad lines and moved to type\,1 state in 
January 2014 and remained there till date \citep{Lu2022}. The \ed was found to vary in the range of $\lambda_{\rm Edd} \sim 0.05-0.12$.

NGC\,5548 entered in type\,1.8 when $\lambda_{\rm Edd}<0.03$. The source was in type\,1 state for a wide range of 
$\lambda_{\rm Edd} \sim 0.05-0.3$. Over the years, NGC\,5548 was observed regularly in X-rays. The source was always found to be unobscured with the \nh ranges $\sim 10^{20-21}$ \pcm. From this, we conclude that the optical state transition in NGC\,5548 is attributed to the change in the accretion rate.

\begin{table*}
\caption{NGC\,5548}
\centering
\begin{tabular}{ccccccccccc}
\hline
Dates & H$\alpha$ & H$\beta$ & Optical & Optical &   $\log L_{\rm X}$  &$\log \lambda_{\rm Edd}$  & $\log N_{\rm H}$ & X--ray & Ref. & $\Delta T$\\
UT    & BEL  & BEL      & type    & Inst.  &  $\log({\rm erg~s^{-1}})$ &       &    $\log({\rm cm}^{-2})$ & Inst.&  &  (Days)   \\
\hline
\hline
1978/06/11& --& --& --       & --  & $43.89\pm0.06$  & $-0.57\pm0.03$ & -- & H-1 & 1 & 37\\
1978/07/18& Y& Y& 1.0 & CrAO & --  & --& --   & -- & 2 & --\\
\hline
1979/06/28& Y& Y& 1.0 & CrAO & --  & --  & --     & -- & 2 & 1\\
1979/06/29& --& --& --       & --  & $43.69\pm0.10$ & $-0.83\pm 0.05$& -- & Ein.  & 3 & --\\
\hline
1984/02/04& Y& Y& 1.0 & CrAO & --  & -- & --      & --  &2 & 29\\
1984/03/02& --& --& --       & --  & $43.85\pm0.04$   &  $-0.63\pm0.03$& $20.00\pm0.91$  & EXO & 4 & --\\
\hline
1984/06/29& Y& Y& 1.0 & CrAO & -- & -- & --   & --  & 2 & 12\\
1984/07/11& --& --& --       & --  &$43.68\pm0.01$   & $-0.85\pm 0.02$& $20.00\pm0.49$ & EXO  & 4 & --\\
\hline
1984/12/20& Y& Y& 1.0 & CrAO & --  & -- & --      & --  & 2 & 45\\
1985/01/14& --& --& --       & --    & $43.75\pm0.05$  &  $-0.76\pm 0.03$&  $21.17\pm0.41$  &EXO & 4 & --\\
\hline
1985/05/13& Y& Y& 1.0 & CrAO & --  & -- & --        & --  & 2 & 26\\
1985/06/08& --& --& --       & --  & $43.30\pm0.07$  & $-1.32\pm0.04$  & $20.00\pm1.36$ & EXO  & 4 & --\\
\hline
1985/06/18& Y& Y& 1.0 & CrAO & -- & -- & --    & --  & 2 & 4\\
1985/06/22& --& --& --       & --  & $43.62\pm0.03$ & $-0.95\pm 0.02$& $21.26\pm0.47$ & EXO & 4 & --\\
\hline
1985/07/14& --& --& --       & --  & $43.31\pm0.08$ & $-1.30\pm 0.04$ &$21.08\pm0.78$   &EXO  & 4 & 2\\
1985/07/16& Y& Y& 1.0 & CrAO & -- & -- & --    & --  & 2 & --\\
\hline
1986/03/03& --& --& --       & --  & $43.81\pm0.10$   & $-0.68\pm 0.05$& $20.10\pm1.55$  &EXO & 4 & 10\\
1985/03/13& Y& Y& 1.0 & CrAO & -- & -- & --  & -- & 2 & --\\
\hline
1988/06/21& Y& Y& 1.0 & CrAO & --  & --  & --    & --  & 2 & 3\\
1988/06/24& --& --& --       & --   & $43.62\pm0.10$  & $-0.92\pm 0.05$& $20.00\pm0.71$ & Ginga & 4 & --\\
\hline
1989/01/28& --& --& --       & --  &$43.73\pm0.07$ &  $-0.78\pm 0.04$& $21.61\pm0.10$  &Ginga  & 4 & 5\\
1989/02/02& Y& Y& 1.0 & CAO & --  & -- & --       & --   & 5, 6 & --\\
\hline
1989/06/08& Y& Y& 1.0 & Mt. HT &  $43.65\pm0.11$ &$-0.88\pm0.06$& $20.95\pm0.37$ & Ginga  & 6, 4  & 0 \\
\hline
1989/07/08& Y& Y& 1.0 & Mt. HT & -- & -- & --  & -- &  6 & 5\\
1989/07/13& --& --& --       & --  & $43.75\pm0.09$  & $-0.76\pm0.05$& $21.08\pm0.24$ &  Ginga & 4 & --\\
\hline
1990/07/16-21& --& --& --   & --   & $43.41\pm0.09$& $-1.18\pm0.05$& $20.11\pm0.40$  & ROSAT & 7 & 0\\
1990/07/17& Y& Y& 1.0 & CAO  &  -- & --  & --    & --  & 5, 6  & --  \\
\hline
1993/07/27& --& --& --   & --  & $43.54\pm0.06$&$-1.02 \pm 0.03$& $20.23\pm0.20$  &  ASCA & 8 & 1\\
1993/07/28& Y& Y& 1.0 & CAO  & -- & -- & -- & -- & 5, 6 & --\\
\hline
1996/07/03& --& --& --       & -- & $43.56\pm0.07$  & $-1.00\pm0.04$ & $20.23\pm0.19$  & ASCA & 9 & 3\\
1996/07/06& Y& Y& 1.0 & SAO  & --  & -- & -- & -- & 10 & --\\
\hline
1997/08/21& --& --& --  & -- & $43.25\pm0.06$  & $-1.38\pm0.03$   &$20.72\pm0.19$  & BS & 12 & 6\\
1997/08/27& Y& Y& 1.0 & SAO  & & -- & --   & -- & 10 & --\\
\hline
1998/07/07& --& --& --       & -- & $43.64\pm0.06$  & $-0.90\pm0.04$ &$20.23^*$ & ASCA &  13 & 7\\
1998/07/14& Y& Y& 1.0 & GHO  &  -- & --    & -- & -- & 10 & --\\
\hline
1999/01/19& Y &Y & 1.0  & GHO  & $43.68\pm0.09$& $-0.85\pm0.05$ & $20.23^*$ & ASCA &  10, 14 & 0\\
\hline
1999/08/16& --& --& --       & --  & $43.76\pm0.09$ &  $-0.74\pm0.05$& $20.23^*$ &RXTE &  14 & 1\\
1999/08/17& Y& Y& 1.0 &  MDM & --   & --   & -- &-- &  5 & --\\
\hline
1999/12/05& Y& Y& 1.0 &  MDM  & --   & --    & -- &--& 5 & 5\\
1999/12/10& --& --& --       & --  & $43.41\pm0.07$ & $-1.18\pm0.04$& $19.48\pm0.64$  & BS & 13 & --\\ 
\hline
2001/07/08& --& --& --       & --  & $43.37\pm0.08$& $-1.23\pm0.04$& $20.58\pm0.10$  & BS  & 13 & 3\\
2001/07/11& Y& Y& 1.0 & SAO  & --  & -- & -- & -- & 10 & --\\
\hline
2001/07/19& --& --& --       & --  & $43.70\pm0.15$ & $-0.82\pm0.08$& $20.05\pm0.84$  & BS  & 13 & 3\\ 
2001/07/22& Y& Y& 1.0 & SAO & --  & -- & -- &-- & 10 & --\\
\hline
2002/01/18& --& --& --       & --  & $43.60\pm0.04$   & $-0.95\pm0.02$& $20.66\pm0.74$   & Ch  & 15 & --\\
\hline
\hline
\end{tabular}
\label{tab:ngc5548}
\end{table*}

\begin{table*}
\centering
\begin{tabular}{ccccccccccc}
\hline
Dates & H$\alpha$ & H$\beta$ & Optical & Optical &   $\log L_{\rm X}$  &$\log \lambda_{\rm Edd}$  & $\log N_{\rm H}$ &X--ray & Ref. & $\Delta T$\\
UT    & BEL       & BEL      & type    & Inst.  &  $\log({\rm erg~s^{-1}})$ &       &    $\log({\rm cm}^{-2})$ & Inst.&  &  (Days)   \\
\hline
\hline
2005/04/15& --& --& --       & --  & $43.12\pm0.04$& $-1.53\pm0.02$ &--& Ch  & 16 & --\\
2005/04/17& Y& Y& 1.8 & SAO   & -- & --       & --&-- & 23 & --\\
\hline
2007/06/18& --& --& --       & --   & $43.96\pm0.03$  & $-0.48\pm0.02$  &$20.23^*$   & Suz &  17 & --\\
\hline
2007/06/24&  Y& Y& 1.8 & SAO & $43.18\pm0.02$ &  $-1.46\pm0.02$  &$20.23^*$ & Suz  & 23, 17 & 0 \\
\hline
2007/07/22&  Y& Y& 1.8 & SAO  & $43.12\pm0.05$  &$-1.53\pm0.03$ & $20.23^*$ & Suz & 23, 17 & 0\\
\hline
2014/01/06& Y& Y& 1.0 & MDM-B   & $43.34\pm0.08$ & $-1.27\pm0.04$ & $21.64\pm0.05$  & Swift   & 18, 19 & 0 \\
\hline
2014/02/04& --& --& --       & --  & $43.53\pm0.02$  & $-1.04\pm0.02$  & $22.04\pm0.35$    & XMM  & 20 & 1\\
2014/02/05& Y& Y& 1.0 & MDM-B & --  & --       & --&-- & 21 & --\\
\hline
2014/02/17& Y& Y& 1.0 & MDM-B &  $43.45\pm0.08$ &$-1.13\pm0.04$ & $22.16\pm0.10$   & Swift   & 18, 19 & 0\\
\hline
2014/03/14& Y& Y& 1.0 & MDM-B & $43.35\pm0.06$ &$-1.26\pm0.03$ & $22.07\pm0.16$  & Swift   & 18, 19 & 0 \\
\hline
2014/04/16& Y& Y& 1.0 & MDM-B & $43.49\pm0.03$ & $-1.09\pm0.02$ & $21.98\pm0.10$  & Swift   & 18, 19 & 0\\
\hline
2014/05/16& Y& Y& 1.0 & MDM-B & $43.63\pm0.03$ & $-0.91\pm0.03$ & $22.18\pm0.11$  & Swift   & 18, 19 & 0\\
\hline
2014/06/16& Y& Y& 1.0 & MDM-B & $43.59\pm0.06$ & $-0.96\pm0.03$ & $22.16\pm0.15$    & Swift   & 18, 19 & 0\\
\hline
2014/07/01& Y& Y& 1.0 & MDM-B &  $43.53\pm0.07$ & $-1.04\pm0.04$ & $21.91\pm0.19$  & Swift &18, 19 & 0\\
\hline
2015/01/15& Y& Y& 1.0 &  Lijang & $43.53\pm0.06$ & $-1.04\pm0.03$ &  $21.95\pm0.13$ & Swift &21, 19 & 0\\
\hline
2015/02/02& Y& Y& 1.0 &  Lijang  & $43.56\pm0.06$& $-1.00\pm0.03$ &  $22.18\pm0.16$   & Swift   & 21, 19 & 0 \\
\hline
2015/06/17& Y& Y& 1.0 &  Lijang   & $43.30\pm0.11$ & $-1.32\pm0.06$  & $22.16\pm0.11$  & Swift   & 21, 19 & 0\\
\hline
2018/08/29& Y& Y& 1.0 &  Lijang  & --  &--   & --  & -- & 21 & 4\\
2018/09/02&--& --& --       & --  &  $43.31\pm0.06$ &   $-1.30\pm0.04$ & $22.01\pm0.14$ & Swift  & 21, 19 & --\\
\hline
2019/12/25& Y& Y& 1.0 &  Lijang  & -- & --  & -- &-- & 21  & 1\\
2019/12/26& --& --& --       & --  & $43.43\pm0.14$  & $-1.16\pm0.07$ &$22.18\pm0.16$ & Swift  &  19 & --\\
\hline
2020/03/18& Y& Y& 1.0 &  Lijang  & $43.45\pm0.10$ & $-1.13\pm0.05$   & $22.23\pm0.19$    & Swift   & 21, 19 & 0 \\
\hline
2020/04/13& --& --& -- & --  & $43.49\pm0.13$ &   $-1.09\pm0.06$ & $21.90\pm0.15$ & Swift   & 21, 19 & 1\\
2020/04/14& Y& Y& 1.0 &  Lijang  & -- & -- & --& -- & 21 & --\\
\hline
2020/05/16& Y& Y& 1.0 &  Lijang  & $43.48\pm0.12$ & $-1.10\pm0.06$ & $21.58\pm0.19$ & Swift   & 21, 19 & 0\\
\hline
2020/06/20& --& --& --       & --  & $43.47\pm0.08$   & $-1.11\pm0.04$ & $22.03\pm0.10$ &  Swift   & 21, 19 & 2\\
2020/06/22& Y& Y& 1.0 &  Lijang &--& -- & -- & -- & 21 & --\\
\hline
2021/01/09& --& --& --       & --   & $43.53\pm0.13$ & $-1.04\pm0.07$ &  $22.01\pm0.14$&Swift  & 21, 19 & 1\\
2021/01/10& Y& Y& 1.0 &  Lijang  & -- & -- & --  & -- & 21 & --\\
\hline
2021/01/26& --& --& --       & --  &$43.53\pm0.02$ & $-1.04\pm0.02$& $22.63\pm0.08$ & Nu  &  22 & 3\\
2021/01/29& Y& Y& 1.0 & Lijang & --  & -- & -- & -- & 21 & --\\
\hline
2021/02/06& Y& Y& 1.0 &  Lijang  & $43.52\pm0.16$& $-1.05\pm0.08$ & $21.84\pm0.22$     & Swift   &21, 19 & 0 \\
\hline
2021/02/13& Y& Y& 1.0 &  Lijang  & $43.33\pm0.12$ & $-1.28\pm0.06$  & $22.06\pm0.16$  & Swift   &21, 19 & 0\\
2021/02/27& Y& Y& 1.0 &  Lijang  & $43.34\pm0.08$ & $-1.27\pm0.04$  & $21.93\pm0.17$ & Swift   &21, 19 & 0\\
2021/03/13& Y& Y& 1.0 &  Lijang  & $43.38\pm0.04$& $-1.22\pm0.02$ & $21.85\pm0.13$ & Swift   &21, 19 & 0\\
2021/03/24& Y& Y& 1.0 &  Lijang  & $43.49\pm0.09$ & $-1.09\pm0.05$ & $21.90\pm0.12$ & Swift   &21, 19 & 0\\
2021/04/24& Y& Y& 1.0 &  Lijang  & $43.45\pm0.09$ &$-1.13\pm0.05$   & $21.91\pm0.18$& Swift   &21, 19 & 0\\
2021/05/22& Y& Y& 1.0 &  Lijang  & $43.41\pm0.09$ & $-1.18\pm0.05$ &$22.02\pm0.12$  & Swift   &21, 19 & 0 \\
2021/06/26& Y& Y& 1.0 &  Lijang  & $43.39\pm0.09$& $-1.21\pm0.05$  & $22.15\pm0.21$ & Swift   &21, 19 & 0\\
\hline
2021/07/17& Y& Y& 1.0 &  Lijang  & -- & -- & -- & --   & 21 & 3\\
2021/07/20& --& --& --       & --  & $43.32\pm0.11$ &  $-1.29\pm0.05$ & $22.09\pm0.18$  & Swift  & 19 & --\\
\hline
\hline
\end{tabular}
\leftline{$^*$ fitted with Galactic absorption.}
\leftline{(1) \citet{Mushotzky1980}, (2) \citet{Sergeev2007}, (3) \citet{Kruper1990}, (4) \citet{Nandra1991},}
\leftline{(5) \citet{Peterson2002}, (6) \citet{Peterson1991}, (7) \citet{Nandra1993}, (8) \citet{Iwasawa1999},}
\leftline{(9) \citet{Yaqoob2001}, (10) \citet{Shapovalova2004}, (11) \citet{Nicastro2000b}, (12) \citet{Dadina2007},}
\leftline{(13) \citet{Dadina2007}, (14) \citet{Chiang2000}, (15) \citet{Steenbrugge2005}, (16) \citet{Detmers2008},}
\leftline{(17) \citet{Liu2010}, (18) \citet{Pei2017}, (19) This work, (20) \citet{Ursini2015}, (21) \citet{Lu2022},}
\leftline{(22) \citet{Pal2022}, (23) \citet{Bon2016}.}
\leftline{Opitcal instruments: CrAO: CrAO 2.6m; CAO: CAO 3.5m/CCDS; Mt. HT: Mt. Hopkins 1.6m; SAO: SA0 6m/UAGS; }
\leftline{GHO: GHO 2.1m/B\&C; MDM: MDM 2.4m; MDM-B: MDM 1.3m/BCS; Lijang: Lijang 2.4m/YFOSC. }
\leftline{X-ray instruments: Ein.: \emph{Einstein}/IPC; EXO: \emph{EXOSAT}/ME; ROSAT: \emph{ROSAT}/HRI; Ginga: \emph{Ginga}/LAC; }
\leftline{BS: \emph{BeppoSAX}/PDS; XMM: \emph{XMM-Newton}/EPIC-PN; Ch: \emph{Chandra}/ACIS-LETG; Suz: \emph{Suzaku}/XIS; Swift: \emph{Swift}/XRT.}
\end{table*}

\subsection{NGC\,6814}\label{subsec:ngc6814}
NGC\,6814 was classified as type\,1.8 Seyfert in 1975 \citep{Yee1980}. It moved to type\,1 state in 1979 and was found in that state until 1984 \citep{Rosenblatt1994}. 
The simultaneous X-ray observations in 1983 and 1984 revealed $\lambda_{\rm Edd}\sim 0.007-0.01$ when the source was in type\,1 state \citep{Mittaz1989}. 
In October 1985, the source moved to type\,1.8 state, with \ed decreased to $\lambda_{\rm Edd} \sim 0.003$. Subsequently, the source recovered and transitioned
to type\,1 state in 1987; before moving back to type\,1.8 state in 1992 \citep{Winkler1992a,Kollatschny2006}. In 1992, the \ed was found to be low with $\lambda_{\rm Edd}\sim 0.001-0.002$. 

The X-ray flux further decreased in 1993 with $\lambda_{\rm Edd}\sim 0.0009$ \citep{Reynolds1997}. 
In March 2008 observation, NGC\,6814 was observed in a type\,1 state \citep{Bentz2009}. Subsequent observations in 2011 and 2015 showed 
type\,1 state spectra of NGC\,6814 \citep{Koss2017,Oh2022}. In 2016, the X-ray flux increased with $\lambda_{\rm Edd}\sim 0.02$.

The \nh is observed in the range of $\sim 10^{20-22}$ \pcm, with no correlation with the spectral state. However, NGC\,6814 showed a clear relation between spectral states and \ed. The source was found generally in type\,1 state when $\lambda_{\rm Edd}>0.007$. The type\,1.8 state was observed at low \ed, with $\lambda_{\rm Edd}<0.002$. This indicated that the change in the accretion rate causes the state transition in NGC\,6814.

\begin{table*}
\caption{NGC\,6814}
\centering
\begin{tabular}{ccccccccccc}
\hline
Dates & H$\alpha$ & H$\beta$ & Optical & Optical &   $\log L_{\rm X}$  &$\log \lambda_{\rm Edd}$  & $\log N_{\rm H}$ & X--ray & Ref. & $\Delta T$\\
UT    & BEL       & BEL      & type    & Inst.  &  $\log({\rm erg~s^{-1}})$ &       &    $\log({\rm cm}^{-2})$ & Inst.&  &  (Days)   \\
\hline
\hline
1975/06/15&Y& Y& 1.8 & SAOO & --  & -- & --  & -- & 1 & --\\
\hline
1979/05/30&Y& Y& 1.0 & Palomar & --  & --   & --   & -- & 2 & -- \\
1979/06/13&Y& Y& 1.0 & Palomar & --  & --   & --   & -- & 2 & -- \\
1981/08/05&Y& Y& 1.0 & Palomar & -- & -- & --    & -- & 2 & -- \\
\hline
1983/11/03& --& --& --       & --  &  $42.14\pm0.07$  & $-1.87\pm0.05$ & $20.77\pm0.19$   & EXO & 3 & 1\\
1983/11/04&Y& Y& 1.0 & SAOO & --  & -- & --    & -- & 4 & --\\
\hline
1984/05/31& --& --& --       & --  & $41.91\pm0.05$ & $-2.13\pm0.04$ & $22.60\pm0.27$  &EXO & 3 & 1\\
1984/06/01&Y& Y& 1.0 & SAOO & --  & -- & --    & -- & 4 & --\\
\hline
1985/10/07&Y& Y& 1.8 & SAOO & --  & --     & --   & -- & 4 & 9\\
1985/10/16& --& --& --       & --  & $41.54\pm0.09$ & $-2.52\pm 0.05$  & $21.27\pm0.19$  & EXO &  3 & --\\
\hline
1987/07/29&Y& Y& 1.0 & SAOO & -- & -- & --    & -- & 5 & -- \\
1989/04/28& --& --& --       & --  & $42.60\pm0.09$ & $-1.34\pm 0.05$  & $22.61\pm0.19$  & Ginga &  6 & --\\
\hline
1992/04/29& --& --& --       & --  &  $41.19\pm0.10$ & $-2.88\pm0.06$ &$20.99^*$ & ROSAT &  7 & 70\\
1992/07/08&Y& Y& 1.8 & CA3   & -- & -- & --     & -- &  8 & --\\
\hline
1992/08/31&Y& Y& 1.8 & CA2   & -- & -- & --     & -- &  8 & 38\\
1992/10/08& --& --& --       & --  &  $41.46\pm0.11$  & $-2.60\pm0.06$ & $20.99^*$ &ROSAT & 7 & --\\
\hline
1993/05/04& --& --& --       & --  & $40.94\pm0.12$ & $-3.12\pm0.07$ & $20.76\pm0.44$ & ASCA &  9 & -- \\
\hline
2008/03/26&Y& Y& 1.0 & Lick & --  & -- & --   & -- & 10 & --\\
2011/05/13&Y& Y& 1.0 & Lick & -- & -- & --     & -- & 11 & --\\
2011/11/02& --& --& -- &--      & $41.73\pm0.08$  & $-2.32\pm0.05$ & $21.17\pm0.22$  & Suz & 12 & -- \\
\hline
2015/05/13&Y& Y& 1.0 & VLT & --  & --    & --   & -- & 13 & --\\
2016/07/04& --& --& --      & --  &$42.31\pm0.08$& $-1.68\pm0.05$ & $20.99^*$  & Nu & 14 & -- \\
\hline
\hline
\end{tabular}
\leftline{References: (1) \citet{Yee1980}, (2) \citet{Rosenblatt1994}, (3) \citet{Mittaz1989},}
\leftline{(4) \citet{Sekiguchi1990}, (5) \citet{Winkler1992}, (6) \citet{Turner1992}, (7) \citet{Koenig1997},}
\leftline{(8) \citet{Kollatschny2006}, (9) \citet{Reynolds1997}, (10) \citet{Bentz2009}, (11) \citet{Koss2017},}
\leftline{(12) \citet{Waddell2020}, (13) \citet{Oh2022}, (14) \citet{Tortosa2018}. }
\leftline{Optical instruments: SAOO: SAOO 1.9m; Palomer: Palomer 1.5m/SIT; CA3: CAO 3.5m/TWIN; CA2: CAO 2.2m/B\&C;}
\leftline{AAT: AAT 3.9m/RGOS; Lick: Lick-Shane 3m/Kast; VLT: ESO-VLT/Xshooter.}
\leftline{X-ray instruments: EXO: \emph{EXOSAT}/ME; Ginga: \emph{Ginga}/LAC; ROSAT: \emph{ROSAT}/PSPC; ASCA: \emph{ASCA}/SIS+GIS;}
\leftline{Suz: \emph{Suzaku}/XIS; Nu: \emph{NuSTAR}/FPMA+FPMB.}
\label{tab:ngc6814}
\end{table*}

\subsection{NGC\,7582}\label{subsec:ngc7582}
NGC\,7582 is one of two CLAGNs in our sample that showed both CS and CO transitions \citep[see,][and references therein]{Lefkir2023}. In 1977, the source was classified as type\,1 state Seyfert \citep{Ward1980}. The X-ray observations in May 1980 showed $\lambda_{\rm Edd}\sim 0.002$ \citep{Morris1985}. Later, the X-ray flux increased over the years, as it was observed by several X-ray satellites, such as {\it EXOSAT}, {\it Ginga}, and {\it ASCA} \citep{Turner1989,Turner1997}. 

NGC\,7582 was found to be in type\,1 state in July 1998. It transitioned to type\,1.9 in October 1998, with broad H$\beta$ line disappearing \citep{Aretxaga1999}. In November 1998, quasi-simultaneous X-ray observation showed a low \ed, with $\lambda_{\rm Edd}\sim 0.003$. At this time, the source was found in Compton thin state with $N_{\rm H}\sim 1.5\times 10^{23}$ \pcm. 

In July 2004, NGC\,7582 showed only NELs in its optical spectra as it moved to type\,2 \citep{RicciTV2018}. The \ed decreased further to $\sim 0.002$ during this time, although the \nh did not change much, with $N_{\rm H} \sim 1.6\times 10^{23}$ \pcm. The source remained in type\,2 state in 2008 and 2016 \citep{RicciTV2018}. The quasi-simultaneous X-ray observation found $\lambda_{\rm Edd}\sim 0.002$ and $N_{\rm H} \sim 3\times 10^{23}$ \pcm in 2016. 

NGC\,7582 showed CO events several times, transiting between Compton-thin and CT states in 2005, 2007, and 2014 \citep{Lefkir2023}. However, we did not have optical observation at that time to study the relation of $N_{\rm H}$ with the optical state. NGC\,7582 showed type\,2 state when $\lambda_{\rm Edd}\sim 0.001$, and transited to type\,1.9 when $\lambda_{\rm Edd}\sim 0.003$. The transition occurred around $\lambda_{\rm Edd}\sim 0.003$. No relation between the \nh and the spectral state is observed in this source. This implied that the variable \ed is the reason for the optical state transition in NGC\,7582.

\begin{table*}
\caption{NGC\,7582}
\centering
\begin{tabular}{ccccccccccccc}
\hline
Dates & H$\alpha$ & H$\beta$ & Optical & Optical &   $\log L_{\rm X}$  &$\log \lambda_{\rm Edd}$  & $\log N_{\rm H}$ &X--ray & Ref. & $\Delta T$\\
UT    & BEL       & BEL      & type    & Inst.  &  $\log({\rm erg~s^{-1}})$ &       &    $\log({\rm cm}^{-2})$ & Inst.&  &  (Days)   \\
\hline
1977/07/21& Y& Y& 1.0 & AAT & -- & -- & -- & --  &  1 & --\\
1977      & --& --& --&  -- & $42.42\pm0.04$  &  $-2.33\pm0.04$ &-- &  A-V  &  2 & --\\
\hline
1980/05/05& --& --& --&  -- & $42.00\pm0.04$ & $-2.77\pm0.04$ & -- & Ein.  &  3 & -- \\
\hline
1998/07/11& Y& Y& 1.0 & Danish & -- & --      & --   &-- & 4 & --\\
1998/10/06& Y& N& 1.9 & Danish & --  & --      & --   &-- & 4 & --\\
\hline
1998/10/21& Y& N& 1.9 & ESO   & --  & --      & --   &-- & 4 & 18\\
1998/11/09& --& --& --& -- & $42.30\pm0.01$ &$-2.46\pm0.03$  & $23.16\pm0.03$   & BS  &  5 & --\\
\hline
2004/07/17& N& N& 2.0 & Gemini    & --  & --      & --   &-- &  6 & 15\\
2004/08/01& --& --& --& -- & $41.99\pm0.01$ & $-2.78\pm0.03$  & $23.20\pm0.12$  & RXTE  &  7 & -- \\
\hline
2008/11/08& N& N& 2.0 & SAAO        & --  &  --      & --   &-- &  8 & -- \\
2016/04/28& --& --& -- & -- & $42.01\pm0.02$   & $-2.76\pm0.03$   & $23.51\pm0.02$   & XMM, Nu   & 9 & 90 \\
2016/07/27& N& N& 2.0 & VLT  & --  &  --      & --   &-- &  8 & --\\
\hline
\end{tabular}
\leftline{References : (1) \citet{Ward1980}, (2) \citet{Ward1978}, (2) \citet{Maccacaro1981}, (3) \citet{Morris1985}}
\leftline{(4) \citet{Aretxaga1999}, (5) \citet{Turner2000}, (6) \citet{RicciTV2018}, (7) \citet{Rivers2015}, }
\leftline{(8) \citet{Oh2022}, (9) \citet{Lefkir2023}. }
\leftline{Optical instruments: AAT: AAT/IPCS; Danish: Danish 1.54m/DFOSC; ESO: ESO 3.6m/EFOSC2; }
\leftline{Gemini: Gemini 8.1m/GMOS-IFU; SAOO: SAOO 1.9m; VLT: ESO-VLT/Xshooter.}
\leftline{X-ray instruments: A-V: \emph{Arial V}/SSI;  Ein.: \emph{Einstein}/MPC; BS: \emph{BeppoSAX}/PDS; RXTE: \emph{RXTE}/PCA;}
\leftline{XMM: \emph{XMM-Newton}/EPIC-PN; Nu: \emph{NuSTAR}/FPMA+FPMB.}
\label{tab:ngc7582}
\end{table*}

\subsection{NGC\,7603}\label{subsec:ngc7603}
NGC\,7603 was found in a type\,1 state in November 1974. Later, in November 1975, NGC\,7603 only showed NELs as it transitioned to a type\,2 state \citep{Tohline1976}. The BELs recovered again, moved to type\,1 state in February 1976, and remained there for three decades \citep{Kollatschny2000,Trippe2010}. In 2012, the source was observed in a type\,1.8 state \citep{Theios2016}. In 2019, the source moved back to a type\,1 state \citep{Koss2022}.

We observed a correlation of spectral type and \ed in NGC\,7603. The source was observed in type\,1.8 state when $\lambda_{\rm Edd}\sim 0.002$. The source was found in a type\,1 state when the \ed was higher, with $\lambda_{\rm Edd}>0.005$. We did not have any information about the obscuration properties when the source was in the type\,2 state.

\begin{table*}
\caption{NGC\,7603}
\centering
\begin{tabular}{cccccccccccc}
\hline
Dates & H$\alpha$ & H$\beta$ & Optical & Optical &   $\log L_{\rm X}$  &$\log \lambda_{\rm Edd}$  & $\log N_{\rm H}$ &X--ray & Ref. & $\Delta T$\\
UT    & BEL       & BEL      & type    & Inst.  &  $\log({\rm erg~s^{-1}})$ &       &    $\log({\rm cm}^{-2})$ & Inst.&  &  (Days)   \\
\hline
1974/11/06& Y& Y& 1.0 & Lick & -- & -- & --   & -- &1 & -- \\
1975/11/08& N& N& 2.0 & Lick & -- & -- & --   & -- &1 & -- \\
1976/02/02& Y& Y& 1.0 & Lick & -- & -- & --   & -- & 1 & -- \\
\hline
1979/10/22& Y& Y& 1.0 & ESO  & -- & -- & --   & -- & 2 & -- \\
1979/12/12& --&-- & -- & --  & $42.86\pm0.05$  & $-2.76\pm0.04$   & --  & Ein. &  3 & --\\ 
1980/06/20&--&-- & -- & --   & $43.33\pm0.05$  & $-2.27\pm0.04$   & --  & Ein.  & 4 & --\\
1981/11/03& Y& Y& 1.0 & ESO & -- & -- & --   & -- & 2 & --\\
\hline
1992/06/16& --&-- & -- & --   & $43.76\pm0.04$ & $-1.79\pm0.05$ & -- & ROSAT  & 5 & 3\\ 
1992/06/19& Y& Y& 1.0 & CA & -- & -- & --   & -- & 2 & --\\
\hline
1998/12/20& Y& Y& 1.0 & CA  & -- & -- & --   & -- & 2 & --\\
2006/06/14& --&-- & -- & --    & $43.60\pm0.01$ & $-1.97\pm0.05$ & $20.61^*$ & XMM &  6 & --\\ 
2007/10/05& Y& Y& 1.0 & CTIO  & -- & --   & --   & -- & 7 & --\\
2009/06/19& Y& Y& 1.0 & KPNO  & -- & --   & --   & -- & 8 & --\\
\hline
2012/06/02& --&-- & -- & --  & $42.92\pm0.02$ & $-2.69\pm0.06$ & $23.34\pm0.10$  & Suz  & 9 & 125\\ 
2012/10/08& Y& Y& 1.8 & Lick-N  & -- & -- & -- & -- & 10 & --\\
\hline
2013/04/23& --&-- & -- & --  & $41.23\pm0.04$& $-4.28\pm0.05$ & $20.61^*$ & Swift & 11& --\\
2019/07/20& Y& Y& 1.0 & Hale  & -- & --   & -- & -- & 12 & --\\
\hline
\end{tabular}
\leftline{References : (1) \citet{Tohline1976}, (2) \citet{Kollatschny2000}, (3) \citet{dellaCeca1990}, }
\leftline{(4) \citet{Kruper1990}, (5) \citet{Moran1996}, (6) \citet{Boissay2016}, (7) \citet{Trippe2010}, (8) \citet{Koss2017}, }
\leftline{(9) \citet{Ehler2018}, (10) \citet{Theios2016}, (11) This Work, (12) \citet{Koss2022}.}
\leftline{Optical instruments: Lick: Lick 3m; ESO: ESO 1.5m; CA: CA 3.5m; CTIO: CTIO 1.5m/RC; KPNO: KPNO 2.1m/Goldcam;}
\leftline{ Lick-N: Lick-Nickel 40inch; Hale: Hale 200inch/DBPS. }
\leftline{X-ray instruments: Ein.: \emph{Einstein}/IPC; ROSAT: \emph{ROSAT}/PSPC; XMM: \emph{XMM-Newton}/EPIC-PN; Suz: \emph{Suzaku}/XIS;}
\leftline{Swift: \emph{Swift}/XRT.}
\label{tab:ngc7603}
\end{table*}

\subsection{Mrk 6}
\label{subsec:mrk6}

Mrk\,6 was identified as a type\,1 AGN in 1976 \citep{Malkan1983}. Later, Mrk\,6 entered in type\,1.8 state with weakening broad lines in June 1977 \citep{Doroshenko2003}. The source remained in the low state for a year before returning to a type\,1 state. Since then, Mrk\,6 has been in type\,1 state for over three decades \citep{Doroshenko2003,Afanasiev2014}.

Mrk\,6 is an obscured AGN, with the obscuring materials reported to be complex. The \nh is found to be in the range of $\sim (0.1-3)\times10^{23}$ \pcm \citep{Layek2024}.
We did not observe any correlation of \nh with the spectral state in the source. On the other hand, Mrk\,6 correlated the spectral state with \ed. In type\,1.8 state, \ed was obtained to be $\lambda_{\rm Edd}\sim 0.002.$ The source was observed in type\,1 state for a wide range of \ed, with $\lambda_{\rm Edd}\sim 0.003-0.02$. No relation between the \nh and the optical spectral state is found in Mrk\,6. This implies that the changing accretion rate led to the CL transition in Mrk\,6.

\begin{table*}
\caption{Mrk\,6}
\centering
\begin{tabular}{cccccccccccc}
\hline
Dates & H$\alpha$ & H$\beta$ & Optical & Optical &   $\log L_{\rm X}$  &$\log \lambda_{\rm Edd}$  & $\log N_{\rm H}$ &X--ray & Ref. & $\Delta T$\\
UT    & BEL       & BEL      & type    & Inst.  &  $\log({\rm erg~s^{-1}})$ &       &    $\log({\rm cm}^{-2})$ & Inst.&  &  (Days)   \\
\hline
\hline
1976/12/19& Y& Y& 1.0& Palomer & -- & --    &  --& -- & 1 & --\\
1977/06/03& Y& Y& 1.8& CrAO   & -- & --     & --& -- & 2 & --\\
1977/09/27&--&--& -- & --  & $42.34\pm0.05$ &  $-2.79\pm0.05$ &--& H-1 & 3 & 76\\
1977/12/12& Y& Y& 1.8& CrAO    & -- & --    &  -- & -- & 2 & --\\
1978/03/22&--&--& -- & -- & $42.51\pm0.04$ & $-2.61\pm0.04$ &--& H-1 &  3 & --\\
1981/11/05& Y& Y& 1.0& Palomer & --   & --  &  -- & -- & 1 & --\\
1990&--&--& -- & --  & $42.97\pm0.05$  & $-2.13\pm0.04$ & $22.84\pm0.06$& ROSAT & 4 & --\\
1990/04/26& Y& Y& 1.0& CrAO    & --    & --  &  -- &-- &  2 & --\\
\hline
1997/04/07&--&--& -- & -- & $43.01\pm0.03$ & $-2.08\pm0.04$ & $22.49\pm0.08$ & ASCA & 5 & 2\\
1997/04/09& Y& Y& 1.0& MMT     & -- & --   &  -- &-- &  5 & --\\
\hline
1999/09/14&--&--& --   &-- & $43.35\pm0.02$ & $-1.70\pm0.04$ & $22.62\pm0.15$ & BS & 6 & -- \\
1999/11/03& Y& Y& 1.0& CrAO    & --   & --  &  -- &-- &  7 & 50\\
\hline
2003/02/05& Y& Y& 1.0& CrAO  & --    & --  &  -- &-- &  7  & 80\\
2003/04/26&--&--& -- & -- & $43.11\pm0.01$  &$-1.97\pm0.03$ & $21.90\pm0.03$ & XMM & 8 & --\\
\hline
2005/10/06& Y& Y& 1.0& CrAO    &-- & --  &  -- & -- & 7 & 21\\
2005/10/27&--&--& --  &-- & $43.22\pm0.01$  & $-1.85\pm0.03$ & $22.38\pm0.01$ & XMM & 8 & --\\
\hline
2005/11/27& Y& Y& 1.0& CrAO    & -- & -- &  -- & -- & 7 & --\\
2006/01/19&--&--& -- &-- & $42.79\pm0.02$ & $-2.32\pm0.03$ &  $22.77\pm0.13$  & Swift & 9 & 19\\
2006/02/08& Y& Y& 1.0& CrAO    & -- & --  &  -- &-- &  7 & --\\
2006/03/24&--&--& -- &-- & $42.69\pm0.03$  & $-2.43\pm0.04$ &  $23.06\pm0.06$ & Swift & 9 & 12\\
2006/04/05& Y& Y& 1.0& CrAO    &-- & --  &  -- & -- & 7 & --\\
2006/04/12&--&--& -- &-- & $42.91\pm0.03$  & $-2.19\pm0.04$ & $22.80\pm0.08$ & Swift & 9 & 38\\
2006/05/20& Y& Y& 1.0& CrAO   & -- & --  & --  &  --  &  7 & --\\
\hline
2008/12/05& Y& Y& 1.0& KPNO& -- & --    & --  & --   & 10 & -- \\
2015/11/09&--&--& -- &-- & $42.65\pm0.02$  & $-2.47\pm0.03$ & $23.00\pm0.16$ & Nu & 11 & --\\
\hline
\end{tabular}
\leftline{References: (1) \citet{Malkan1983}, (2) \citet{Doroshenko2003}, (3) \citet{dellaCeca1990}, (4) \citet{Boller1992}, }
\leftline{(5) \citet{Feldmeier1999}, (6) \citet{Malizia2003}, (7) \citet{Sergeev1999}, (8) \citet{Mingo2011}, (9) This Work,}
\leftline{(10) \citet{Oh2022}, (11) \citet{Molina2019}. }
\leftline{Optical instruments: Palomer: Palomer 1.5m/SIT; CrAO: CrAO 2.6m; MMT: MMT/BCS; KPNO: KPNO 2.1m/Goldcam.}
\leftline{X-ray instruments: H1: \emph{HEAO-1}; ROSAT: \emph{ROSAT}/HRI; ASCA: \emph{ASCA}/SIS+GIS; BS: \emph{BeppoSAX}/PDS;}
\leftline{ XMM: \emph{XMM-Newton}/EPIC-PN; Swift: \emph{Swift}/XRT; Nu: \emph{NuSTAR}/FPMA+FPMB.}
\label{tab:mrk6}
\end{table*}

\subsection{Mrk\,590}
\label{subsec:mrk590}

Mrk\,590 has undergone several CL transitions in the last $\sim 40 $ years. 
In 1982, it was found in a type\,1 state \citep{Peterson1984}.
It remained in a type\,1 state for $\sim 25$ years with a $\lambda_{\rm Edd}\sim 0.02-0.7$ \citep{Peterson1993,Koss2017}. 
The broad lines weakened, and the source moved to a type\,1.8 state in September 2006 \citep{Denney2014}. 
Mrk\,590 lost its broad H$\beta$ line in February 2013 and moved to type\,1.9 state \citep{Denney2014}. 
The source remained in this low state till January 2014. The X-ray flux was also low in this period, with $\lambda_{\rm Edd}<0.004$. 
From late 2014, the X-ray flux increased; by 2017, it increased over 100 times. In December 2017, Mrk\,590 moved to type\,1 state, with $\lambda_{\rm Edd}\sim 0.2$ \citep{Oh2022}.

Mrk\,590 showed an increasing \nh as the source moved towards the type\,2 state, though the \nh was always $N_{\rm H}<10^{21}$ \pcm. 
Hence, the obscuration is unlikely to cause the CL transition. Mrk\,590 showed a clear correlation of \ed with the spectral state throughout the years. 
Mrk\,590 was found in type\,1.9 state with very low \ed, with $\lambda_{\rm Edd}\sim 0.004$. The type\,1 state was observed when \ed varied in the range of $\lambda_{\rm Edd}>0.02$. This indicates that the variation in the accretion rate is responsible for the CL transition.

\begin{table*}
\caption{Mrk\,590}
\centering
\begin{tabular}{cccccccccccc}
\hline
Dates & H$\alpha$ & H$\beta$ & Optical & Optical &   $\log L_{\rm X}$  &$\log \lambda_{\rm Edd}$  & $\log N_{\rm H}$ &X--ray & Ref. & $\Delta T$\\
UT    & BEL       & BEL      & type    & Inst.  &  $\log({\rm erg~s^{-1}})$ &       &    $\log({\rm cm}^{-2})$ & Inst.&  &  (Days)   \\
\hline
\hline
1979/07/10&--&--& -- &-- & $43.61\pm0.05$ &  $-0.74\pm0.04$ & --  & Ein. & 1 & 103 \\
1982/11/16& Y& Y& 1.0& Perkins & --    & --  &  -- & --&2 & --\\
\hline
1983/12/04& Y& Y& 1.0& Perkins & --    & --  &  -- & --&3 & 30\\
1984/01/03&--&--& -- &-- & $43.93\pm0.04$   & $-0.32\pm0.04$ & $20.69\pm0.30$& EXO & 4 & --\\
\hline
1993/01/12& Y& Y& 1.0 & Perkins & --   & --  &  -- & --&5 & 2\\
1993/01/14&--&--& -- &-- & $44.14\pm0.05$  &$-0.16\pm0.05$ &$20.40\pm0.15$& ROSAT & 6 & --\\
\hline
2002/01/01&--&--& -- &-- & $42.82\pm0.04$ & $-1.70\pm0.03$ & $20.45^*$& XMM & 7 & -- \\
\hline
2003/01/08& Y& Y& 1.0& SDSS & -- & --  &  -- & --&8 & --\\
\hline
2004/04/07&--&--& -- &-- & $42.96\pm0.03$  & $-1.54\pm0.04$ &$20.45^*$& XMM & 9 & --\\
\hline
2006/09/22& Y& Y& 1.8& MDM & --   & --  &  -- &--& 10 & --\\
\hline
2011/01/23&--&--& -- &-- & $43.02\pm0.02$ & $-1.47\pm0.03$ & $20.45^*$ & Suz & 9 & --\\
\hline
2013/02/18& Y& Y& 1.9& LBT & --     & --  &  -- &--& 10 & 87\\
2013/06/16&--&--& -- &-- & $42.25\pm0.03$  & $-2.33\pm0.03$ &$20.45^*$& Ch & 12 & --\\
\hline
2013/12/14&--&--& --  &-- & $42.21\pm0.08$ & $-2.37\pm0.06$ &$20.45^*$& Swift & 13 & 4\\
2013/12/18& Y& Y& 1.9& Mayall & --  & --  &  -- & --&10 & --\\
\hline
2014/01/07& Y& Y& 1.9& MDM & --   & --  &  -- & --&11 & 4\\
2014/01/11&--&--& -- &-- & $42.23\pm0.08$ & $-2.34\pm0.05$&$20.45^*$ & Swift & 13 & --\\
\hline
2016/02/05&--&--& -- &-- & $42.64\pm0.03$ & $-1.91\pm0.04$ &$20.45^*$& Swift, Nu & 7 & --\\
\hline
2017/12/04&--&--& -- &-- & $44.67\pm0.09$ & $-0.67\pm0.05$ & $20.45^*$& Swift & 13 & 1\\ 
2017/12/05& Y& Y& 1.0& VLT & --  & --  &  -- & --&14 & --\\
\hline
2018/10/27& Y& Y& 1.0& Subaru& $43.15\pm0.01$  & $-1.32\pm0.04$ & $20.45^*$ & Nu & 15, 7 & 0\\
\hline
2021/01/10&--&--& --&-- & $42.98\pm0.01$ & $-1.52\pm0.03$ & $20.45^*$ &Swift, Nu & 7 & -- \\
\hline
\hline
\end{tabular}
\leftline{$^*$ Spectra are fitted with Galactic absorption.}
\leftline{References: (1) \citet{Kriss1980}, (2) \citet{Peterson1984}, (3) \citet{Ferland1990}, (4) \citet{Turner1989}, }
\leftline{(5) \citet{Peterson1993}, (6) \citet{Boller1992}, (7) \citet{Ghosh2022}, (8) \citet{Koss2017},}
\leftline{(9) \citet{Rivers2012}, (10) \citet{Denney2014}, (11) \citet{Grier2012}, (12)  \citet{Longinotti2007}, (13) This Work,}
\leftline{(14) \citet{Oh2022}, (15) \citet{Mandal2021}.}
\leftline{Optical instruments: LCO: LCO 0.6m; Perkins: Lowell-Perkins 1.8m/IDS; MDM: MDM 1.3m; LBT: LBT/MODS1; }
\leftline{Mayall: Mayall 4m/KOSMOS; VLT: VLT-Xshooter; Subaru: Subaru 8.2m/HDS. }
\leftline{X-ray instruments: Ein.: \emph{Einstein}/IPC; EXO: \emph{EXOSAT}/ME; ROSAT: \emph{ROSAT}/PSPC; XMM: \emph{MM-Newton}/EPIC-PN;}
\leftline{Suz: \emph{Suzaku}/XIS; Ch: \emph{Chandra}/ACIS; Swift: \emph{Swift}/XRT; Nu: \emph{NuSTAR}/FPMA+FPMB.}
\label{tab:mrk590}
\end{table*}

\subsection{Mrk\,1018}
\label{subsec:mrk1018}
Mrk\,1018 was observed to be in a type\,1.9 state in 1979 with a weak broad H$\alpha$ line \citep{Osterbrock1981}. 
The source transitioned to type\,1 state between 1979 and 1984 \citep{Cohen1986}. The source remained in the high type\,1 state for three decades, 
with a high \ed, $\lambda_{\rm Edd}\sim 0.06-0.09$. After 2008, the X-ray flux declined, and Mrk\,1018 transitioned to a type\,1.9 state in January 
2015 \citep{McElroy2016}. An X-ray observation six months earlier revealed a low \ed, with $\lambda_{\rm Edd}\sim 0.01$. In October 2019, the source 
was observed in a type\,1.8 state, with increasing X-ray flux \citep{Hutsemekers2020}, with quasi-simultaneous X-ray observation showed $\lambda_{\rm Edd}\sim 0.02$.

Mrk\,1018 clearly showed a correlation of \ed with the spectral state. The \ed was observed to be $\lambda_{\rm Edd}\sim 0.06-0.09$ when the source was in the type\,1 state. Mrk\,1018 was observed in type\,1.8 state at $\lambda_{\rm Edd}\sim 0.02$. The type\,1.9 state was observed at $\lambda_{\rm Edd}\sim 0.01$. It is clear that Mrk\,1018 lost its broad H$\beta$ line at $\lambda_{\rm Edd}<0.02$. Mrk\,1018 did not show any signature of the obscuration in the X-ray spectra, suggesting that the \ed is the reason for the CL transition in the source.

\begin{table*}
\caption{Mrk\,1018}
\centering
\begin{tabular}{ccccccccccccc}
\hline
Dates & H$\alpha$ & H$\beta$ & Optical & Optical &   $\log L_{\rm X}$  &$\log \lambda_{\rm Edd}$  & $\log N_{\rm H}$ &X--ray & Ref. & $\Delta T$\\
UT    & BEL       & BEL      & type    & Inst.  &  $\log({\rm erg~s^{-1}})$ &       &    $\log({\rm cm}^{-2})$ & Inst.&  &  (Days)   \\
\hline
\hline
1979/09/14& Y& N& 1.9& Lick  &  -- & --   & --   & -- & 1& --\\
\hline
1984/01/27& Y& Y& 1.0 & UCDS  &  --&  -- & --   & -- & 2& --\\
\hline
2004/08/15& Y& Y& 1.0& 6dF  &  -- & --    & --   & -- & 3 & --\\
2005/01/15&--&--& -- &-- & $43.58\pm0.02$ & $-1.09\pm0.03$ & $20.38^*$ & XMM & 4 & --\\
\hline
2007/06/11&--&--& --&-- & $43.59\pm0.06$  & $-1.07\pm0.04$ &$20.38^*$ & Swift & 5 &  120\\
2007/10/05& Y& Y& 1.0& CTIO  & -- & --    & --   & -- & 6& --\\
\hline
2008/08/07&--&--& --  &-- & $43.62\pm0.01$  & $-1.04\pm0.03$ &$20.38^*$ & XMM & 4 & -- \\
2009/01/22& Y& Y& 1.0 & Keck  & -- & --      & --   & -- &7 & 161\\
2009/07/03&--&-- &-- & -- & $43.56\pm0.02$  & $-1.11\pm0.03$ & $20.38^*$ & Suz & 8 & --\\
\hline
2010/11/27&--&--& --  &-- & $43.53\pm0.01$  & $-1.15\pm0.03$ &$20.38^*$ & Ch & 4 & 12 \\
\hline
2013/06/07&--&--& -- &-- & $43.46\pm0.07$ & $-1.23\pm0.03$ & $20.38^*$ & Swift & 5 & 140\\
2013/10/25& Y& Y& 1.0 & SDSS  & -- & --      & --   & -- &9 & --\\
\hline
2014/06/09&--&--& --&-- & $42.83\pm0.13$ &  $-1.96\pm0.07$ &$20.38^*$ & Swift & 5 & 216\\
2015/01/11& Y& Y& 1.9 & VLT-M & --& --    & --   & -- &10& --\\
\hline
2016/02/10&--&--& -- &-- & $42.82\pm0.02$ & $-1.97\pm0.03$ &$20.38^*$ & Nu & 4 & -- \\
2019/06/29&--&--& --&-- & $43.06\pm0.02$ &  $-1.70\pm0.03$ &$20.38^*$ & Swift & 5 & 119\\
2019/10/26& Y& Y& 1.8& VLT-F & -- & --  & --   & -- &8 & --\\
\hline
\end{tabular}
\leftline{References: (1) \citet{Osterbrock1981}, (2) \citet{Cohen1986}, (3) \citet{Jones2009}, (4) \citet{Lyu2021}, (5) This Work,}
\leftline{(6) \citet{Trippe2010}, (7) \citet{Bennert2011}, (7) \citet{Hutsemekers2020}, (9) \citet{Oh2022},}
\leftline{(10) \citet{McElroy2016}.}
\leftline{Otical instruments: Lick: Lick-Shane 3m; UCSD: UCSD/UMinn 1.5m; CTIO: CTIO 1.5m/RC; Keck: Keck/LRIS;}
\leftline{Lowell: Lowell-Perkins/Deveny; VLT-M: VLT/MUSE; VLT-F: VLT/FORS2. }
\leftline{X-ray instruments: XMM: \emph{XMM-Newton}/EPIC-PN; Suz: \emph{Suzaku}/XIS; Ch: \emph{Chandra}/ACIS; Swift: \emph{Swift}/XRT;}
\leftline{Nu: \emph{NuSTAR}/FPMA+FPMB.}
\label{tab:mrk1018}
\end{table*}

\subsection{Mrk\,1393}
\label{subsec:mrk1393}
Mrk\,1393 was observed in type\,1.9 state in 1993 \citep{Owen1995}. The March 2001 observation also found the source remained in type\,1.9 state \citep{Wang2009}.
Later, the source brightened and entered in a type\,1.8 state in 2005 \citep{Wang2009}. From quasi-simultaneous X-ray observations, the \ed was estimated to be 
$\lambda_{\rm Edd}\sim 0.015$. In May 2022, Mrk\,1393 was found to move in a type\,1 state with increasing X-ray flux. The quasi-simultaneous X-ray observation estimated
the $\lambda_{\rm Edd}\sim 0.06$.

Mrk\,1393 was found in type\,1.8 state at $\lambda_{\rm Edd}\sim 0.012$. It moved to type\,1 state with increasing \ed, with $\lambda_{\rm Edd}\sim 0.05$. 
This source showed a clear correlation of spectral state with \ed. In Mrk\,1393, no relation of the \nh with the spectral state was observed. In fact, 
a higher \nh is observed in the type\,1 state than in the type\,1.8/1.9 state. The change in the accretion rate is likely to be the reason for the CL transition.

\begin{table*}
\caption{Mrk\,1393}
\centering
\begin{tabular}{ccccccccccccc}
\hline
Dates & H$\alpha$ & H$\beta$ & Optical & Optical &   $\log L_{\rm X}$  &$\log \lambda_{\rm Edd}$  & $\log N_{\rm H}$ &X--ray & Ref. & $\Delta T$\\
UT    & BEL       & BEL      & type    & Inst.  &  $\log({\rm erg~s^{-1}})$ &       &    $\log({\rm cm}^{-2})$ & Inst.&  &  (Days)   \\
\hline
1991/08/06&--&--& --  &-- & $43.00\pm0.08$ & $-2.63\pm0.16$ & -- & ROSAT & 1 & --\\
1993/04/24& Y& N& 1.9& SDSS  &  -- & --     & --   & -- & 2 & --\\
2001/03/22& Y& N& 1.9& SDSS  &  -- & --     & --   & -- & 1 & --\\
\hline
2005/07/20&--&--& -- & -- & $43.68\pm0.02$  & $-1.91\pm0.15$ &$21.48\pm0.07$ & XMM & 1 & --\\
2005/09/04& Y& Y& 1.8& BAO  &  -- & -- & --  & -- & 1 & 6\\
2005/09/10&--&--& -- &-- & $43.67\pm0.04$  & $-1.92\pm0.15$ & $20.52\pm0.74$& Swift & 1 & --\\
\hline
2006/01/03& Y& Y& 1.8& BAO  & -- & -- &  --& --  & 1      & --\\
\hline
2022/05/31& Y& Y& 1.0& LCO&  -- & --   & -- & -- & 3 & 195\\
2022/12/22&--&--& --  &-- & $44.19\pm0.14$  & $-1.32\pm0.17$ & $21.23\pm0.14$ & Swift & 4  & --\\
\hline
\hline
\end{tabular}
\leftline{References: (1) \citet{Wang2009}, (2) \citet{Owen1995}, (3) \citet{Temple2023}, (4) This Work. }
\leftline{Optical instruments: AAT: AAT/IPCS; BAO: BAO 2.16m/OMS; LCO: LCO/FLOYD.}
\leftline{X-ray instruments: ROSAT: \emph{ROSAT}/PSPC; XMM: \emph{XMM-Newton}/EPIC-PN; Swift: \emph{Swift}/XRT.}
\label{tab:mrk1393}
\end{table*}

\subsection{3C\,390.3}
\label{subsec:3c390}
3C\,390.3 is a changing-look radio galaxy. It was first classified as a broad-line radio galaxy in the 1970s \citep{Yee1981}. 
The optical observations in 1978 showed a type\,1 spectra of the source. In June 1979, the broad lines weakened as the 
source entered in a type\,1.8 state \citep{Netzer1982}. Subsequent optical observation suggested that 3C\,390.3 remained in the 
type\,1.8 state until 1984 \citep{Veilleux1991}. The source moved to type\,1 state in 1985 \citep{Veilleux1991}, with increasing 
X-ray flux \citep{Inda1994}. The source remained in the high state till the observation in 2014 \citep{Shapovalova2010,Sergeev2017}. 

3C\,390.3 showed type\,1 spectra when the X-ray flux was high, with $\lambda_{\rm Edd}>0.04$. The type\,1.8 state was seen for 
$\lambda_{\rm Edd}< 0.01$. This indicated that 3C\,390.3 moved towards type\,2 as the \ed decreased. 3C\,390.3 is an unobscured 
AGN with no intrinsic absorption. This fact suggested that the changing \ed is behind the CL transition.

\begin{table*}
\caption{3C\,390.3}
\centering
\begin{tabular}{ccccccccccccc}
\hline
Dates & H$\alpha$ & H$\beta$ & Optical & Optical & $\log L_{\rm X}$ &$\log \lambda_{\rm Edd}$ & $\log N_{\rm H}$ &X--ray & Ref. & $\Delta T$\\
UT    & BEL       & BEL      & type    & Inst.  & $\log({\rm erg~s^{-1}})$ & & $\log({\rm cm}^{-2})$ & Inst.& & (Days)\\
\hline
\hline
1978/11/02& Y& Y& 1.0& Hale & -- & -- & -- &--& 1 & 19\\
1978/11/21&--&--& -- &-- & $44.31\pm0.10$ & $-1.21\pm0.06$ &--& Ein. & 2 & --\\
\hline
1979/10/26& Y& Y& 1.8& Hale & -- & -- & -- &--& 1 & 57\\
1980/01/01&--&--& -- &-- & $43.42\pm0.19$  & $-2.22\pm0.10$ &--& Ein. & 2 & --\\
\hline
1980/04/08&--&--& --  &-- & $43.63\pm0.10$  & $-2.00\pm0.06$ &--& Ein. & 2 & 30\\
1980/05/08& Y& Y& 1.8& Lick & -- & -- & -- &--& 2 & --\\
\hline
1984/05/28& Y& Y& 1.8& Lick & -- & -- & -- & --& 3 & 4\\
1984/06/01&--&--& -- &-- & $43.43\pm0.04$  & $-2.21\pm0.03$ &$20.78^*$& EXO & 4 & --\\
\hline
1985/08/18& Y& Y& 1.0& Lick & -- & -- & -- & --&2 & 91\\
1985/11/17&--&--& --   &-- & $44.24\pm0.02$  & $-1.29\pm0.04$ &$20.78^*$& EXO & 4 & --\\
\hline
1986/03/18&--&--& --  &-- & $44.20\pm0.02$ &$-1.34\pm0.04$ &$20.78^*$& EXO & 4 & 140\\
1986/08/06& Y& Y& 1.0& Lick & -- & -- & -- &--& 2 & --\\
\hline
1995/01/15&--&--& -- &-- & $44.13\pm0.01$  & $-1.42\pm0.03$ & $21.08\pm0.04$& ASCA & 5 & --\\
1995/04/25& Y& Y& 1.0& SAO & -- & -- & -- & --& 6 & 10\\
1995/05/05&--&--& -- &-- & $44.30\pm0.01$  & $-1.22\pm0.03$ & $21.08\pm0.04$& ASCA & 5 & 21\\
1995/05/26& Y& Y& 1.0& SAO & -- & -- & -- & --& 6 & --\\
\hline
1996/03/21& Y& Y& 1.0& SAO & -- & -- & -- & --& 6 & --\\
1996/05/19&--&--& -- &-- & $44.44\pm0.05$  & $-1.05\pm0.04$ &--& RXTE & 7 & --\\
\hline
1996/06/13&--&--& -- &-- & $44.17\pm0.06$  & $-1.38\pm0.04$ &$20.60^*$ & RXTE & 7 & 2\\
1996/06/15& Y& Y& 1.0& SAO & -- & -- & -- & --& 6 & --\\
\hline
1996/07/10&--&--& -- &-- & $44.23\pm0.08$  & $-1.30\pm0.03$ &$20.60^*$& RXTE & 7 & 2\\
1996/07/12& Y& Y& 1.0& SAO & -- & -- & -- & --& 6 & --\\
\hline
1997/01/09&--&--& -- &-- & $44.16\pm0.02$ & $-1.39\pm0.04$ &$21.00\pm0.08$ &BS & 8 & 54 \\
1997/03/04& Y& Y& 1.0& SAO & -- & -- & -- & --& 6 & --\\
\hline
2004/09/06& Y& Y& 1.0& SAO & -- & -- & -- &--& 6 & 32\\
2004/10/08&--&--& --  &-- & $44.47\pm0.01$ & $-1.01\pm0.03$ &$20.30\pm0.15$& XMM & 9 & --\\
2004/10/17&--&--& --  &-- & $44.37\pm0.01$ & $-1.13\pm0.03$ &$20.48\pm0.04$ &XMM & 9 & 62\\
2004/12/18& Y& Y& 1.0& SAO & -- & -- & -- & --& 6 & --\\
\hline
2006/12/14&--&--& -- &-- & $44.45\pm0.02$  & $-1.05\pm0.04$ &$20.30\pm0.15$ &Suz & 9 & 4\\
2006/12/18& Y& Y& 1.0& SAO & -- & -- & -- & --& 6 & --\\
\hline
2008/05/30&--&--& --  &-- & $44.42\pm0.01$ & $-1.07\pm0.03$ & $20.85\pm0.11$&Swift & 10 & 1\\
2008/05/31& Y& Y& 1.0 & CrAO & -- & -- & -- & --&11& --\\
\hline
2008/06/07& Y& Y& 1.0 & CrAO & $44.53\pm0.01$  & $-0.94\pm0.03$ &$21.00\pm0.11$ & Swift & 11, 10 & 0\\
\hline
2013/05/24&--&--& --  &-- & $44.48\pm0.02$ & $-1.00\pm0.04$ & $20.60^*$&Nu & 12& 8 \\
2013/06/02& Y& Y& 1.0& CrAO & -- & -- & -- &--& 11& --\\
\hline
2014/05/30& Y& Y& 1.0 & CrAO & -- & -- & -- &--& 11 & 16\\
2014/06/15&--&--& --  &-- & $44.61\pm0.02$ &  $-0.83\pm0.03$ & $20.70\pm0.01$&Ch & 13 & --\\
\hline
\hline
\end{tabular}
\leftline{$^*$ fitted with the fixed value.}
\leftline{References: (1) \citet{Yee1981}, (2) \citet{Wamsteker1997}, (3) \citet{Veilleux1991}, (4) \citet{Inda1994},}
\leftline{(5) \citet{Leighly1997}, (6) \citet{Shapovalova2010}, (7) \citet{Gliozzi2003}, (8) \citet{Grandi1999}, }
\leftline{(9) \citet{Sambruna2009}, (10) This Work, (11) \citet{Sergeev2017}, (12) \citet{Lohfink2015}, (13) \citet{Tombesi2016}.  }
\leftline{Optical instruments: Hale: Hale/5.08m; Lick: Lick-Shane 3m; SAO: SAO 1m; CrAO: CrAO 2.6m.}
\leftline{X-ray instruments: Ein.: \emph{Einstein}/MPC; EXO: \emph{EXOSA}T/ME; ASCA: \emph{ASCA}/GIS+SIS; RXTE: \emph{RXTE}/PCA;}
\leftline{BS: \emph{BeppoSAX}/PDS; XMM: \emph{XMM-Newton}/EPIC-PN; Suz: \emph{Suzaku}/XIS; Swift: \emph{Swift}/XRT; Ch: \emph{Chandra}/ACIS.}
\leftline{}
\label{tab:3c390}
\end{table*}

\subsection{Fairall\,9}
\label{subsec:f9}

Fairall\,9 was classified as type\,1 AGN in the 1980 \citep{Lub1992}, with strong broad Balmer lines. The source remained in type\,1 state until December 1983 \citep{Lub1992}. The quasi-simultaneous X-ray observation in 1979 and 1983 found a high \ed, with $\lambda_{\rm Edd}\sim 0.04-0.11$. 

In December 1984, the broad H$\beta$ line disappeared, and the source was found in type\,1.9 state \citep{Kollatschny1985,Lub1992}. The X-ray flux diminished at this time, with $\lambda_{\rm Edd}\sim 0.013$. Fairall\,9 recovered its broad lines in December 1985 and moved to the type\,1 state \citep{Lub1992}. Since then, the source was found to be in a high state over three decades \citep{Winge1996,Koss2022}. The X-ray flux was high in this period, with $\lambda_{\rm Edd} \sim 0.05-0.1$.

Over the years, Fairall\,9 has been observed to be an unobscured AGN with $N_{\rm H}<10^{22}$ \pcm. In type\,1.9 state, Fairall\,9 showed $N_{\rm H}\sim 5.5\times 10^{21} $ \pcm, which is greater than in type\,1 state. In the type\,1 state, only Galactic absorption was observed. On the other hand, Fairall\,9 showed a clear correlation between \ed and spectral state. Most of the time, the source was observed in type\,1 state with a high \ed, with $\lambda_{\rm Edd} > 0.04$. The type\,1.9 state was observed at a low \ed, with $\lambda_{\rm Edd}\sim 0.013$. Given the unobscured nature of the source, the accretion is likely to be the reason for the CL transition.

\begin{table*}
\caption{Fairall\,9}
\centering
\begin{tabular}{ccccccccccccc}
\hline
Dates & H$\alpha$ & H$\beta$ & Optical & Optical &   $\log L_{\rm X}$  &$\log \lambda_{\rm Edd}$  & $\log N_{\rm H}$ &X--ray & Ref. & $\Delta T$\\
UT    & BEL       & BEL      & type    & Inst.  &  $\log({\rm erg~s^{-1}})$ &       &    $\log({\rm cm}^{-2})$ & Inst.&  &  (Days)   \\
\hline
1979/06/26&--&--&--  & -- & $44.19\pm0.04$   &$-0.94\pm0.04$ & -- & Ein. &  1  & 413 \\
1980/08/16& Y& Y& 1.0 & ESO1  &  -- & --    & --  &  -- & 2 & -- \\
1983/10/17&--&--&--  & -- & $43.86\pm0.05$  & $-1.34\pm0.05$ & $20.56\pm0.08$ & EXO & 3 & 51 \\
1983/12/07& Y& Y& 1.0& ESO2 & -- & --   & --  &  --  &  2 & -- \\
\hline
1984/10/23&--&--&--  & --    &  $43.39\pm0.03$  & $-1.88\pm0.04$ & $20.48\pm0.15$  & EXO & 3 & 33 \\
1984/11/25& Y& N& 1.9& ESO2 & -- & --    & --  &  --  &  4 & -- \\
1984/12/24& Y& N& 1.9& ESO2 & -- & --    & --  &  --  &  2 & -- \\
\hline
1985/12/16& Y& Y& 1.0& ESO2 & -- & --    & --  &  --  &  2 & -- \\
1986/11/25& Y& Y& 1.0& ESO2 & -- & --    & --  &  --  &  2 & -- \\
\hline
1992/07/30& Y& Y& 1.0 & CTIO & -- & --    & --  &  --  &  5 & --\\
1993/11/21&--&--&--  & --    & $44.26\pm0.05$  & $-0.85\pm0.05$ & $20.48^*$& ASCA  & 6 & --\\
\hline
2008/08/05& Y& Y& 1.0 & SAAO & -- & --  & --  &  --  & 7 & 36\\
2008/09/10&--&--&-- & -- & $44.12\pm0.03$ & $-1.02\pm0.04$ & $20.48^*$ & Swift & 8 & --\\
\hline
2014/05/09&--&--&-- & -- & $44.05\pm0.02$  & $-1.11\pm0.04$ & $20.48^*$ & XMM, Nu & 9 & --\\
2016/09/12& Y& Y& 1.0 & duPont & --  & --  & -- & -- &  9 & --\\
\hline
\end{tabular}
\leftline{References: (1) \citet{Petre1984}, (2) \citet{Lub1992}, (3) \citet{Morini1986}, (4) \citet{Kollatschny1985},}
\leftline{(5) \citet{Winge1996}, (6) \citet{Reynolds1997}, (7) \citet{Koss2017}, (8) This Work, (9) \citet{Lohfink2016},}
\leftline{(10) \citet{Oh2022}.}
\leftline{Optical instruments: ESO1: ESO3.6m/IDS; ESO2: ESO1.5m/IDS; CTIO: CTIO 1.5m/R-C; SAOO: SAOO 1.9m;}
\leftline{duPONT: duPONT/B\&C.}
\leftline{X-ray instruments: Ein.: \emph{Einstein}/MPC; EXO: \emph{EXOSAT}/ME; ASCA: \emph{ASCA}/GIS+SIS; Swift: \emph{Swift}/XRT;}
\leftline{XMM: \emph{XMM-Newton}/EPIC-PN.}
\label{tab:f9}
\end{table*}

\subsection{HE\,1136--2304}
\label{subsec:he1136}

HE\,1136--2304 was observed in type\,1.9 state in 1993 \citep{Reimers1996}. The X-ray flux was very low with $\lambda_{\rm Edd} \sim 0.0014$. The May 2002 observation also found HE\,1136--2304 in type\,1.9 state. In 2014, an outburst was detected, as the source moved to type\,1 state with increasing flux \citep{Parker2016,Kollatschny2018,Zetzl2018}. The X-ray flux also increased, with $\lambda_{\rm Edd}\sim 0.02-0.2$ in 2014-2015. HE\,1136--2304 also showed a clear correlation of \ed and the spectral state, as it moved towards type\,1 state with increasing \ed. 

HE\,1136--2304 showed a variable \nh, however, the \nh varied in the range of $\sim 1-5 \times 10^{21}$ \pcm. Hence, the obscuration is not likely the reason for the CL transition in this source. The accretion rate is the most likely responsible for the CL event.

\begin{table*}
\caption{HE\,1136--2304}
\centering
\begin{tabular}{ccccccccccccc}
\hline
Dates & H$\alpha$ & H$\beta$ & Optical & Optical &   $\log L_{\rm X}$  &$\log \lambda_{\rm Edd}$  & $\log N_{\rm H}$ &X--ray & Ref. & $\Delta T$\\
UT    & BEL       & BEL      & type    & Inst.  &  $\log({\rm erg~s^{-1}})$ &       &    $\log({\rm cm}^{-2})$ & Inst.&  &  (Days)   \\
\hline
\hline
1990      &--&--& --  &-- & $41.81\pm0.18$ & $-2.84\pm0.31$ &--& ROSAT & 1, 2 & --\\
1993/03/20& Y& N& 1.9& ESO  &  -- & --    &  --  &--& 3& --\\
2002/05/16& Y& N& 1.9& UKS  &  --    & --  &  -- & --&4 & --\\
\hline
2014/07/02&--&--& -- & --  & $43.23\pm0.01$  & $-1.28\pm0.30$ &  $20.97\pm0.01$ & XMM, Nu & 2 & 5\\
2014/07/07& Y& Y& 1.0& SALT  &  -- & --    &--  & --&5, 6 & --\\
\hline
2014/12/25& Y& Y& 1.0& SALT  &  -- & --    &--  & --&5, 6& --\\
\hline
2015/02/28& Y& Y& 1.0& SALT  &  -- & --    & --  & --&5, 6 & 4\\
2015/03/03&--&--& -- & -- & $43.41\pm0.05$ & $-1.06\pm0.30$ & $21.45\pm0.07$ &Swift & 7 & --\\
\hline
2015/03/06&--&--& -- & -- & $43.62\pm0.05$ & $-0.80\pm0.30$ & $21.25\pm0.23$ &Swift & 7 & --\\
2015/03/07& Y& Y& 1.0& SALT  &  -- & --      & --  & --&5, 6 & --\\
\hline
2015/04/18& Y& Y& 1.0& SALT  &  -- & --      & --  & --&5, 6 & 1\\
2015/04/19&--&--& -- & -- & $43.08\pm0.06$  & $-1.46\pm0.30$ &$21.56\pm0.20$ &Swift & 7 & --\\
\hline
2015/05/24& Y& Y& 1.0& SALT  &   $43.35\pm0.04$ & $-1.13\pm0.30$ &  $21.29\pm0.18$ &Swift & 5, 6& 0\\
\hline
2015/06/07&--&--& -- & -- & $42.93\pm0.08$  & $-1.63\pm0.30$ & $21.59\pm0.15$&Swift & 7 & --\\
2015/06/08& Y& Y& 1.0& SALT  &  -- & --    & --  & --&5, 6 & --\\
\hline
2015/06/13& Y& Y& 1.0& SALT  &  -- & --      & --  & --& 5, 6 & 1\\
2015/06/14&--&--& -- & -- & $42.82\pm0.09$  & $-1.76\pm0.29$ &$21.43\pm0.17$ & Swift & 7 & --\\
\hline
2015/06/21&--&--& -- & -- & $42.86\pm0.08$  & $-1.71\pm0.29$ &$21.34\pm0.19$ & Swift & 7 & 1\\
2015/06/22& Y& Y& 1.0& SALT  &  -- & --      & --  & --&5, 6 & --\\
\hline
2015/06/27& Y& Y& 1.0& SALT  &  -- & --      & --  &  --&5, 6 & 1\\
2015/06/28&--&--& -- & -- & $42.92\pm0.06$  & $-1.65\pm0.30$ & $21.30\pm0.20$&Swift & 7& -- \\
\hline
2015/07/13& Y& Y& 1.0& SALT  &  -- & --     & --  & --&5, 6 & 1\\
2015/07/14&--&--& -- & -- & $43.05\pm0.07$ & $-1.49\pm0.28$& $21.50\pm0.17$ & Swift & 7 & --\\
\hline
\end{tabular}
\leftline{References: (1) \citet{Voges2000}, (2) \citet{Parker2016}, (3) \citet{Reimers1996}, (4) \citet{Koss2017},}
\leftline{(5) \citet{Kollatschny2018}, (6) \citet{Zetzl2018}, (7) This Work. }
\leftline{Optical instrument: ESO: ESO 1.52m/BCS; UKS: UK Schmidt 1.2m (6dFS), SALT: SALT 10m.}
\leftline{X-ray instruments: ROSAT: \emph{ROSAT}/HRI; XMM: \emph{XMM-Newton}/EPIC-PN; Nu: \emph{NuSTAR}/FPMA+FPMB; Swift: \emph{Swift}/XRT.}
\label{tab:he1136}
\end{table*}

\subsection{IRAS\,23226--3843}
\label{subsec:iras23226}

IRAS 23226--3809 was observed in type\,1.9 state in 1997 \citep{Kollatschny2023}. 
It was found in type\,1.9 state in 2005 observation \citep{Koss2017}.
The 2016 observation revealed that the source transitioned to a type\,1 state \citep{Oh2022}, with $\lambda_{\rm Edd}\sim 0.01$. 
In May 2017, IRAS 23226--3809 lost its broad H$\beta$ line as it transitioned to type\,1.9 state \citep{Kollatschny2020}. 
On September 1, 2019, the optical observation revealed that the source was in a type\,1.9 state. With increasing X-ray flux, 
the source returned to the type\,1 state on September 24, 2019. The spectral state change was associated with an optical outburst in the source. 
The source remained in type\,1 state until the end of 2019 \citep{Kollatschny2023}. In May 2020, IRAS 23226--3809 fainted and entered in a low type\,1.9 state \citep{Kollatschny2023}. The X-ray flux was observed to be low in this state; however, we did not have any simultaneous observation to estimate the \ed at this time.

Similar to most CLAGNs in our sample, IRAS 23226--3809 moved towards type\,2 with decreasing \ed. 
The source was found in type\,1.9 state when $\lambda_{\rm Edd} \sim 0.002$. The type\,1 state was observed a higher \ed, 
with $\lambda_{\rm Edd}>0.01 $. In IRAS\,23226--3843, the \nh was found very low, with $ N_{\rm H}<4\times10^{20}$ \pcm. 
This indicates that the accretion rate is the reason for the CL transition in the source.

\begin{table*}
\caption{IRAS\,23226--3843}
\centering
\begin{tabular}{ccccccccccccc}
\hline
Dates & H$\alpha$ & H$\beta$ & Optical & Optical &   $\log L_{\rm X}$  &$\log \lambda_{\rm Edd}$ & $\log N_{\rm H}$ &X--ray & Ref. & $\Delta T$\\
UT    & BEL       & BEL      & type    & Inst.  &  $\log({\rm erg~s^{-1}})$ &       &        $\log({\rm cm}^{-2})$ & Inst.&  &  (Days)   \\
\hline
1997/10/03& Y& N& 1.9& SAAO  &  -- & --      & --   & --& 1& --\\
1999/06/21& Y& N& 1.9& CTIO  &  -- & --      & --   & --& 2& --\\
\hline
2005/08/05& Y& N& 1.9& SAAO  &  -- & --      & --     & --&3& --\\
\hline
2016/07/08&--&--& -- &-- & $42.82\pm0.04$  & $-2.00\pm0.04$  & $20.30^*$&Swift, Nu & 4 & 64 \\
2016/09/09& Y& N& 1.0& duPont  &  -- & --& --   & --  & 5 & --\\
\hline
2017/05/07&--&--& --   &-- & $42.49\pm0.03$  & $-2.35\pm0.04$ & $20.30^*$ & Swift & 4 & --\\
2017/05/10& Y& N& 1.9& SALT  &  -- & --      & --   & --  & 2 & --\\
\hline
2017/06/11&--&--& --  &-- & $42.01\pm0.14$ &$-2.85\pm0.07$  & $20.60\pm0.18$ & Swift, Nu & 2 & 1\\
2017/06/12& Y& N& 1.9& SALT  & -- & --       & --   & -- & 2 & --\\
\hline
2019/09/01& Y& N& 1.9& SAOO  &  -- & --      & --   &  --&1& --\\
2019/09/10& Y& N& 1.9& SALT  &  -- & --     & --    & --&1& 4\\
2019/09/14&--&--& --  &-- & $43.31\pm0.06$ &$-1.44\pm0.04$ & $20.30^*$& Swift & 4 & --\\
2019/09/20&--&--& --  &-- & $43.52\pm0.04$ & $-1.18\pm0.04$ & $20.30^*$& Swift & 4 & 4\\
2019/09/24& Y& Y& 1.0& SALT  &  -- & --      & --   & --& 1& --\\
\hline
2019/11/07& Y& Y& 1.0& SALT  & $43.67\pm0.06$ & $-1.00\pm0.04$ &  $20.30^*$& Swift, Nu & 1 & 0\\
\hline
2020/01/10&--&--& --  &-- & $43.39\pm0.08$ & $-1.34\pm0.05$ & $20.30^*$& Swift & 4 & -- \\
2020/07/23& N& N& 1.9& SAOO  & -- & --       & --   & -- & 1 & --\\
2021/04/08&--&--& --  &-- & $42.92\pm0.06$ & $-1.88\pm0.04$ & $20.30^*$ & Swift & 4 & --\\
\hline
\end{tabular}
\leftline{$^*$ Fitted with Galactic absorption.}
\leftline{References: (1) \citet{Kollatschny2023}, (2) \citet{Kollatschny2020}, (3) \citet{Koss2017}, (4) This Work,}
\leftline{(5) \citet{Oh2022}.  }
\leftline{Optical telescope: CTIO: CTIO 4m/R-C; duPONT: duPONT/B\&C; SAOO : SAOO 1.9m.}
\leftline{X-ray instruments: XMM: \emph{XMM-Newton}/EPIC-PN; Nu: \emph{NuSTAR}/FPMA+FPMB; Swift: \emph{Swift}/XRT.}
\label{tab:iras23226}
\end{table*}



\end{document}